\documentclass[12pt,nohyper,letterpaper]{JHEP3}         
\usepackage{amssymb,amsfonts}

\usepackage{epsfig}

\def\be{\begin{eqnarray}}
\def\ee{\end{eqnarray}}
\newcommand{\nn}{\nonumber}
\newcommand\para{\paragraph{}}
\newcommand{\ft}[2]{{\textstyle\frac{#1}{#2}}}
\newcommand{\eqn}[1]{(\ref{#1})}
\newcommand{\p}{\partial}

\newcommand{\ppp}[2]{\frac{\partial {#1}}{\partial {#2}}}

\newcommand\bomega{\mbox{\boldmath $\omega$}}

\newcommand\bphi{\mbox{\boldmath $\phi$}}

\def\Dslash{\,\,{\raise.15ex\hbox{/}\mkern-12mu D}}
\def\Dbarslash{\,\,{\raise.15ex\hbox{/}\mkern-12mu {\bar D}}}
\def\delslash{\,\,{\raise.15ex\hbox{/}\mkern-9mu \partial}}
\def\delbarslash{\,\,{\raise.15ex\hbox{/}\mkern-9mu {\bar\partial}}}
\def\pslash{\,\,{\raise.15ex\hbox{/}\mkern-9mu p}}
\def\calDslash{\,\,{\raise.15ex\hbox{/}\mkern-12mu {\cal D}}}

\newcommand\Tr{{\rm Tr}}
\newcommand\tr{{\rm tr}}

\newcommand{\N}{{\bf N}}

\newcommand{\Z}{{\bf Z}}

\newcommand{\R}{{\bf R}}

\newcommand{\C}{{\bf C}}
\newcommand{\CC}{{\mathbb C}}
\newcommand{\PP}{{\mathbb P}}

\newcommand{\CP}{{\CC\PP}}
\newcommand{\bra}{\langle}
\newcommand{\ket}{\rangle}

\newcommand{\I}{{\cal I}}
\newcommand{\ikn}{{\cal I}_{k,N}}
\newcommand{\M}{{\cal M}}
\newcommand{\vg}{\vec{g}}
\newcommand{\mg}{{\cal M}_{\vg}}
\newcommand{\V}{{\cal V}}
\newcommand{\vkn}{{\cal V}_{k,N}}
\newcommand{\W}{{\cal W}}
\newcommand{\wg}{{\cal W}_{\vg}}
\newcommand{\D}{{\cal D}}
\renewcommand{\d}{\delta}
\newcommand{\hk}{hyperK\"ahler\ }
\renewcommand{\S}{{\bf S}}
\renewcommand{\N}{{\cal N}}
\newcommand{\diag}{{\rm diag}}
\newcommand{\valpha}{\vec{\alpha}}

\newcommand{\starf}{{}^\star F}

\title{{\huge TASI Lectures on Solitons}\\
{\large Instantons, Monopoles, Vortices and Kinks}}

\author{David Tong\\
Department of Applied Mathematics and Theoretical Physics, \\
Centre for Mathematical Sciences, \\
Wilberforce Road, \\
Cambridge, CB3 OBA, UK \\
\email{d.tong@damtp.cam.ac.uk}
}

\preprint{June 2005}

\abstract{These lectures cover aspects of solitons with focus on
applications to the quantum dynamics of supersymmetric gauge
theories and string theory. The lectures consist of four sections,
each dealing with a different soliton. We start with instantons
and work down in co-dimension to monopoles, vortices and,
eventually, domain walls. Emphasis is placed on the moduli space
of solitons and, in particular, on the web of connections that links
solitons of different types. The D-brane realization of the ADHM
and Nahm construction for instantons and monopoles is reviewed,
together with related constructions for vortices and domain walls.
Each lecture ends with a series of vignettes detailing the roles
solitons play in the quantum dynamics of supersymmetric gauge
theories in various dimensions. This includes applications to the
AdS/CFT correspondence, little string theory, S-duality, cosmic
strings, and the quantitative correspondence between 2d sigma
models and 4d gauge theories.}

\begin{document}

\newpage
\setcounter{section}{-1}
\section{Introduction}
\label{sec.0}

170 years ago, a Scotsman on horseback watched a wave travelling
down Edinburgh's Union canal. He was so impressed that he followed
the wave for several miles, described the day of observation as
the happiest of his life,  and later attempted to recreate the
experience in his own garden. The man's name was John Scott
Russell and he is generally credited as the first person to
develop an unhealthy obsession with the "singular and beautiful
phenomenon" that we now call a soliton.

\para
Russell was ahead of his time. The features of stability and
persistence that so impressed him were not appreciated by his
contemporaries, with Airy arguing that the "great primary wave"
was neither great nor primary\footnote{More background on Russell and his
wave can be found at http://www.ma.hw.ac.uk/ $\tilde{\ }$chris/scott\_russell.html and
http://www-groups.dcs.st-and.ac.uk/$\tilde{\ }$history/Mathematicians/Russell \_Scott.html.}. It wasn't until the following
century that solitons were understood to play an important role in
areas ranging from engineering to biology, from condensed matter
to cosmology.

\para
The purpose of these lectures is to explore the properties of
solitons in gauge theories. There are four leading characters: the
instanton, the monopole, the vortex, and the domain wall (also
known as the kink). Most reviews of solitons start with kinks and
work their way up to the more complicated instantons. Here we're
going to do things backwards and follow the natural path:
instantons are great and primary, other solitons follow. A major
theme of these lectures is to flesh out this claim by describing
the web of inter-relationships connecting our four solitonic
characters.

\para
Each lecture will follow a similar pattern. We start by deriving
the soliton equations and examining the basic features of the
simplest solution. We then move on to discuss the interactions of
multiple solitons, phrased in terms of the moduli space. For each type
of soliton,
D-brane techniques are employed to gain a better understanding of
the relevant geometry. Along the way, we shall discuss various
issues including fermionic zero modes, dyonic excitations and
non-commutative solitons. We shall also see the earlier solitons
reappearing in surprising places, often nestling within the
worldvolume of a larger soliton, with interesting consequences.
Each lecture concludes with a few brief descriptions of the roles
solitons play in supersymmetric gauge theories in various
dimensions.

\para
These notes are aimed at advanced graduate students who have some
previous awareness of solitons. The basics will be covered, but
only very briefly. A useful primer on solitons can be found  in
most modern field theory textbooks (see for example
\cite{weinberg}). More details are contained in the recent book by
Manton and Sutcliffe \cite{mansut}. There are also a number of
good reviews dedicated to solitons of a particular type and these
will be mentioned at the beginning of the relevant lecture. Other
background material that will be required for certain sections
includes a basic knowledge of the structure of supersymmetric
gauge theories and D-brane dynamics. Good reviews of these
subjects can be found in \cite{poldbrane,givkut,cliffprimer}.

\newpage
\section{Instantons}

30 years after the discovery of Yang-Mils instantons \cite{bpst},
they continue to fascinate both physicists and mathematicians
alike. They have lead to new insights into a wide range of
phenomenon, from the structure of the Yang-Mills vacuum
\cite{thooft,jacreb,cdg} to the classification of four-manifolds
\cite{don}. One of the most powerful uses of instantons in recent
years is in the analysis of supersymmetric gauge dynamics where
they play a key role in unravelling the plexus of entangled
dualities that relates different theories. The purpose of this
lecture is to review the classical properties of instantons,
ending with some applications to the quantum dynamics of
supersymmetric gauge theories.

\para
There exist many good reviews on the subject of instantons. The
canonical reference for basics of the subject remains the
beautiful lecture by Coleman \cite{coleman}. More recent
applications to supersymmetric theories are covered in detail in
reviews by Shifman and Vainshtein \cite{sv} and by Dorey,
Hollowood, Khoze and Mattis \cite{dhkm}. This latter review
describes the ADHM construction of instantons and overlaps with
the current lecture.

\subsection{The Basics}

The starting point for our journey is four-dimensional, pure
$SU(N)$ Yang-Mills theory with action\footnote{Conventions: We
pick Hemitian generators $T^m$ with Killing form
$\Tr\,T^mT^n=\ft12\delta^{mn}$. We write $A_\mu=A_\mu^mT^m$ and
$F_{\mu\nu}=\partial_\mu A_\nu-\partial_\nu A_\mu-i[A_\mu,A_\nu]$.
Adjoint covariant derivatives are ${\cal D}_\mu X=\partial_\mu
X-i[A_\mu,X]$. In this section alone we work with Euclidean
signature and indices will wander from top to bottom with
impunity; in the following sections we will return to Minkowski
space with signature $(+,-,-,-)$.}
\be S=\frac{1}{2e^2}\int\,d^4x\
\Tr\,F_{\mu\nu}F^{\mu\nu}\label{lag1}\ee
Motivated by the semi-classical evaluation of the path integral,
we search for finite action solutions to the Euclidean equations
of motion,
\be {\cal D}_\mu F^{\mu\nu}=0 \label{eom1}\ee
which, in the imaginary time formulation of the theory, have the
interpretation of mediating quantum mechanical tunnelling events.

\para
The requirement of  finite action means that the potential $A_\mu$
must become pure gauge as we head towards the boundary
$r\rightarrow\infty$ of spatial $\R^4$,
\be A_\mu\rightarrow ig^{-1}\,\p_\mu g\label{puregauge}\ee
with $g(x)=e^{iT(x)}\in SU(N)$. In this way, any finite action
configuration provides a map from $\partial\R^4\cong {\bf
S}^3_\infty$ into the group $SU(N)$. As is well known, such maps
are classified by homotopy theory. Two maps are said to lie in the
same homotopy class if they can be continuously deformed into each
other, with different classes labelled by the third homotopy
group,
\be \Pi_3(SU(N))\cong\Z \ee
The integer $k\in\Z$ counts how many times the group wraps itself
around spatial ${\bf S}^3_\infty$ and is known as the Pontryagin
number, or second Chern class. We will sometimes speak simply of
the "charge" $k$ of the instanton. It is measured by the surface
integral
\be k=\frac{1}{24\pi^2}\int_{{\bf S}^3_\infty}d^3S_\mu\ \Tr\,
(\partial_\nu g)g^{-1}\,(\partial_\rho g)g^{-1}\,(\partial_\sigma
g)g^{-1}\,\epsilon^{\mu\nu\rho\sigma}\label{winding}\ee
The charge $k$ splits the space of field configurations into
different sectors. Viewing $\R^4$ as a foliation of concentric
${\bf S}^3$'s, the homotopy classification tells us that we cannot
transform a configuration with non-trivial winding $k\neq 0$ at
infinity into one with trivial winding on an interior $\S^3$ while
remaining in the pure gauge ansatz \eqn{puregauge}. Yet, at the
origin, obviously the gauge field must be single valued,
independent of the direction from which we approach. To reconcile
these two facts,  a configuration with $k\neq 0$ cannot remain in
the pure gauge form \eqn{puregauge} throughout all of $\R^4$: it
must have non-zero action.

\subsubsection*{An Example: $SU(2)$}

The simplest case to discuss is the gauge group $SU(2)$ since, as
a manifold, $SU(2)\cong {\bf S}^3$ and it's almost possible to
visualize the fact that $\Pi_3({\bf S}^3)\cong \Z$. (Ok, maybe
${\bf S}^3$ is a bit of a stretch, but it is possible to visualize
$\Pi_1({\bf S}^1)\cong {\bf Z}$ and $\Pi_2({\bf S}^2)\cong\Z$ and
it's not the greatest leap to accept that, in general, $\Pi_n({\bf
S}^n)\cong \Z$). Examples of maps in the different sectors are
\begin{itemize}
\item $g^{(0)}=1$, the identity map has winding $k=0$ \item
$g^{(1)}=(x_4+ix_i\sigma^i)/r$ has winding number $k=1$. Here
$i=1,2,3$, and the $\sigma^i$ are the Pauli matrices \item
$g^{(k)}=[g^{(1)}]^k$ has winding number $k$.
\end{itemize}
To create a non-trivial configuration in $SU(N)$, we could try to
embed the maps above into a suitable $SU(2)$ subgroup, say the
upper left-hand corner of the $N\times N$ matrix. It's not obvious
that if we do this they continue to be a maps with non-trivial
winding since one could envisage that they now have space to slip
off. However, it turns out that this doesn't happen and the above
maps retain their winding number when embedded in higher rank
gauge groups.

\subsubsection{The Instanton Equations}

We have learnt that the space of configurations splits into
different sectors, labelled by their winding $k\in\Z$ at infinity.
The next question we want to ask is whether solutions actually
exist for different $k$. Obviously for $k=0$ the usual vacuum
$A_\mu=0$ (or gauge transformations thereof) is a solution. But
what about higher winding with $k\neq 0$? The first step to
constructing solutions is to derive a new set of equations that
the instantons will obey, equations that are first order rather
than second order as in \eqn{eom1}. The trick for doing this is
usually referred to as the Bogomoln'yi bound  \cite{bog} although,
in the case of instantons, it was actually introduced in the
original  paper \cite{bpst}. From the above considerations, we
have seen that any configuration with $k\neq 0$ must have some
non-zero action. The Bogomoln'yi bound quantifies this. We rewrite
the action by completing the square,
\be S_{\rm inst}&=& \frac{1}{2e^2}\int d^4x\ \Tr\, F_{\mu\nu}F^{\mu\nu} \nn\\
&=&\frac{1}{4e^2}\int d^4x\ \Tr\, (F_{\mu\nu}\mp {}^\star
F^{\mu\nu})^2 \pm 2\Tr\ F_{\mu\nu}{}^\star F^{\mu\nu} \nn\\
&\geq&\pm\frac{1}{2e^2}\int d^4x\ \p_\mu\left(A_\nu F_{\rho\sigma}
+\ft{2i}{3}A_\nu A_\rho A_\sigma\right)\epsilon^{\mu\nu\rho\sigma}
\ee
where the dual field strength is defined as ${}^\star
F_{\mu\nu}=\ft12\epsilon_{\mu\nu\rho\sigma}F^{\rho\sigma}$ and, in
the final line, we've used the fact that $F_{\mu\nu}{}^\star
F^{\mu\nu}$ can be expressed as a total derivative. The final
expression is a surface term which measures some property of the
field configuration on the boundary ${\bf S}^3_\infty$. Inserting
the asymptotic form $A_\nu\rightarrow ig^{-1}\partial_\nu g$ into
the above expression and comparing with \eqn{winding}, we learn
that the action of the instanton in a topological sector $k$ is
bounded by
\be S_{\rm inst}\geq \frac{8\pi^2}{e^2}\,|k|\ee
with equality if and only if
\be F_{\mu\nu}={}^\star F_{\mu\nu}\ \ \ \ \ \ \ &&(k>0)\nn\\
F_{\mu\nu}=-{}^\star F_{\mu\nu}\ \ \ \ \ \ \ &&(k<0)\nn\ee
Since parity maps $k\rightarrow -k$, we can focus on the self-dual
equations $F=\starf$. The Bogomoln'yi argument (which we shall see
several more times in later sections) says that a solution to the
self-duality equations must necessarily solve the full equations
of motion since it minimizes the action in a given topological
sector. In fact, in the case of instantons, it's trivial to see
that this is the case since we have
\be {\cal D}_\mu F^{\mu\nu}={\cal D}_\mu\starf^{\mu\nu}=0 \ee
by the Bianchi identity.

\subsubsection{Collective Coordinates}

So we now know the equations we should be solving to minimize the
action. But do solutions exist? The answer, of course, is yes!
Let's start by giving an example, before we move on to examine
some of its properties, deferring discussion of the general
solutions to the next subsection.

\para
The simplest solution is the $k=1$ instanton in $SU(2)$ gauge
theory. In singular gauge, the connection is given by
\be
A_\mu=\frac{\rho^2(x-X)_\nu}{(x-X)^2((x-X)^2+\rho^2)}\,\bar{\eta}^i_{\mu\nu}\,(g\sigma^i
g^{-1})\label{sol1}\ee
The $\sigma^i$, $i=1,2,3$ are the Pauli matrices and carry the
$su(2)$ Lie algebra indices of $A_\mu$. The $\bar{\eta}^i$ are three
$4\times 4$ anti-self-dual 't Hooft matrices which intertwine the group
structure of the index $i$ with the spacetime structure of the
indices $\mu,\nu$. They are given by
\be \bar{\eta}^1=\scriptsize{\left(\begin{array}{cccc}0 &0 &0 & -1
\\0 & 0& 1 & 0
\\ 0& -1 & 0&0 \\ 1 &0 &0 &0 \end{array}\right)} \ \ \ ,\ \ \ \
\bar{\eta}^2={\left(\begin{array}{cccc}0 &0 &-1 & 0 \\0 & 0& 0 &
-1
\\ 1& 0 & 0&0 \\ 0 &1 &0 &0 \end{array}\right)}\ \ \ \ ,\ \ \ \
\bar{\eta}^3={\left(\begin{array}{cccc}0 &1 &0 & 0 \\-1 & 0& 0 & 0
\\ 0& 0 & 0&-1 \\ 0 &0 &1 &0 \end{array}\right)}
\label{antithooft}\ee
It's a useful exercise to compute the field strength to see how it
inherits its self-duality from the anti-self-duality
of the $\bar{\eta}$ matrices. To
build an anti-self-dual field strength, we need to simply exchange
the $\bar{\eta}$ matrices in \eqn{sol1} for their self-dual counterparts,
\be {\eta}^1=\scriptsize{\left(\begin{array}{cccc}0 &0 &0 & 1
\\0 & 0& 1 & 0
\\ 0& -1 & 0&0 \\ -1 &0 &0 &0 \end{array}\right)} \ \ \ ,\ \ \ \
{\eta}^2={\left(\begin{array}{cccc}0 &0 &-1 & 0 \\0 & 0& 0 & 1
\\ 1& 0 & 0&0 \\ 0 &-1 &0 &0 \end{array}\right)}\ \ \ \ ,\ \ \ \
{\eta}^3={\left(\begin{array}{cccc}0 &1 &0 & 0 \\-1 & 0& 0 & 0
\\ 0& 0 & 0&1 \\ 0 &0 &-1 &0 \end{array}\right)}
\label{thooft}\ee
For our immediate purposes, the most important feature of the
solution \eqn{sol1} is that it is not unique: it contains a number
of parameters. In the context of solitons, these are known as {\it
collective coordinates}. The solution \eqn{sol1} has eight such
parameters. They are of three different types:
\newcounter{bean}
\begin{list}{\roman{bean})}{\usecounter{bean}}
\item 4 translations $X_\mu$: The instanton is an object localized
in $\R^4$, centered around the point $x_\mu=X_\mu$.
\item 1 scale size $\rho$: The interpretation of  $\rho$ as the
size of the instanton can be seen by rescaling $x$ and $X$ in the
above solution to demote $\rho$ to an overall constant.
\item 3 global gauge transformations $g\in SU(2)$: This determines
how the instanton is embedded in the gauge group.
\end{list}
At this point it's worth making several comments about the
solution and its collective coordinates.
\begin{itemize}
\item For the $k=1$ instanton,  each of the collective coordinates
described above is a  Goldstone mode, arising because the
instanton configuration breaks a symmetry of the Lagrangian
\eqn{lag1}. In the case of $X_\mu$ and $g$ it is clear that the
symmetry is translational invariance and $SU(2)$ gauge invariance
respectively. The parameter $\rho$ arises from broken conformal
invariance. It's rather common that all the collective coordinates
of a single soliton are Goldstone modes. It's not true for higher
$k$.

\item The apparent singularity at $x_\mu=X_\mu$ is merely a gauge
artifact (hence the name "singular gauge"). A plot of a gauge
invariant quantity, such as the action density, reveals a smooth
solution. The exception is when the instanton shrinks to zero size
$\rho\rightarrow 0$. This singular configuration is known as the
small instanton. Despite its singular nature, it plays an
important role in computing the contribution to correlation
functions in supersymmetric theories. The small instanton lies at
finite distance in the space of classical field configurations (in
a way which will be made precise in Section 1.2).

 \item You may be surprised that we are counting the gauge modes
 $g$ as physical parameters of the solution. The key point is that they
arise from the {\it global} part of the gauge symmetry, meaning
transformations  that don't die off asymptotically. These are
physical symmetries of the system rather than redundancies. In the
early days of studying instantons the 3 gauge modes weren't
included, but it soon became apparent that many of the nicer
mathematical properties of instantons (for example,
hyperK\"ahlerity of the moduli space) require us to include them,
as do certain physical properties (for example,  dyonic instantons
in five dimensions)

\end{itemize}

%
%

\para
The $SU(2)$ solution \eqn{sol1} has $8$ collective coordinates.
What about $SU(N)$ solutions? Of course, we should keep the $4+1$
translational and scale parameters but we would expect more
orientation parameters telling us how the instanton sits in the
larger $SU(N)$ gauge group. How many? Suppose we embed the above
$SU(2)$ solution in the upper left-hand corner of an $N\times  N$
matrix. We can then rotate this into other embeddings by acting
with $SU(N)$, modulo the stabilizer which leaves the configuration
untouched. We have
\be SU(N)/S[U(N-2)\times U(2)] \ee
where the $U(N-2)$ hits the lower-right-hand corner and doesn't
see our solution, while the $U(2)$ is included in the denominator
since it acts like $g$ in the original solution \eqn{sol1} and we
don't want to overcount. Finally, the notation $S[U(p)\times
U(q)]$ means that we lose the overall central $U(1)\subset
U(p)\times U(q)$. The coset space above has dimension $4N-8$. So,
within the ansatz \eqn{sol1} embedded in $SU(N)$, we see that the
$k=1$ solution has $4N$ collective coordinates. In fact, it turns
out that this is all of them and the solution \eqn{sol1}, suitably
embedded, is the most general $k=1$ solution in an $SU(N)$ gauge
group. But what about solutions with higher $k$? To discuss this,
it's useful to introduce the idea of the moduli space.

\subsection{The Moduli Space}

We now come to one of the most important concepts of these
lectures: the {\it moduli space}. This is defined to be the space
of all solutions to $F=\starf$, modulo gauge transformations,
in a given winding sector $k$ and
gauge group $SU(N)$. Let's denote this space as $\I_{k,N}$. We
will define similar moduli spaces for the other solitons and much
of these lectures will be devoted to understanding the different
roles these moduli spaces play and the relationships between them.

\para
Coordinates on $\I_{k,N}$ are given by the collective coordinates
of the solution. We've seen above that the $k=1$ solution has $4N$
collective coordinates or, in other words, $\dim(\I_{1,N})=4N$.
For higher $k$, the number of collective coordinates can be
determined by index theorem techniques. I won't give all the
details, but will instead simply tell you the answer.
\be \dim(\I_{k,N})=4kN \label{ikn}\ee
This has a very simple interpretation. The charge $k$ instanton
can be thought of as $k$ charge $1$ instantons, each with its own
position, scale, and gauge orientation. When the instantons are
well separated, the solution does indeed look like this. But when
instantons start to overlap, the interpretation of the collective
coordinates can become more subtle.

\para
Strictly speaking, the index theorem which tells us the result
\eqn{ikn} doesn't count the number of collective coordinates, but
rather  related quantities known as {\it zero modes}. It works as
follows. Suppose we have a solution $A_\mu$ satisfying $F=\starf$.
Then we can perturb this solution $A_\mu\rightarrow A_\mu+\delta
A_\mu$ and ask how many other solutions are nearby. We require the
perturbation $\delta A_\mu$ to satisfy the linearized self-duality
equations,
\be \D_\mu\d A_\nu-\D_\nu\d
A_\mu=\epsilon_{\mu\nu\rho\sigma}\D^\rho\d
A^\sigma\label{inlin}\ee
where the covariant derivative $\D_\mu$ is evaluated on the
background solution. Solutions to \eqn{inlin} are called zero
modes. The idea of zero modes is that if we have a general
solution $A_\mu=A_\mu(x_\mu,X^\alpha)$, where $X^\alpha$ denote
all the collective coordinates, then for each collective
coordinate we can define the zero mode $\d_\alpha A_\mu=\partial
A_\mu/\partial X^\alpha$ which will satisfy \eqn{inlin}. In
general however, it is not guaranteed that any zero mode can be
successfully integrated to give a corresponding collective
coordinate. But it will turn out that all the solitons discussed
in these lectures do have this property (at least this is true for
bosonic collective coordinates; there is a subtlety with the
Grassmannian collective coordinates arising from fermions which
we'll come to shortly).

\para
Of course, any local gauge transformation will also solve the
linearized equations \eqn{inlin} so we require a suitable gauge
fixing condition. We'll write each zero mode to include an
infinitesimal gauge transformation $\Omega_\alpha$,
\be\d_\alpha A_\mu=\ppp{A_\mu}{X^\alpha}+{\cal
D}_\mu\Omega_\alpha\label{inomega}\ee
and choose $\Omega_\alpha$ so that $\d_\alpha A_\mu$ is orthogonal
to any other gauge transformation, meaning
\be\int d^4x\ \Tr\,(\d_\alpha A_\mu)\,\D_\mu\eta = 0\ \ \ \ \
\forall\ \eta \ee
which, integrating by parts, gives us our gauge fixing condition
\be \D_\mu\,(\d_\alpha A_\mu) =0 \label{ingf}\ee
This gauge fixing condition does not eliminate the collective
coordinates arising from global gauge transformations which, on an
operational level, gives perhaps the clearest reason why we must
include them. The Atiyah-Singer index theorem counts the number of
solutions to \eqn{inlin} and \eqn{ingf} and gives the answer
\eqn{ikn}.

\para
So what does the most general solution, with its $4kN$ parameters,
look like? The general explicit form  of the solution is not
known. However, there are rather clever ansatz\"e which give rise
to various subsets of the solutions. Details can be found in the
original literature \cite{inssolwit,inssol} but, for now, we head
in a different, and ultimately more important, direction and study
the geometry of the moduli space.

\subsubsection{The Moduli Space Metric}

A priori, it is not obvious that $\ikn$ is a manifold. In fact, it
does turn out to be a smooth space apart from  certain localized
singularities corresponding to small instantons at
$\rho\rightarrow 0$ where the field configuration itself also
becomes singular.

\para
The moduli space $\I_{k,N}$ inherits a natural metric from the
field theory, defined by the overlap of zero modes. In the
coordinates $X^\alpha$, $\alpha=1,\ldots, 4kN$, the metric is
given by
\be g_{\alpha\beta}=\frac{1}{2e^2}\int d^4x\ \Tr\ (\delta_\alpha
A_\mu)\ (\delta_\beta A_\mu)\label{inmetric}\ee
It's hard to overstate the importance of this metric. It distills
the information contained in the solutions to $F=\starf$ into a
more manageable geometric form. It turns out that for many
applications, everything we need to know about the instantons is
contained in the metric $g_{\alpha\beta}$, and this remains true
of similar metrics that we will define for other solitons.
Moreover, it is often much simpler to determine the metric
\eqn{inmetric} than it is to determine the explicit solutions.

\para
The metric has a few rather special properties. Firstly, it
inherits certain isometries from the symmetries of the field
theory. For example, both the $SO(4)$ rotation symmetry of
spacetime and the $SU(N)$ gauge action will descend to give
corresponding isometries of the metric $g_{\alpha\beta}$ on
$\ikn$.

\para
Another important property of the metric \eqn{inmetric} is that it
is  {\it hyperK\"ahler}, meaning that the manifold has reduced
holonomy $Sp(kN)\subset SO(4kN)$. Heuristically, this means that
the manifold admits something akin to a quaternionic
structure\footnote{Warning: there is also something called a
quaternionic manifold which arises in $\N=2$ supergravity theories
\cite{wb} and is different from a hyperK\"ahler manifold. For a
discussion on the relationship see \cite{dewit}.}. More precisely,
a \hk manifold admits three complex structures $J^{i}$, $i=1,2,3$
which obey the relation
\be J^i\,J^j=-\delta^{ij}+\epsilon^{ijk}\,J^k\ee
The simplest example of a \hk manifold is $\R^4$, viewed as the
quaternions. The three complex structures can be taken to be the
anti-self-dual 't Hooft matrices $\bar{\eta}^i$ that we defined in
\eqn{antithooft}, each of which gives a different complex pairing
of $\R^4$. For example, from $\bar{\eta}^3$ we get $z^1=x^1+ix^2$
and $z^2=x^3-ix^4$.
%
%
%
%
%
\para
The instanton moduli space $\ikn$ inherits its complex structures
$J^{i}$ from those of $\R^4$. To see this, note if $\d A_\mu$ is a
zero mode, then we may immediately write down three other zero
modes $\bar{\eta}^i_{\nu\mu}\,\d A_\mu$, each of which satisfy the
equations \eqn{inlin} and \eqn{ingf}. It must be possible to
express these three new zero modes as a linear combination of the
original ones, allowing us to define three matrices $J^i$,
\be \bar{\eta}^i_{\mu\nu}\ \d_\beta A_\nu=(J^i)^{\alpha}_{\
\beta}\ [\d_\alpha A_\mu] \ee
These matrices $J^i$  then descend to three complex structures on
the moduli space $\ikn$ itself which are given by
\be (J^i)^\alpha_{\ \beta}=g^{\alpha\gamma}\ \int d^4x\
\bar{\eta}^i_{\mu\nu}\ \Tr\ \d_\beta A_\mu\ \d_\gamma A_\nu\ee
So far we have shown only that $J^i$ define almost complex
structures. To prove hyperK\"ahlerity, one must also show
integrability which, after some gymnastics, is possible using the
formulae above. A more detailed discussion of the geometry of the
moduli space in this language can be found in \cite{hs,jerome} and
more generally in \cite{as,mac}. For physicists the simplest proof
of hyperK\"ahlerity follows from supersymmetry as we shall review
in section 1.3.

\para
It will prove useful to return briefly to discuss the isometries.
In K\"ahler and \hk manifolds, it's often important to state
whether isometries are compatible with the complex structure $J$.
If the complex structure doesn't change as we move along the
isometry, so that the Lie derivative ${\cal L}_{k}J=0$, with $k$
the Killing vector, then the isometry is said to be {\it
holomorphic}. In the instanton moduli space $\ikn$, the $SU(N)$
gauge group action is tri-holomorphic, meaning it preserves all
three complex structures. Of the $SO(4)\cong SU(2)_L\times
SU(2)_R$ rotational symmetry, one half, $SU(2)_L$, is
tri-holomorphic, while the three complex structures are rotated
under the remaining $SU(2)_R$ symmetry.

\subsubsection{An Example: A Single Instanton in $SU(2)$}

In the following subsection we shall show how to derive metrics on
$\ikn$ using the powerful ADHM technique. But first, to get a
flavor for the ideas, let's take a  more pedestrian route for the
simplest case of a $k=1$ instanton in $SU(2)$. As we saw above,
there are three types of collective coordinates.
\newcounter{sean}
\begin{list}{\roman{sean})}{\usecounter{sean}}
\item The four translational modes are $\d_{(\nu)} A_\mu=\partial
A_\mu/\partial X^\nu+{\cal D}_\mu\Omega_\nu$ where $\Omega_\nu$
must be chosen to satisfy \eqn{ingf}. Using the fact that
$\partial/\partial X^\nu = -\partial/\partial x^\nu$, it is simple
to see that the correct choice of gauge is $\Omega_\nu=A_\nu$, so
that the zero mode is simply given by $\d_\nu A_\mu = F_{\mu\nu}$,
which satisfies the gauge fixing condition by virtue of the
original equations of motion \eqn{eom1}. Computing the overlap of
these translational zero modes then gives
\be \int d^4x\ \Tr\ (\d_{(\nu)}A_\mu\,\d_{(\rho)}A_\mu) = S_{\rm
inst}\ \delta_{\nu\rho}\label{norm1}\ee
\item One can check that the scale zero mode $\d A_\mu = \partial
A_\mu/\partial \rho$ already satisfies the gauge fixing condition
\eqn{ingf} when the solution is taken in singular gauge
\eqn{sol1}. The overlap integral in this case is simple to
perform, yielding
\be \int d^4x\ \Tr\ (\d A_\mu \,\d A_\mu) = 2S_{\rm
inst}\label{norm2}\ee
\item Finally, we have the gauge orientations. These are simply of
the form $\d A_\mu = \D_\mu\Lambda$, but where $\Lambda$ does not
vanish at infinity, so that it corresponds to a global gauge
transformation. In singular gauge it can be checked that the three
$SU(2)$ rotations $\Lambda^i=[(x-X)^2/((x-X)^2+\rho^2)]\sigma^i$
satisfy the gauge fixing constraint. These give rise to an
$SU(2)\cong {\bf S}^3$ component of the moduli space with radius
given by the norm of any one mode, say, $\Lambda^3$
\be \int d^4x\ \Tr\ (\d A_\mu\,\d A_\mu)=2S_{\rm inst}\,\rho^2
\label{norm3}\ee
\end{list}
Note that, unlike the others, this component of the metric depends
on the collective coordinate $\rho$, growing as $\rho^2$. This
dependance means that the ${\bf S}^3$ arising from $SU(2)$ gauge
rotations combines with the $\R^+$ from scale transformations to
form the space $\R^4$. However, there is a discrete subtlety.
Fields in the adjoint representation are left invariant under the
center $Z_2\subset SU(2)$, meaning that the gauge rotations give
rise to $\S^3/\Z_2$ rather than $\S^3$. Putting all this together,
we learn that the moduli space of a single instanton is
\be \I_{1,2} \cong \R^4\times \R^4/\Z_2\ee
where the first factor corresponds to the position of the
instanton, and the second factor determines its scale size and
$SU(2)$ orientation. The normalization of the flat metrics on the
two $\R^4$ factors is given by \eqn{norm1} and \eqn{norm2}. In
this case, the \hk structure on $\I_{1,2}$ comes simply by viewing
each $\R^4\cong\mathbb{H}$, the quaternions. As is clear from our
derivation, the singularity at the origin of the orbifold
$\R^4/\Z_2$ corresponds to the small instanton $\rho\rightarrow
0$.

\subsection{Fermi Zero Modes}

So far we've only concentrated on the pure Yang-Mills theory
\eqn{lag1}. It is natural to wonder about the possibility of other
fields in the theory: could they also have non-trivial solutions
in the background of an instanton, leading to further collective
coordinates? It turns out that this doesn't happen for bosonic
fields (although they do have an important impact if they gain a
vacuum expectation value as we shall review in later sections).
Importantly, the fermions do contribute zero modes.

\para
Consider a single Weyl fermion $\lambda$ transforming in the
adjoint representation of $SU(N)$, with kinetic term
$i\Tr\,\bar{\lambda}\bar{\Dslash}\lambda$. In Euclidean space, we
treat $\lambda$ and $\bar{\lambda}$ as independent variables, a
fact which leads to difficulties in defining a real action. (For
the purposes of this lecture, we simply ignore the issue -  a
summary of the problem and its resolutions can be found in
\cite{dhkm}). The equations of motion are
\be \bar{\Dslash}\lambda \equiv \bar{\sigma}^\mu\D_\mu\lambda=0 \
\ \ \ ,\ \ \ \ \
\Dslash\bar{\lambda}\equiv\sigma^\mu\D_\mu\bar{\lambda}=0\ee
where $\Dslash=\sigma^\mu\D_\mu$ and the $2\times 2$ matrices are
$\sigma^\mu=(\sigma^i,-i1_2)$. In the background of an instanton
$F=\starf$, only $\lambda$ picks up zero modes. $\bar{\lambda}$
has none. This situation is reversed in the background of an
anti-instanton $F=-\starf$. To see that $\bar{\lambda}$ has no
zero modes in the background of an instanton, we look at
\be \bar{\Dslash}\Dslash = \bar{\sigma}^\mu\sigma^\nu
D_\mu\D_\nu=\D^2\,1_2+F^{\mu\nu}\bar{\eta}^i_{\mu\nu}\sigma^i \ee
where $\bar{\eta}^i$ are the anti-self-dual 't Hooft matrices
defined in \eqn{antithooft}. But a self-dual matrix $F_{\mu\nu}$
contracted with  an anti-self-dual matrix $\bar{\eta}_{\mu\nu}$
vanishes, leaving us with $\bar{\Dslash}\Dslash=\D^2$. And the
positive definite operator $\D^2$ has no zero modes. In contrast,
if we try to repeat the calculation for $\lambda$, we find
\be
{\Dslash}\bar{\Dslash}=\D^2\,1_2+F^{\mu\nu}{\eta}^i_{\mu\nu}\sigma^i
\ee
where $\eta^i$ are the self-dual 't Hooft matrices \eqn{thooft}.
Since we cannot express the operator $\Dslash\bar{\Dslash}$ as a
total square, there's a chance that it has zero modes. The index
theorem tells us that each Weyl fermion $\lambda$ picks up $4kN$
zero modes in the background of a charge $k$ instanton. There are
corresponding Grassmann collective coordinates, which we shall
denote as $\chi$, associated to the most general solution for the
gauge field and fermions. But these Grassmann collective
coordinates occasionally have subtle properties. The quick way to
understand this is in terms of supersymmetry. And often the quick
way to understand the full power of supersymmetry is to think in
higher dimensions.

\subsubsection{Dimension Hopping}

It will prove useful to take a quick break in order to make a few
simple remarks about instantons in higher dimensions. So far we've
concentrated on solutions to the self-duality equations in
four-dimensional theories, which are objects localized in
Euclidean spacetime. However, it is a simple matter to embed the
solutions in higher dimensions simply by insisting that all fields
are independent of the new coordinates. For example, in $d=4+1$
dimensional theories one can set $\partial_0=A_0=0$, with the
spatial part of the gauge field satisfying $F=\starf$. Such
configurations have finite energy and the interpretation of
particle like solitons. We shall describe some of their properties
when we come to applications. Similarly, in $d=5+1$, the
instantons are string like objects, while in $d=9+1$, instantons
are five-branes. While this isn't a particularly deep insight,
it's a useful trick to keep in mind when considering the fermionic
zero modes of the soliton in supersymmetric theories as we shall
discuss shortly.

\para
When solitons have a finite dimensional worldvolume, we can
promote the collective coordinates to fields which depend on the
worldvolume directions. These correspond to massless excitations
living on the solitons. For example, allowing the translational
modes to vary along the instanton string simply corresponds to
waves propagating along the string. Again, this simple observation
will become rather powerful when viewed in the context of
supersymmetric theories.

\para
A note on terminology: Originally the term "instanton" referred to
solutions to the self-dual Yang-Mills equations $F=\starf$. (At
least this was true once Physical Review lifted its censorship of
the term!). However, when working with theories in spacetime
dimensions other than four, people often refer to the relevant
finite action configuration as an instanton. For example, kinks in
quantum mechanics are called instantons. Usually this doesn't lead
to any ambiguity but in this review we'll consider a variety of
solitons in a variety of dimensions. I'll try to keep the phrase
"instanton" to refer to (anti)-self-dual Yang-Mills instantons.

\subsubsection{Instantons and Supersymmetry}

Instantons share an intimate relationship with supersymmetry.
Let's consider an instanton in a $d=3+1$ supersymmetric theory
which could be either $\N=1$, $\N=2$ or $\N=4$ super Yang-Mills.
The supersymmetry transformation for any adjoint Weyl fermion
takes the form
\be \d\lambda=F^{\mu\nu}\sigma_\mu\bar{\sigma}_\nu\epsilon \ \ \
,\ \ \
\d\bar{\lambda}=F^{\mu\nu}\bar{\sigma}_\mu\sigma_\nu\bar{\epsilon}\ee
where, again, we treat the infinitesimal supersymmetry parameters
$\epsilon$ and $\bar{\epsilon}$ as independent. But we've seen
above that in the background of a self-dual solution $F=\starf$
the combination $F^{\mu\nu}\bar{\sigma}_\mu\sigma_\nu=0$. This
means that the instanton is annihilated by half of the
supersymmetry transformations $\bar{\epsilon}$, while the other
half, $\epsilon$, turn on the fermions $\lambda$. We say that the
supersymmetries arising from $\epsilon$ are broken by the soliton,
while those arising from $\bar{\epsilon}$ are preserved.
Configurations in supersymmetric theories which are annihilated by
some fraction of the supersymmetries are known as BPS states
(although the term Witten-Olive state would be more appropriate
\cite{wo}).

\para
Both the broken and preserved supersymmetries play an important
role for solitons. The broken ones are the simplest to describe,
for they generate fermion zero modes
$\lambda=F^{\mu\nu}\sigma_\mu\bar{\sigma}_\nu\epsilon$. These
"Goldstino" modes are a subset of the $4kN$ fermion zero modes
that exist for each Weyl fermion $\lambda$. Further modes can also
be generated by acting on the instanton with superconformal
transformations.

\para
The unbroken supersymmetries $\bar{\epsilon}$ play a more
important role: they descend to a supersymmetry on the soliton
worldvolume, pairing up bosonic collective coordinates $X$ with
Grassmannian  collective coordinates $\chi$. There's nothing
surprising here. It's simply the statement that if a symmetry is
preserved in a vacuum (where, in this case, the "vacuum" is the
soliton itself) then all excitations above the vacuum fall into
representations of this symmetry. However, since supersymmetry in
$d=0+0$ dimensions is a little subtle, and the concept of
"excitations above the vacuum" in $d=0+0$ dimensions even more so,
this is one of the places where it will pay to lift the instantons
to higher dimensional objects. For example, instantons in theories
with 8 supercharges (equivalent to $\N=2$ in four dimensions) can
be lifted to instanton strings in six dimensions, which is the
maximum dimension in which Yang-Mills theory with eight
supercharges exists. Similarly, instantons in theories with 16
supercharges (equivalent to $\N=4$ in four dimensions) can be
lifted to instanton five-branes in ten dimensions. Instantons in
$\N=1$ theories are stuck in their four dimensional world.

\para
Considering Yang-Mills instantons as solitons in higher dimensions
allows us to see this relationship between bosonic and fermionic
collective coordinates. Consider exciting a long-wavelength mode
of the soliton in which a bosonic collective coordinate $X$
depends on the worldvolume coordinate of the instanton $s$, so
$X=X(s)$. Then if we hit this configuration with the unbroken
supersymmetry $\bar{\epsilon}$, it will no longer annihilate the
configuration, but will turn on a fermionic mode proportional to
$\partial_s X$. Similarly, any fermionic excitation will be
related to a bosonic excitation.

\para
The observation that the unbroken supersymmetries descend to
supersymmetries on the worldvolume of the soliton saves us a lot
of work in analyzing fermionic zero modes: if we understand the
bosonic collective coordinates and the preserved supersymmetry,
then the fermionic modes pretty much come for free. This includes
some rather subtle interaction terms.

\para
For example, consider instanton five-branes in ten-dimensional
super Yang-Mills. The worldvolume theory must preserve 8 of the 16
supercharges. The only such theory in $5+1$ dimensions is a
sigma-model on a \hk target space \cite{st} which, for instantons,
is the manifold $\ikn$. The Lagrangian is
\be {\cal L}=g_{\alpha\beta}\partial X^\alpha\partial X^\beta +
i\bar{\chi}^\alpha D_{\alpha\beta}\chi^\beta +\ft14
R_{\alpha\beta\gamma\delta} \bar{\chi}^\alpha{\chi}^\beta
\bar{\chi}^\gamma\chi^\delta\label{instleea}\ee
where $\partial$ denotes derivatives along the soliton worldvolume
and the covariant derivative is
$D_{\alpha\beta}=g_{\alpha\beta}\partial +
\Gamma_{\alpha\beta}^\gamma (\partial X_\gamma)$. This is the
slick proof that the instanton moduli space metric must be
hyperK\"ahler: it is dictated by the 8 preserved supercharges.

\para
The final four-fermi term couples the fermionic collective
coordinates to the Riemann tensor. Suppose we now want to go back
down to instantons in four dimensional $\N=4$ super Yang-Mills. We
can simply dimensionally reduce the above action. Since there are
no longer worldvolume directions for the instantons, the first two
terms vanish, but we're left with the term
\be S_{\rm inst}=\ft14 R_{\alpha\beta\gamma\delta}
\bar{\chi}^\alpha{\chi}^\beta
\bar{\chi}^\gamma\chi^\delta\label{riemann}\ee
This term reflects the point we made earlier: zero modes cannot
necessarily be lifted to collective coordinates. Here we see this
phenomenon for fermionic zero modes. Although each such mode
doesn't change the action of the instanton, if we turn on four
Grassmannian collective coordinates at the same time then the
action does increase! One can derive this term without recourse to
supersymmetry but it's a bit of a pain \cite{blum}. The term is
very important in applications of instantons.

\para
Instantons in four-dimensional $\N=2$ theories can be lifted to
instanton strings in six dimensions. The worldvolume theory must
preserve half of the 8 supercharges. There are two such
super-algebras in two dimensions, a non-chiral $(2,2)$ theory and
a chiral $(0,4)$ theory, where the two entries correspond to left
and right moving fermions respectively. By analyzing the fermionic
zero modes one can show that the instanton string preserves
$(0,4)$ supersymmetry. The corresponding sigma-model doesn't
contain the term \eqn{riemann}. (Basically because the
$\bar{\chi}$ zero modes are missing). However, similar terms can
be generated if we also consider fermions in the fundamental
representation.

\para
Finally, instantons in $\N=1$ super Yang-Mills preserve $(0,2)$
supersymmetry on their worldvolume.

\para
In the following sections, we shall pay scant attention to the
fermionic zero modes, simply stating the fraction of supersymmetry
that is preserved in different theories. In many cases this is
sufficient to fix the fermions completely: the beauty of
supersymmetry is that we rarely have to talk about fermions!

\subsection{The ADHM Construction}

In this section we describe a powerful method to solve the
self-dual Yang-Mills equations $F=\starf$ due to Atiyah, Drinfeld,
Hitchin and Manin and known as the ADHM construction \cite{adhm}.
This will also give us a new way to understand the moduli space
$\ikn$ and its metric. The natural place to view the ADHM
construction is twistor space. But, for a physicist, the simplest
place to view the ADHM construction is type II string theory
\cite{doug,doug1,witadhm}. We'll do things the simple way.

\EPSFIGURE{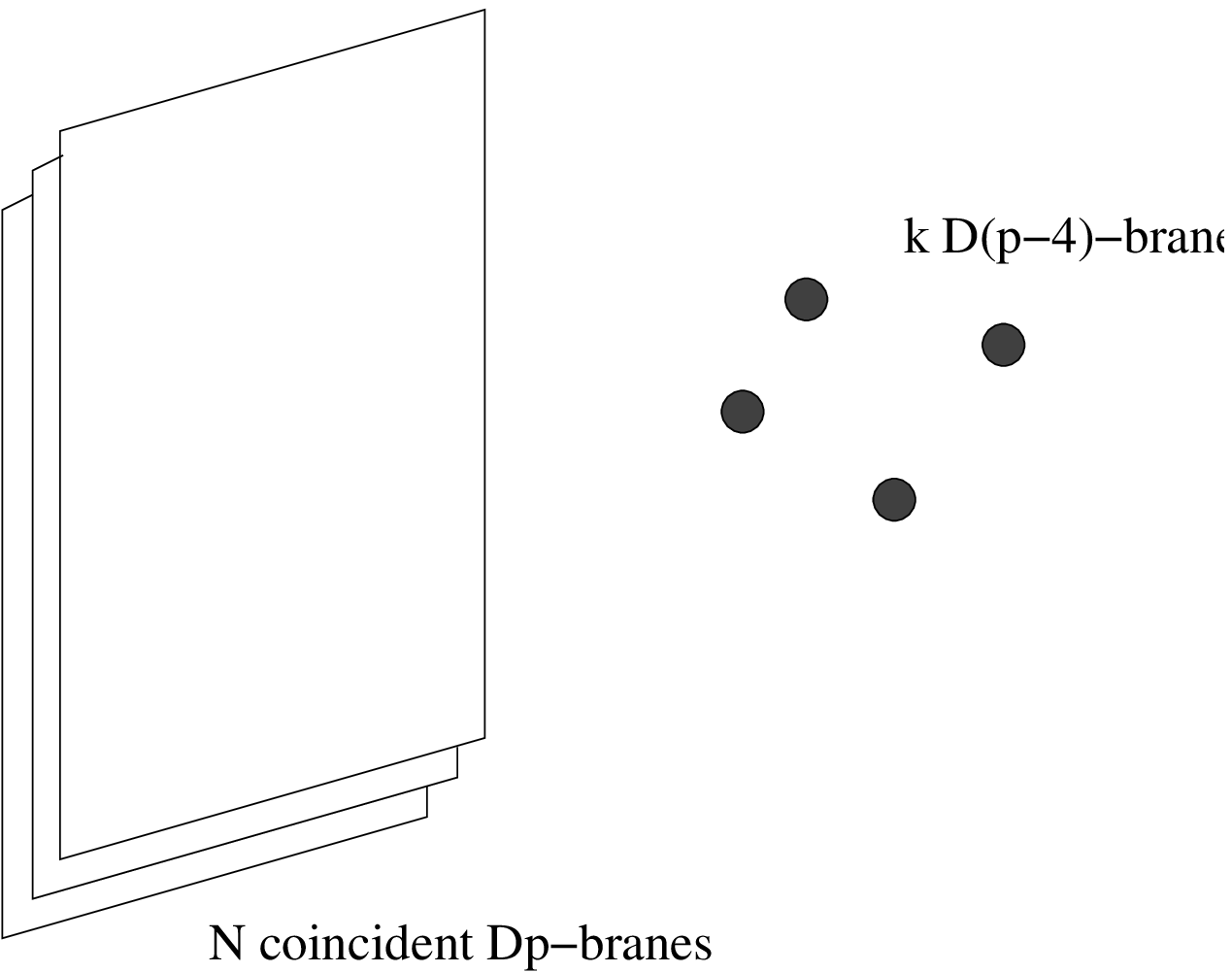,height=140pt}{Dp-branes as instantons.}
\para
The brane construction is another place where it's useful to
consider Yang-Mills instantons embedded as solitons in a $p+1$
dimensional theory with $p\geq 3$. With this in mind, let's
consider a configuration of $N$ D$p$-branes, with $k$
D$(p-4)$-branes in type II string theory (Type IIB for $p$ odd;
type IIA for $p$ even). A typical configuration is drawn in figure
1. We place all $N$ D$p$-branes on top of each other so that, at
low-energies, their worldvolume dynamics is described by
\be\mbox{$d=p+1$ $U(N)$ Super Yang-Mills with 16 Supercharges}
\nn\ee
For example, if $p=3$ we have the familiar ${\cal N}=4$ theory in
$d=3+1$ dimensions. The worldvolume theory of the D$p$-branes also
includes couplings to the various RR-fields in the bulk. This
includes the term
\be \Tr\ \int_{Dp}d^{p+1} x\ \ C_{p-3}\wedge F\wedge F \ee
where $F$ is the $U(N)$ gauge field, and $C_{p-3}$ is the RR-form
that couples to D$(p-4)$-branes. The importance of this term lies
in the fact that it relates instantons on the D$p$-branes to
D$(p-4)$ branes. To see this, note that an instanton with non-zero
$F\wedge F$ gives rise to a source $(8\pi^2/e^2)\int d^{p-3}x\
C_{p-3}$ for the RR-form. This is the same source induced by a
D$(p-4)$-brane. If you're careful in comparing the factors of 2
and $\pi$ and such like, it's not hard to show that the instanton
has precisely the mass and charge of the D$(p-4)$-brane
\cite{poldbrane,cliffprimer}. They are the same object! We have
the important result that
\be \mbox{Instanton in D$p$-Brane\ } \equiv \mbox{\
D$(p-4)$-Brane} \ee
The strategy to derive the ADHM construction from branes is to
view this whole story from the perspective of the D$(p-4)$-branes
\cite{doug,doug1,witadhm}. For definiteness, let's revert back to
$p=3$, so that we're considering D-instantons interacting with
$D3$-branes. This means that we have to write down the $d=0+0$
dimensional theory on the D-instantons. Since supersymmetric
theories in no dimensions may not be very familiar, it will help
to keep in mind that the whole thing can be lifted to higher $p$.

\para
Suppose firstly that we don't have the D3-branes. The theory on
the D-instantons in flat space is simply the dimensional reduction
of $d=3+1$ ${\cal N}=4$ $U(k)$ super Yang-Mills to zero
dimensions. We will focus on the bosonic sector, with the fermions
dictated by supersymmetry as explained in the previous section. We
have $10$ scalar fields, each of which is a $k\times k$ Hermitian
matrix. For later convenience, we split them into two batches:
\be (X^\mu,\hat{X}^m)\ \ \ \ \mu=1,2,3,4; \ \ m=5,\ldots,10
\label{bob}\ee
where we've put hats on directions transverse to the D3-brane.
We'll use the index notation $(X^\mu)^\alpha_{\ \beta}$ to denote
the fact that each of these is a $k\times k$ matrix. Note that
this is a slight abuse of notation since, in the previous section,
$\alpha=1,\ldots,4k$ rather than $1,\ldots,k$ here. We'll also
introduce the complex notation
\be Z=X_1+iX_2\ \ \ \ ,\ \ \ \ W=X_3-iX_4\ee
When $X_\mu$ and $\hat{X}_m$ are all mutually commuting, their
$10k$ eigenvalues have the interpretation of the positions of the
$k$ D-instantons in flat ten-dimensional space.

\para
What effect does the presence of the $D3$-branes have? The answer
is well known. Firstly, they reduce the supersymmetry on the lower
dimensional brane by half, to eight supercharges (equivalent to
${\cal N}=2$ in $d=3+1$). The decomposition \eqn{bob} reflects
this, with the $\hat{X}_m$ lying in a vector multiplet and the
$X_\mu$ forming an adjoint hypermultiplet. The new fields which
reduce the supersymmetry are $N$ hypermultiplets, arising from
quantizing strings stretched between the D$p$-branes and
D$(p-4)$-branes. Each hypermultiplet carries an $\alpha=1,\ldots
k$ index, corresponding to the D$(p-4)$-brane on which the string
ends, and an $a=1,\ldots,N$ index corresponding to the D$p$-brane
on which the other end of the string sits.. Again we ignore
fermions. The two complex scalars in each hypermultiplet are
denoted
\be \psi^\alpha_{\ a}\ \ \ ,\ \ \ \ \tilde{\psi}^a_{\ \alpha} \ee
\EPSFIGURE{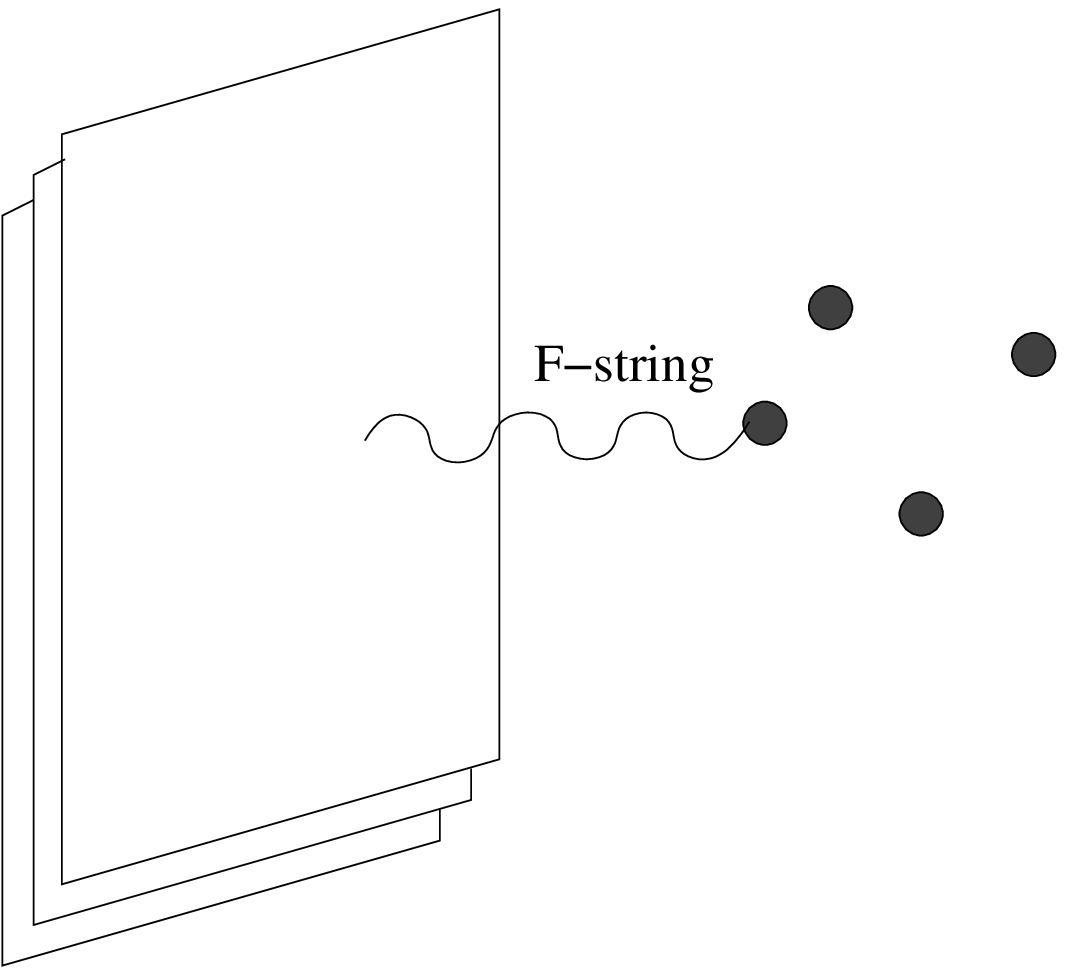,height=140pt}{F-strings give rise to
hypermultiplets.}
\noindent where the index structure reflects the fact that $\psi$
transforms in the ${\bf k}$ of the $U(k)$ gauge symmetry, and the
$\bar{\bf N}$ of a $SU(N)$ flavor symmetry. In contrast
$\tilde{\psi}$ transforms in the $(\bar{\bf k},{\bf N})$ of
$U(k)\times SU(N)$. (One may wonder about the difference between a
gauge and flavor symmetry in zero dimensions; again the reader is
invited to lift the configuration to higher dimensions where such
nasty questions evaporate. But the basic point will be that we
treat configurations related by $U(k)$ transformations as
physically equivalent). These hypermultiplets can be thought of as
the dimensional reduction of ${\cal N}=2$ hypermultiplets in
$d=3+1$ dimensions which, in turn, are composed of two chiral
multiplets $\psi$ and $\tilde{\psi}$.

\para
The scalar potential for these fields is fixed by supersymmetry
(Actually, supersymmetry in $d=0+0$ dimensions is rather weak; at
this stage we should lift up to, say $p=7$, where so we can figure
out the familiar ${\cal N}=2$ theory on the D$(p-3)$=D3-branes,
and then dimensionally reduce back down to zero dimensions). We
have
\be V&=&\frac{1}{g^2} \sum_{m,n=5}^{10}[\hat{X}_m,\hat{X}_n]^2
+\sum_{m=5}^{10}\sum_{\mu=1}^4[\hat{X}_m,X_\mu]^2 +\sum_{a=1}^N
(\psi^{a\dagger}
\hat{X}^2_m\psi_a+\tilde{\psi}^a\hat{X}_m^2\tilde{\psi}^\dagger_a) \label{inpot}\\
&& + g^2\,\Tr\,(\sum_{a=1}^N \psi_a\psi^{a\dagger}-
\tilde{\psi}^\dagger_a\tilde{\psi}^a +[Z,Z^\dagger]
+[W,W^\dagger])^2
+g^2\,\Tr\,|\sum_{a=1}^N\psi_a\tilde{\psi}^a+[Z,W]|^2 \nn\ee
The terms in the second line are usually referred to as D-terms
and F-terms respectively (although, as we shall shall review
shortly, they are actually on the same footing in theories with
eight supercharges). Each is a $k\times k$ matrix. The third term
in the first line ensures that the hypermultiplets get a mass if
the $\hat{X}_m$ get a vacuum expectation value. This reflects the
fact that, as is clear from the picture, the D$p$-D$(p-4)$ strings
become stretched if the branes are separated in the $\hat{X}^m$,
$m=5,\ldots,10$ directions. In contrast, there is no mass for the
hypermultiplets if the D$(p-4)$ branes are separated in the
$X_\mu$, $\mu=1,2,3,4$ directions. Finally, note that we've
included an auxiliary coupling constant $g^2$ in \eqn{inpot}.
Strictly speaking we should take the limit $g^2\rightarrow\infty$.

\para
We are interested in the ground states of the D-instantons,
determined by the solutions to $V=0$. There are two possibilities
\begin{enumerate}
\item The second line vanishes if $\psi=\tilde{\psi}=0$ and
$X_\mu$ are diagonal. The first two terms vanish if $\hat{X}_m$
are also diagonal. The eigenvalues of $X_\mu$ and $\hat{X}_m$ tell
us where the $k$ D-instantons are placed in flat space. They are
unaffected by the existence of the D3-branes whose presence is
only felt at the one-loop level when the hypermultiplets are
integrated out. This is known as the "{\it Coulomb branch}", a
name inherited from the structure of gauge symmetry breaking:
$U(k)\rightarrow U(1)^k$. (The name is, of course, more
appropriate in dimensions higher than zero where particles charged
under $U(1)^k$ experience a Coulomb interaction).

\item The first line vanishes if $\hat{X}_m=0$, $m=5,\ldots, 10$.
This corresponds to the D$(p-4)$ branes lying on top of the
D$p$-branes. The remaining fields $\psi$, $\tilde{\psi}$, $Z$ and
$W$ are constrained by the second line in \eqn{inpot}. Since these
solutions allow  $\psi, \tilde{\psi}\neq 0$ we will generically
have the $U(k)$ gauge group broken completely, giving the name
"{\it Higgs branch}" to this class of solutions. More precisely,
the Higgs branch is defined to be the space of solutions
\be \M_{\rm Higgs}\cong \{\hat{X}_m=0, V=0\}/U(k) \ee
where we divide out by $U(k)$ gauge transformations. The Higgs
branch describes the D$(p-4)$ branes nestling inside the larger
D$p$-branes. But this is exactly where they appear as instantons.
So we might expect that the Higgs branch knows something about
this. Let's start by computing its  dimension. We have $4kN$ real
degrees of freedom in $\psi$ and $\tilde{\psi}$ and a further
$4k^2$ in $Z$ and $W$. The D-term imposes $k^2$ real constraints,
while the F-term imposes $k^2$ complex constraints. Finally we
lose a further $k^2$ degrees of freedom when dividing by $U(k)$
gauge transformations. Adding, subtracting,  we have
\be \dim(\M_{\rm Higgs})=4kN \ee
\end{enumerate}
Which should look familiar \eqn{ikn}. The first claim of the ADHM
construction is that we have an isomorphism between manifolds,
\be\M_{\rm Higgs}\cong \ikn \ee

\subsubsection{The Metric on the Higgs Branch}

To summarize, the D-brane construction has lead us to identify the
instanton moduli space $\ikn$ with the Higgs branch of a gauge
theory with 8 supercharges (equivalent to ${\cal N}=2$ in
$d=3+1$). The field content of this gauge theory is
\be \mbox{$U(k)$ Gauge Theory} &+& \mbox{Adjoint Hypermultiplet
$Z,W$} \nn\\ &+& \mbox{$N$ Fundamental Hypermultiplets
$\psi_a,\tilde{\psi}^a$}\ee
This auxiliary $U(k)$ gauge theory defines its own metric on the
Higgs branch. This metric arises in the following manner: we start
with the flat metric on $\R^{4k(N+k)}$, parameterized by $\psi$,
$\tilde{\psi}$, $Z$ and $W$. Schematically,
\be ds^2=|d\psi|^2 + |d\tilde{\psi}|^2 + |dZ|^2 + |dW|^2 \ee
This metric looks somewhat more natural if we consider higher
dimensional D-branes where it arises from the canonical kinetic
terms for the hypermultiplets. We now pull back this metric to the
hypersurface $V=0$, and subsequently quotient by the $U(k)$ gauge
symmetry, meaning that we only consider tangent vectors to $V=0$
that are orthogonal to the $U(k)$ action. This procedure defines a
metric on $\M_{\rm Higgs}$. The second important result of the
ADHM construction is that this metric coincides with the one
defined in terms of solitons in \eqn{inmetric}.

\para
I haven't included a proof of the equivalence between the metrics
here, although it's not too hard to show (for example, using
Macocia's hyperK\"ahler potential \cite{mac} as reviewed in
\cite{dhkm}). However, we will take time to show that the
isometries of the metrics defined in these two different ways
coincide. From the perspective of the auxiliary $U(k)$ gauge
theory, all isometries appear as flavor symmetries. We have the
$SU(N)$ flavor symmetry rotating the hypermultiplets; this is
identified with the $SU(N)$ gauge symmetry in four dimensions. The
theory also contains an $SU(2)_R$ R-symmetry, in which
$(\psi,\tilde{\psi}^\dagger)$ and $(Z, W^\dagger)$ both transform
as doublets (this will become more apparent in the following
section in equation \eqn{triplet}). This coincides with the
$SU(2)_R \subset SO(4)$ rotational symmetry in four dimensions.
Finally, there exists an independent $SU(2)_L$ symmetry rotating
just the $X_\mu$.

\para
The method described above for constructing \hk metrics is an
example of a technique known as the \hk quotient \cite{hk}. As we
have seen, it arises naturally in gauge theories with 8
supercharges. The D- and F-terms of the potential \eqn{inpot} give
what are called the triplet of "moment-maps" for the $U(k)$
action.

\subsubsection{Constructing the Solutions}

As presented so far, the ADHM construction relates the moduli
space of instantons $\ikn$ to the Higgs branch of an auxiliary
gauge theory. In fact, we've omitted the most impressive part of
the story: the construction can also be used to give solutions to
the self-duality equations. What's more, it's really very easy!
Just a question of multiplying a few matrices together. Let's see
how it works.

\para
Firstly, we need to rewrite the vacuum conditions in a more
symmetric fashion. Define
\be \omega_{a}=\left(\begin{array}{c}\psi^\alpha_{\ a} \\
\tilde{\psi}^{\dagger\alpha}_{\ \
a}\end{array}\right)\label{omega}\ee
Then the real D-term and complex F-term which lie in the second
line of \eqn{inpot} and define the Higgs branch can be combined in
to the triplet of constraints,
\be
\sum_{a=1}^N\omega^{\dagger}_{a}\sigma^i\,\omega_{a}-i[X_\mu,X_\mu]
\bar{\eta}^i_{\mu\nu}=0\label{triplet}\ee
where $\sigma^i$ are, as usual, the Pauli matrices and
$\bar{\eta}^i$ the 't Hooft matrices \eqn{antithooft}. These give
three $k\times k$ matrix equations. The magic of the ADHM
construction is that for each solution to the algebraic equations
\eqn{triplet}, we can build a solution to the set of non-linear
partial differential equations $F=\starf$. Moreover, solutions to
\eqn{triplet} related by $U(k)$ gauge transformations give rise to
the same field configuration in four dimensions. Let's see how
this remarkable result is achieved.

\para
The first step is to build the $(N+2k)\times 2k$ matrix $\Delta$,
\be \Delta = \left(\begin{array}{c} \omega^T \\
X_\mu\sigma^\mu\end{array}\right)+\left(\begin{array}{c} 0 \\
x_\mu\sigma^\mu\end{array}\right)\ee
where $\sigma_\mu=(\sigma^i,-i{\bf 1}_2)$. These have the
important property that $\sigma_{[\mu}\bar{\sigma}_{\nu]}$ is
self-dual, while $\bar{\sigma}_{[\mu}\sigma_{\nu]}$ is
anti-self-dual, facts that we also used in Section 1.3 when
discussing  fermions. In the second matrix we've re-introduced the
spacetime coordinate $x_\mu$ which, here, is to be thought of as
multiplying the $k\times k$ unit matrix. Before proceeding, we
need a quick lemma:
\\{}\\
\noindent {\bf Lemma:} $\Delta^\dagger\Delta = f^{-1}\otimes 1_2$
\\{}\\
 where
$f$ is a $k\times k$ matrix, and $1_2$ is the unit $2\times 2$
matrix. In other words, $\Delta^\dagger\Delta$ factorizes and is
invertible.
\\{}\\
\noindent {\bf Proof:} Expanding out, we have (suppressing the
various indices)
\be \Delta^\dagger\Delta=\omega^\dagger\omega +X^\dagger X+
(X^\dagger x+x^\dagger X)+x^\dagger x 1_k\label{dddagger}\ee
Since the factorization happens for all $x\equiv x_\mu\sigma^\mu$,
we can look at three terms separately. The last is $x^\dagger
x=x_\mu\bar{\sigma}^\mu x_\nu\sigma^\nu=x^2\,1_2$. So that works.
For the term linear in $x$, we simply need the fact that
$X_\mu=X_\mu^\dagger$ to see that it works. What's more tricky is
the term that doesn't depend on $x$. This is where the triplet of
D-terms \eqn{triplet} comes in. Let's write the relevant term from
\eqn{dddagger} with all the indices, including an $m,n=1,2$ index
to denote the two components we introduced in \eqn{omega}. We
require
\be && \omega^{\dagger\alpha}_{m a}\,\omega_{a \beta
n}+(X_\mu)^\alpha_{\ \gamma}(X_\nu)^\gamma_{\
\beta}\,\bar{\sigma}^{\mu\, mp}\sigma^\nu_{\ pn} \sim \delta^m_n
\label{consterms}\\ &\Leftrightarrow\ \ \ \ \ & \tr_2\
\sigma^i\left[\omega\omega^\dagger +X^\dagger X\right] = 0\ \ \ \
i=1,2,3 \nn\\
&\Leftrightarrow\ \ \ \ \ & \omega^\dagger\sigma^i\omega + X_\mu
X_\nu \bar{\sigma}^\mu\sigma^i\sigma^\nu =0 \nn\ee
But, using the identity
$\bar{\sigma}^\mu\sigma^i\sigma^\nu=2i\bar{\eta}^i_{\mu\nu}$, we
see that this last condition is simply the D-terms \eqn{triplet}.
This concludes our proof of the lemma.\hfill$\Box$

\para
The rest is now plain sailing. Consider the matrix $\Delta$ as
defining  $2k$ linearly independent vectors in $\C^{N+2k}$. We
define $U$ to be the $(N+2k)\times N$ matrix containing the $N$
normalized, orthogonal vectors. i.e
\be \Delta^\dagger U =0\ \ \ \ ,\ \ \ \ U^\dagger U =1_N\ee
Then the potential for a charge $k$ instanton in $SU(N)$ gauge
theory is given by
\be A_\mu = i U^\dagger\, \partial_\mu U \ee
Note firstly that if $U$ were an $N\times N$ matrix, this would be
pure gauge. But it's not, and it's not. Note also that $A_\mu$ is
left unchanged by auxiliary $U(k)$ gauge transformations.

\para
We need to show that $A_\mu$ so defined gives rise to a self-dual
field strength with winding number $k$. We'll do the former, but
the latter isn't hard either: it just requires more matrix
multiplication. To help us in this, it will be useful to construct
the projection operator $P=UU^\dagger$ and notice that this can
also be written as $P=1-\Delta f \Delta^\dagger$. To see that
these expression indeed coincide, we can check that $PU=U$ and
$P\Delta=0$ for both. Now we're almost there:
\be F_{\mu\nu}&=&\partial_{[\mu}A_{\nu]}-iA_{[\mu}A_{\nu]} \nn\\
&=&
\partial_{[\mu}\,iU^\dagger\partial_{\nu]}U+iU^\dagger(\partial_{[\mu}U)
U^\dagger(\partial_{\nu]}U) \nn\\ &=& i(\partial_{[\mu}U^\dagger)
(\partial_{\nu]}U)-i(\partial_{[\mu}U^\dagger)UU^\dagger(\partial_{\nu]}U)
\nn\\&=&
i(\partial_{[\mu}U^\dagger)(1-UU^\dagger)(\partial_{\nu]}U)\nn\\
&=&
i(\partial_{[\mu}U^\dagger)\Delta\,f\,\Delta^\dagger(\partial_{\nu]}U)
\nn\\ &=&
iU^\dagger(\partial_{[\mu}\Delta)\,f\,(\partial_{\nu]}U)\nn\\ &=&
iU^\dagger\sigma_{[\mu}f\bar{\sigma}_{\nu]}U \nn\ee
At this point we use our lemma. Because $\Delta^\dagger\Delta$
factorizes, we may commute $f$ past $\sigma_\mu$. And that's it!
We can then write
\be F_{\mu\nu}=iU^\dagger f\sigma_{[\mu}\bar{\sigma}_{\nu]}U =
\starf_{\mu\nu}\ee
since, as we mentioned above,
$\sigma_{\mu\nu}=\sigma_{[\mu}\bar{\sigma}_{\nu]}$ is self-dual.
Nice huh! What's harder to show is that the ADHM construction
gives all solutions to the self-dualily equations. Counting
parameters, we see that we have the right number and it turns out
that we can indeed get all solutions in this manner.

\para
The construction described above was first described in ADHM's
original paper, which weighs in at a whopping 2 pages.
Elaborations and extensions to include, among other things,
$SO(N)$ and $Sp(N)$ gauge groups, fermionic zero modes,
supersymmetry and constrained instantons, can be found in
\cite{christ,corrigan,osborn,dkm}.

\subsubsection*{An Example: The Single $SU(2)$ Instanton
Revisited}

Let's see how to re-derive the $k=1$ $SU(2)$ solution \eqn{sol1}
from the ADHM method. We'll set $X_\mu=0$ to get a solution
centered around the origin. We then have the $4\times 2$ matrix
\be \Delta=\left(\begin{array}{c} \omega^T \\
x_\mu\sigma^\mu\end{array}\right) \ee
where the D-term constraints \eqn{triplet} tell us that
$\omega^{\dagger a}_{\ \ m}(\sigma^{i})^m_{\ n}\omega^n_{\ b}=0$.
We can use our $SU(2)$ flavor rotation, acting on the indices
$a,b=1,2$, to choose the solution
\be \omega^{\dagger a}_{\ \ m}\omega^m_{\ b}=\rho^2\delta^a_{\
b}\label{omerho}\ee
in which case the matrix $\Delta$ becomes $\Delta^T=(\rho 1_2,
x_\mu\sigma^\mu)$. Then solving for the normalized zero
eigenvectors $\Delta^\dagger U=0$, and $U^\dagger U=1$, we have
\be U=\left(\begin{array}{c} \sqrt{x^2/(x^2+\rho^2)}\,1_2 \\
-\sqrt{\rho^2/x^2(x^2+\rho^2)}\,x_\mu\bar{\sigma}^\mu
\end{array}\right) \ee
From which we calculate
\be A_\mu=iU^\dagger\partial_{\mu} U =
\frac{\rho^2x_\nu}{x^2(x^2+\rho^2)}\,\bar{\eta}^i_{\mu\nu}\,\sigma^i \ee
which is indeed the solution \eqn{sol1} as promised.

\subsubsection{Non-Commutative Instantons}

There's an interesting deformation of the ADHM construction
arising from studying instantons on a non-commutative space,
defined by
\be [x_\mu,x_\nu]=i\theta_{\mu\nu}\ee
The most simple realization of this deformation arises by
considering functions on the space $\R^4_\theta$, with
multiplication given by the $\star$-product
\be f(x)\star g(x)=\exp\left.\left(\frac{i}{2} \theta_{\mu\nu}
\frac{\partial}{\partial y^\mu}\frac{\partial}{\partial
x^\nu}\right) f(y)g(x)\right|_{x=y}\ee
so that we indeed recover the commutator $x_\mu\star x_\nu -
x_\nu\star x_\mu = i\theta_{\mu\nu}$. To define gauge theories on
such a non-commutative space, one must extend the gauge symmetry
from $SU(N)$ to $U(N)$. When studying instantons, it is also
useful to decompose the non-commutivity parameter into self-dual
and anti-self-dual pieces:
\be \theta_{\mu\nu}=
\xi^i\,\eta^i_{\mu\nu}+\zeta^i\,\bar{\eta}^i_{\mu\nu}\ee
where $\eta^i$ and $\bar{\eta}^i$ are defined in \eqn{thooft} and
\eqn{antithooft} respectively. At the level of solutions, both
$\xi$ and $\zeta$ affect the configuration. However, at the level
of the moduli space, we shall see that the self-dual instantons
$F=\starf$ are only affected by the anti-self-dual part of the
non-commutivity, namely $\zeta^i$. (A similar statement holds for
$F=-\starf$ solutions and $\xi$). This change to the moduli space
appears in a beautifully simple fashion in the ADHM construction:
we need only add a constant term to the right hand-side of the
constraints \eqn{triplet}, which now read
\be
\sum_{a=1}^N\omega^{\dagger}_{a}\sigma^i\,\omega_a-i[X_\mu,X_\mu]
\bar{\eta}^i_{\mu\nu}=\zeta^i\,1_k\label{noncom}\ee
From the perspective of the auxiliary $U(k)$ gauge theory, the
$\zeta^i$ are Fayet-Iliopoulous (FI) parameters.

\para
The observation that the FI parameters $\zeta^i$ appearing in the
D-term give the correct deformation for non-commutative instantons
is due to Nekrasov and Schwarz \cite{ns}. To see how this works,
we can repeat the calculation above, now in non-commutative space.
The key point in constructing the solutions is once again the
requirement that we have the factorization
\be \Delta^\dagger\star\Delta=f^{-1}\,1_2\ee
The one small difference from the previous derivation is that in
the expansion \eqn{dddagger}, the $\star$-product means we have
\be x^\dagger\star x=x^2\,1_2-\zeta^i\sigma^i\ee
Notice that only the anti-self-dual part contributes. This extra
term combines with the constant terms \eqn{consterms} to give the
necessary factorization if the D-term with FI parameters
\eqn{noncom} is satisfied. It is simple to check that the rest of
the derivation proceeds as before, with $\star$-products in the
place of the usual commutative multiplication.

\para
The addition of the FI parameters in \eqn{noncom} have an
important effect on the moduli space $\ikn$: they resolve the
small instanton singularities. From the ADHM perspective, these
arise when $\psi=\tilde{\psi}=0$, where the $U(k)$ gauge symmetry
does not act freely. The FI parameters remove these points from
the moduli space, $U(k)$ acts freely everywhere on the Higgs
branch, and the deformed instanton moduli space $\ikn$ is smooth.
This resolution of the instanton moduli space was considered by
Nakajima some years before the relationship to non-commutivity was
known \cite{nak}. A related fact is that non-commutative
instantons occur even for $U(1)$ gauge theories. Previously such
solutions were always singular, but the addition of the FI
parameter stabilizes them at a fixed size of order
$\sqrt{\theta}$. Reviews of instantons and other solitons on
non-commutative spaces can be found in \cite{nek,ncharv}.

\subsubsection{Examples of Instanton Moduli Spaces}

\subsubsection*{A Single Instanton}

Consider a single $k=1$ instanton in a $U(N)$ gauge theory, with
non-commutivity turned on. Let us choose $\theta_{\mu\nu}=\zeta
\bar{\eta}^3_{\mu\nu}$. Then the ADHM gauge theory consists of a
$U(1)$ gauge theory with $N$ charged hypermultiplets, and a
decoupled neutral hypermultiplet parameterizing the center of the
instanton. The D-term constraints read
\be \sum_{a=1}^N|\psi_a|^2-|\tilde{\psi}_a|^2=\zeta\ \ \ \ ,\ \ \
\ \sum_{a=1}^N\tilde{\psi}_a\psi_a&=&0\ee
To get the moduli space we must also divide out by the $U(1)$
action $\psi_a\rightarrow e^{i\alpha}\psi_a$ and
$\tilde{\psi}_a\rightarrow e^{-i\alpha}\tilde{\psi}_a$. To see
what the resulting space is, first consider setting
$\tilde{\psi}_a=0$. Then we have the space
\be \sum_{a=1}^{N}|\psi_a|^2=\zeta\ee
which is simply ${\bf S}^{2N-1}$. Dividing out by the $U(1)$
action then gives us the complex projective space $\CP^{N-1}$ with
size (or K\"ahler class) $\zeta$. Now let's add the $\tilde{\psi}$
back. We can turn them on but the F-term insists that they lie
orthogonal to $\psi$, thus defining the co-tangent bundle of
$\CP^{N-1}$, denoted $T^\star\CP^{N-1}$. Including the decoupled
$\R^4$, we have \cite{swd1d5}
\be \I_{1,N}\cong\R^4\times T^\star\CP^{N-1} \ee
where the size of the zero section $\CP^{N-1}$ is $\zeta$. As
$\zeta\rightarrow 0$, this cycle lying in the center of the space
shrinks and $\I_{1,N}$ becomes singular at this point.

\para
For a single instanton in $U(2)$, the relative moduli space is
$T^\star{\bf S}^2$. This is the smooth resolution of the $A_1$
singularity $\C^2/\Z_2$ which we found to be the moduli space in
the absence of non-commutivity. It inherits a well-known
hyperK\"ahler metric known as the Eguchi-Hanson metric \cite{eh},
\be ds^2_{EH}=\left(1-{4\zeta^2}/{\rho^4}\right)^{-1}d\rho^2 +
\frac{\rho^2}{4}\left(\sigma_1^2+\sigma_2^2+\left(1-{4\zeta^2}/{\rho^4}\right)
\sigma_3^2\right)\label{eh}\ee
Here the $\sigma_i$ are the three left-invariant $SU(2)$ one-forms
which, in terms of polar angles $0\leq\theta\leq \pi$,
$0\leq\phi\leq 2\pi$ and $0\leq \psi\leq 2\pi$, take the form
\be \sigma_1&=&-\sin\psi\ d\theta+ \cos\psi\sin\theta\ d\phi\nn\\
\sigma_2&=&\cos\psi\ d\theta + \sin\psi\sin\theta\ d\phi \nn\\
\sigma_3&=&d\psi + \cos\theta\ d\phi\ee
As $\rho\rightarrow\infty$, this metric tends towards the cone
over ${\bf S}^3/{\bf Z}_2$. However, as we approach the origin,
the scale size is truncated at $\rho^2=2\zeta$, where the apparent
singularity is merely due to the choice of coordinates and hides
the zero section ${\bf S}^2$.

\subsubsection*{Two $U(1)$ Instantons}

Before resolving by a non-commutative deformation, there is no
topology to support a $U(1)$ instanton. However, it is perhaps
better to think of the $U(1)$ theory as admitting small, singular,
instantons with moduli space given by the symmetric product ${\rm
Sym}^k(\C^{\, 2})$, describing the positions of $k$ points. Upon
the addition of a non-commutivity parameter, smooth $U(1)$
instantons exist with moduli space given by a resolution of ${\rm
Sym}^k(\C^{\, 2})$. To my knowledge, no explicit metric is known
for $k\geq 3$ $U(1)$ instantons, but in the case of two $U(1)$
instantons, the metric is something rather familiar, since ${\rm
Sym}^2{\C^2}\cong \C^2\times \C^2/Z_2$ and we have already met the
resolution of this space above. It is
\be \I_{k=2,N=1}\cong \R^4\times T^\star\S^2 \ee
endowed with the Eguchi-Hanson metric \eqn{eh} where $\rho$ now
has the interpretation of the separation of two instantons rather
than the scale size of one. This can be checked explicitly by
computing the metric on the ADHM Higgs branch using the \hk
quotient technique \cite{mekim}. Scattering of these instantons was
studied in \cite{begohta}. So, in this particular case we
have $\I_{1,2}\cong \I_{2,1}$. We shouldn't get carried away
though as this equivalence doesn't hold for higher $k$ and $N$
(for example, the isometries of the two spaces are different).

\subsection{Applications}

Until now we've focussed exclusively on classical aspects of the
instanton configurations. But, what we're really interested in is
the role they play in various quantum field theories. Here we
sketch a two examples which reveal the importance of instantons in
different dimensions.

\subsubsection{Instantons and the AdS/CFT Correspondence}

We start by considering instantons where they were meant to be: in
four dimensional gauge theories. In a semi-classical regime,
instantons give rise to  non-perturbative contributions to
correlation functions and there exists a host of results in the
literature, including exact results in both ${\cal N}=1$
\cite{ads,nsvz} and ${\cal N}=2$ \cite{finpol,dkm,nek}
supersymmetric gauge theories. Here we describe the role
instantons play in $\N=4$ super Yang-Mills and, in particular,
their relationship to the AdS/CFT correspondence \cite{adscft}.
Instantons were first considered in this context in
\cite{banksgreen,colour}. Below we provide only a sketchy
description of the material covered in the paper of Dorey et al
\cite{dhkmv}. Full details can be found in that paper or in the
review \cite{dhkm}.

\para
In any instanton computation, there's a number of things we need
to calculate \cite{thooft}. The first is to count the zero modes
of the instanton to determine both the bosonic collective
coordinates $X$ and their fermionic counterparts $\chi$. We've
described this in detail above. The next step is to perform the
leading order Gaussian integral over all modes in the path
integral. The massive (i.e. non-zero) modes around the background
of the instanton leads to the usual determinant operators which
we'll denote as $\det\Delta_B$ for the bosons, and $\det\Delta_F$
for the fermions. These are to be evaluated on the background of
the instanton solution. However, zero modes must be treated
separately. The integration over the associated collective
coordinates is left unperformed, at the price of introducing a
Jacobian arising from the transformation between field variables
and collective coordinates. For the bosonic fields, the Jacobian
is simply $J_B=\sqrt{\det\,g_{\alpha\beta}}$, where
$g_{\alpha\beta}$ is the metric on the instanton moduli space
defined in \eqn{inmetric}. This is the role played by the
instanton moduli space metric in four dimensions: it appears in
the measure when performing the path integral. A related factor
$J_F$ occurs for fermionic zero modes. The final ingredient in an
instanton calculation is the action $S_{\rm inst}$ which includes
both the constant piece $8\pi k/g^2$, together with terms quartic
in the fermions \eqn{riemann}. The end result is summarized in the
instanton measure
\be d\mu_{\rm inst}= d^{n_B}X\,d^{n_F}\chi\ J_BJ_F\ \frac{{\rm det}\Delta_F}
{{\rm det}^{1/2}\Delta_B}\,e^{-S_{\rm inst}}\ee
where there are $n_B=4kN$ bosonic and $n_F$ fermionic collective
coordinates. In supersymmetric theories in four dimensions, the
determinants famously cancel \cite{thooft} and we're left only
with the challenge of evaluating the Jacobians and the action. In
this section, we'll sketch how to calculate these objects for
${\cal N}=4$ super Yang-Mills.

\para
As is well known, in the limit of strong 't Hooft coupling, $\N=4$
super Yang-Mills is dual to type IIB supergravity on $AdS_5\times
\S^5$. An astonishing fact, which we shall now show, is that we
can see this geometry even at weak 't Hooft coupling by studying
the $d=0+0$ ADHM gauge theory describing instantons. Essentially,
in the large $N$ limit,  the instantons live in $AdS_5\times
\S^5$. At first glance this looks rather unlikely! We've seen that
if the instantons live anywhere it is in $\ikn$, a $4kN$
dimensional space that doesn't look anything like $AdS_5\times
{\bf S}^5$. So how does it work?

\para
While the calculation can be performed for an arbitrary number of
$k$ instantons, here we'll just stick with a single instanton as a
probe of the geometry. To see the $AdS_5$ part is pretty easy and,
in fact, we can do it even for an instanton in $SU(2)$ gauge
theory. The trick is to integrate over the orientation modes of
the instanton, leaving us with a five-dimensional space
parameterized by $X_\mu$ and $\rho$. The rationale for doing this
is that if we want to compute gauge invariant correlation
functions, the $SU(N)$ orientation modes will only give an overall
normalization. We calculated the metric for a single instanton in
equations \eqn{norm1}-\eqn{norm3}, giving us $J_B\sim \rho^3$
(where we've dropped some numerical factors and factors of $e^2$).
So integrating over the $SU(2)$ orientation to pick up an overall
volume factor, we get the bosonic measure for the instanton to be
\be d\mu_{\rm inst}\sim{\rho^3}\,d^4Xd\rho\ee
We want to interpret this measure as a five-dimensional space in
which the instanton moves, which means thinking of it in the form
$d\mu=\sqrt{G}\,d^4X\,d\rho$ where $G$ is the metric on the
five-dimensional space. It would be nice if it was the metric on
$AdS_5$. But it's not! In the appropriate coordinates, the $AdS_5$
metric is,
\be ds^2_{AdS}=\frac{R^2}{\rho^2}(d^4X+d\rho^2)\ee
giving rise to a measure $d\mu_{AdS}=(R/\rho)^5 d^4Xd\rho$.
However, we haven't finished with the instanton yet since we still
have to consider the fermionic zero modes. The fermions are
crucial for quantum conformal invariance so we may suspect that
their zero modes are equally crucial in revealing the $AdS$
structure, and this is indeed the case. A single $k=1$ instanton
in the ${\cal N}=4$ $SU(2)$ gauge theory has 16 fermionic zero
modes. 8 of these, which we'll denote as $\xi$ are from broken
supersymmetry while the remaining 8, which we'll call $\zeta$
arise from broken superconformal transformations. Explicitly each
of the four Weyl fermions $\lambda$ of the theory has a profile,
\be \lambda =
\sigma^{\mu\nu}F_{\mu\nu}(\xi-\sigma^\rho\zeta\,(x_\rho-X_\rho))\ee
One can compute the overlap of these fermionic zero modes in the
same way as we did for bosons. Suppressing indices, we have
\be \int d^4x\ \frac{\partial\lambda}{\partial\xi}\frac{\partial
\lambda}{\partial \xi}=\frac{32\pi^2}{e^2}\ \ \ , \ \ \ \int d^4x\
\frac{\partial\lambda}{\partial\zeta}\frac{\partial
\lambda}{\partial \zeta}=\frac{64\pi^2\rho^2}{e^2} \ee
So, recalling that Grassmannian integration is more like
differentiation, the fermionic Jacobian is $J_F\sim 1/\rho^8$.
Combining this with the bosonic contribution above, the final
instanton measure is
\be d\mu_{\rm inst}= \left(\frac{1}{\rho^5}\,d^4X d\rho\right)\,
d^8\xi d^8\zeta=d\mu_{Ads}\,d^8\xi d^8\zeta\ee
So the bosonic part does now look like $AdS_5$. The presence of
the 16 Grassmannian variables reflects the fact that the instanton
only contributes to a 16 fermion correlation function. The
counterpart in the AdS/CFT correspondence is that D-instantons
contribute to $R^4$ terms and their 16 fermion superpartners and
one can match the supergravity and gauge theory correlators
exactly.

\para
So we see how to get $AdS_5$ for $SU(2)$ gauge theory. For
$SU(N)$, one has  $4N-8$ further orientation modes and  $8N-16$
further fermi zero modes. The factors of $\rho$ cancel in their
Jacobians, leaving the $AdS_5$ interpretation intact. But there's
a problem with these extra fermionic zero modes since we must
saturate them in the path integral in some way even though we
still want to compute a 16 fermionic correlator. This is achieved
by the four-fermi term in the instanton action \eqn{riemann}.
However, when looked at in the right way, in the large $N$ limit
these extra fermionic zero modes will generate the ${\bf S}^5$ for
us. I'll now sketch how this occurs.

\para
The important step in reforming these fermionic zero modes is to
introduce auxiliary variables $\hat{X}$ which allows us to split
up the four-fermi term \eqn{riemann} into terms quadratic in the
fermions. To get the index structure right, it turns out that we
need six such auxiliary fields, let's call them $\hat{X}^m$, with
$m=1,\ldots, 6$. In fact we've met these guys before: they're the
scalar fields in the vector multiplet of the ADHM gauge theory. To
see that they give rise to the promised four fermi term, let's
look at how they appear in the ADHM Lagrangian. There's already a
term quadratic in $\hat{X}$ in \eqn{inpot}, and another couples
this to the surplus fermionic collective coordinates $\chi$ so
that, schematically,
\be {\cal L}_{\hat{X}}\sim \hat{X}^2\omega^\dagger\omega +
\bar{\chi} \hat{X}\chi \ee
where, as we saw in Section 1.4, the field $\omega$ contains the
scale and orientation collective coordinates, with
$\omega^\dagger\omega\sim \rho^2$. Integrating out $\hat{X}$ in
the ADHM Lagrangian does indeed result in a four-fermi term which
is identified with \eqn{riemann}. However, now we perform a famous
trick: we integrate out the variables we thought we were
interested in, namely the $\chi$ fields, and focus on the ones we
thought were unimportant, the $\hat{X}$'s. After dealing correctly
with all the indices we've been dropping, we find that this
results in the contribution to the measure
\be d\mu_{\rm auxiliary} =d^6\hat{X}\
(\hat{X}^m\hat{X}^m)^{2N-4}\,\exp\left(-2\rho^2\hat{X}^m\hat{X}^m\right)
\ee
In the large $N$ limit, the integration over the radial variable
$|\hat{X}|$  may be performed using the saddle point approximation
evaluated at $|\hat{X}|=\rho$. The resulting powers of $\rho$ are
precisely those mentioned above that are needed to cancel the
powers of $\rho$ appearing in the bosonic Jacobian. Meanwhile, the
integration over the angular coordinates in $\hat{X}^m$ have been
left untouched. The final result for the instanton measure becomes
\be d\mu_{\rm inst} = \left(\frac{1}{\rho^5}\,d^4X\, d\rho
\,d^5\hat{\Omega}\right)\ d^8\xi d^8\zeta\ee
And the instanton indeed appears as if its moving in $AdS_5\times
\S^5$ as promised.

\para
The above discussion is a little glib. The invariant meaning of the measure
alone is not clear: the real meaning is that when integrated against
correlators, it gives results in agreement with gravity calculations
in AdS${}_5\times{\bf S}^5$. This, and several further results,
were shown in \cite{dhkmv}.
Calculations of this type were later performed for instantons in
other four-dimensional gauge theories, both conformal and
otherwise \cite{orb0,orb1,orb2,orb3,greencoul}. Curiously, there
appears to be an unresolved  problem with performing the
calculation for instantons in non-commutative gauge theories.

\subsubsection{Instanton Particles and the $(2,0)$ Theory}

There exists a rather special superconformal quantum field theory
in six dimensions known as the $(2,0)$ theory. It is the theory
with 16 supercharges which lives on $N$ M5-branes in M-theory and
it has some intriguing and poorly understood properties. Not least
of these is the fact that it appears to have $N^3$ degrees of
freedom. While it's not clear what these degrees of freedom are,
or even if it makes sense to talk about "degrees of freedom" in a
strongly coupled theory, the $N^3$ behavior is seen when computing
the free energy $F\sim N^3 T^6$ \cite{kt}, or anomalies whose
leading coefficient also scales as $N^3$ \cite{hmm}.

\para
If the $(2,0)$ theory is compactified on a circle of radius $R$,
it descends to $U(N)$ $d=4+1$ super Yang-Mills with 16
supercharges, which can be thought of as living on D4-branes in
Type IIA string theory. The gauge coupling $e^2$, which has
dimension of length in five dimensions, is given by
\be e^2= 8\pi^2 R \label{kkgauge}\ee
As in any theory compactified on a spatial circle, we expect to
find Kaluza-Klein modes, corresponding to momentum modes around
the circle with mass $M_{\rm KK}=1/R$. Comparison with the gauge
coupling constant \eqn{kkgauge} gives a strong hint what these
particles should be, since
\be M_{\rm kk}=M_{\rm inst}\ee
and, as we discussed in section 1.3.1, instantons are
particle-like objects in $d=4+1$ dimensions. The observation that
instantons are  Kaluza-Klein modes is clear from the IIA
perspective: the instantons in the D4-brane theory are D0-branes
which are known to be the Kaluza-Klein modes for the lift to
M-theory.

\para
The upshot of this analysis is a remarkable conjecture: the
maximally supersymmetric $U(N)$ Yang-Mills theory in five
dimensions is really a six-dimensional theory in disguise, with
the size of the hidden dimension given by $R\sim e^2$
\cite{rozali,brs,seiberg16}. As $e^2\rightarrow\infty$, the
instantons become light. Usually as solitons become light, they
also become large floppy objects, losing their interpretation as
particle excitations of the theory. But this isn't necessarily
true for instantons because, as we've seen, their scale size is
arbitrary and, in particular, independent of the gauge coupling.

\para
Of course, the five-dimensional theory is non-renormalizable and
we can only study questions that do not require the introduction
of new UV degrees of freedom. With this caveat, let's see how we
can test the conjecture using instantons. If they're really
Kaluza-Klein modes, they should exhibit Kaluza-Klein-eqsue
behavior which includes a characteristic spectrum of threshold
bound state of particles with $k$ units of momentum going around
the circle. This means that if the five-dimensional theory
contains the information about its six dimensional origin, it
should exhibit a threshold bound state of $k$ instantons for each
$k$. But this is something we can test in the semi-classical
regime by solving the low-energy dynamics of $k$ interacting
instantons. As we have seen, this is given by supersymmetric
quantum mechanics on $\ikn$, with the Lagrangian given by
\eqn{instleea} where $\partial=\partial_t$ in this equation.

\para
Let's review how to solve the ground states of  $d=0+1$
dimensional supersymmetric sigma models of the form
\eqn{instleea}. As explained by Witten, a beautiful connection to
de Rahm cohomology emerges after quantization \cite{constraints}.
Canonical quantization of the fermions leads to operators
satisfying the algebra
\be
\{\chi_\alpha,\chi_\beta\}=\{\bar{\chi}_\alpha,\bar{\chi}_\beta\}=0
\ \ {\rm and} \ \
\{\chi_\alpha,\bar{\chi}_\beta\}=g_{\alpha\beta}\ee
which tells us that we may regard $\bar{\chi}_\alpha$ and
$\chi_\beta$ as creation and annihilation operators respectively.
The states of the theory are described by wavefunctions
$\varphi(X)$ over the moduli space $\ikn$, acted upon by some
number $p$ of fermion creation operators. We write
$\varphi_{\alpha_1,\ldots,\alpha_p}(X)\equiv
\bar{\chi}_{\alpha_1}\ldots\bar{\chi}_{\alpha_p}\,\varphi(X)$. By
the Grassmann nature of the fermions, these states are
anti-symmetric in their $p$ indices, ensuring that the tower stops
when $p={\rm dim}(\ikn)$. In this manner, the states can be
identified with the space of all $p$-forms on $\ikn$.

\para
The Hamiltonian of the theory has a similarly natural geometric
interpretation. One can check that the Hamiltonian arising from
\eqn{instleea} can be written as
\be H=QQ^\dagger + Q^\dagger Q\ee
where $Q$ is the supercharge which takes the form
$Q=-i\bar{\chi}_\alpha p_\alpha$ and $Q^\dagger =-i\chi_\alpha
p_\alpha$, and $p_\alpha$ is the momentum conjugate to $X^\alpha$.
Studying the action of $Q$ on the states above, we find that
$Q=d$, the exterior derivative on forms, while $Q^\dagger
=d^\dagger$, the adjoint operator. We can therefore write the
Hamiltonian is the Laplacian acting on all $p$-forms,
\be H=dd^\dagger + d^\dagger d \ee
We learn that the space of ground states $H=0$ coincide with the
harmonic forms on the target space.

\para
 There are two subtleties in applying this analysis to
instantons. The first is that the instanton moduli space $\ikn$ is
singular. At these points, corresponding to small instantons, new
UV degrees of freedom are needed. Presumably this reflects the
non-renormalizability of the five-dimensional gauge theory.
However, as we have seen, one can resolve the singularity by
turning on non-commutivity. The interpretation of the instantons
as KK modes only survives if there is a similar non-commutative
deformation of the $(2,0)$ theory which appears to be the case.

\para
The second subtlety is an infra-red effect: the instanton moduli
space is non-compact. For compact target spaces, the ground states
of the sigma-model coincide with the space of harmonic forms or,
in other words, the cohomology. For non-compact target spaces such
as $\ikn$, we have the further requirement that any putative
ground state wavefunction must be normalizable and we need to
study cohomology with compact support. With this in mind, the
relationship between the five-dimensional theory and the
six-dimensional $(2,0)$ theory therefore translates into the
conjecture
\be\mbox{There is a unique normalizable harmonic form on $\ikn$
for each $k$ and $N$} \nn\ee
Note that even for a single instanton, this is non-trivial. As we
have seen above, after resolving the small instanton singularity,
the moduli space for a $k=1$ instanton in $U(N)$ theory is
$T^\star ({\bf CP}^{N-1})$, which has Euler character $\chi=N$.
Yet, there should be only a single groundstate. Indeed, it can be
shown explicitly that of these $N$ putative ground states, only a
single one has sufficiently compact support to provide an $L^2$
normalizable wavefunction \cite{leeyiinst}. For an arbitrary
number of $k$ instantons in $U(N)$ gauge theory, there is an index
theorem argument that this unique bound state exists
\cite{dkmbound}.

\para
So much for the ground states. What about the $N^3$ degrees of
freedom. Is it possible to see this from the five-dimensional
gauge theory? Unfortunately, so far, no one has managed this. Five
dimensional gauge theories become strongly coupled in the
ultra-violet where their non-renormalizability becomes an issue
and we have to introduce new degrees of freedom. This occurs at an
energy scale $E\sim 1/e^2N$, where the 't Hooft coupling becomes
strong. This is parametrically lower than the KK scale $E\sim
1/R\sim 1/e^2$. Supergravity calculations reveal that the $N^3$
degrees of freedom should also become apparent at the lower scale
$E\sim 1/e^2N$ \cite{imsy}. This suggests that perhaps the true
degrees of freedom of the theory are "fractional instantons", each
with mass $M_{\rm inst}/N$. Let me end this section with some
rampant speculation along these lines. It seems possible that the
$4kN$ moduli of the instanton may rearrange themselves into the
positions of $kN$ objects, each living in ${\bf R}^4$ and each,
presumably, carrying the requisite mass $1/e^2N$. We shall see a
similar phenomenon occurring for vortices in Section 3.8.2. If
this speculation is true, it would also explain why a naive
quantization of the instanton leads to a continuous spectrum,
rather strange behavior for a single particle: it's because the
instanton is really a multi-particle state. However, to make sense
of this idea we would need to understand why the fractional
instantons are confined to lie within the instanton yet, at the
same time, are also able to wander freely as evinced by the $4kN$
moduli. Which, let's face it, is odd! A possible explanation for
this strange behavior may lie in the issues of non-normalizability
of non-abelian modes discussed above, and related issues described
in \cite{colcol}.

\para
While it's not entirely clear what a fractional instanton means on
${\bf R}^4$, one can make rigorous sense of the idea when the
theory is compactified on a further ${\bf S}^1$ with a Wilson line
\cite{fractional,fractional2}. Moreover, there's evidence from
string dualities \cite{dijkgraaf,swd1d5} that the moduli space of
instantons on compact spaces ${\bf M}={\bf T}^4$ or $K3$ has the
same topology as the symmetric product ${\rm Sym}^{kN}({\bf M})$,
suggesting an interpretation in terms of $kN$ entities (strictly
speaking, one needs to resolve these spaces into an object known
as the Hilbert scheme of points over ${\bf M}$).

\newpage
\section{Monopoles}

The tale of magnetic monopoles is well known. They are postulated
particles with long-range, radial, magnetic field $B_i$,
$i=1,2,3$,
\be B_i=\frac{g\,\hat{r}_i}{4\pi r^2}\ee
where $g$ is the magnetic charge. Monopoles have never been
observed and one of Maxwell's equations, $\nabla\cdot B=0$,
insists they never will be. Yet they have been a recurrent theme
in high energy particle physics for the past 30 years! Why?

\para
The study of monopoles began with Dirac \cite{dirac} who showed
how one could formulate a theory of monopoles consistent with a
gauge potential $A_\mu$. The requirement that the electron doesn't
see the inevitable singularities in $A_\mu$ leads to the famed
quantization condition
\be eg=2\pi n\ \ \ \ \ n\in\Z\label{dirac}\ee
However, the key step in the rehabilitation of magnetic monopoles
was the observation  of 't Hooft \cite{tmon} and Polyakov
\cite{polyakov} that monopoles naturally occur in non-abelian
gauge theories, making them a robust prediction of grand unified
theories based on semi-simple groups. In this lecture we'll review
the formalism of 't Hooft-Polyakov monopoles in $SU(N)$ gauge
groups, including the properties of the solutions and the D-brane
realization of the Nahm construction. At the end we'll cover
several applications to quantum gauge theories in various
dimensions.

\para
There are a number of nice reviews on monopoles in the literature.
Aspects of the classical solutions are dealt with by Sutcliffe
\cite{paul} and Shnir \cite{shnir}; the mathematics  of monopole
scattering can be found in the book by Atiyah and Hitchin
\cite{ah}; the application to S-duality of quantum field theories
is covered in the review by Harvey \cite{harveymon}. A
comprehensive review of magnetic monopoles by Weinberg and Yi will
appear shortly \cite{erickpiljin}.

\subsection{The Basics}

To find monopoles, we first need to change our theory from that of
Lecture 1. We add a single real scalar field $\phi\equiv\phi^a_{\
b}$, transforming in the adjoint representation of $SU(N)$. The
action now reads
\be S=\Tr\ \int d^4x\ \frac{1}{2e^2}
F_{\mu\nu}F^{\mu\nu}+\frac{1}{e^2}(\D_\mu\phi)^2\label{monact}\ee
where we're back in Minkowski signature $(+,-,-,-)$. The spacetime
index runs over $\mu=0,1,2,3$ and we'll also use the purely
spatial index $i=1,2,3$. Actions of this type occur naturally as a
subsector of $\N=4$ and $\N=2$ super Yang-Mills theories. There is
no potential for $\phi$ so, classically, we are free to chose the
vacuum expectation value (vev) as we see fit. Gauge inequivalent
choices correspond to  different ground states of the theory. By
use of a suitable gauge transformation, we may set
\be
\bra\phi\ket=\diag(\phi_1,\ldots,\phi_N)=\vec{\phi}\cdot\vec{H}
\label{phivev}\ee
where the fact we're working in $SU(N)$ means that
$\sum_{a=1}^N\phi_a=0$. We've also introduced the notation of the
root vector $\vec{\phi}$, with $\vec{H}$  a basis for the
$(N-1)$-dimensional Cartan subalgebra of $su(N)$. If you're not
familiar with roots of Lie algebras and the Cartn-Weyl basis then
you can simply think of
 $\vec{H}$ as the set of $N$ matrices, each with a single entry $1$ along the
diagonal. (This is actually the Cartan subalgebra for $u(N)$
rather than $su(N)$ but this will take care of itself if we
remember that $\sum_a\phi_a=0$). Under the requirement that
$\phi_a\neq \phi_b$ for $a\neq b$ the gauge symmetry breaks to the
maximal torus,
\be SU(N)\rightarrow U(1)^{N-1}\label{monbreaking}\ee
The spectrum of the theory consists of $(N-1)$ massless photons
and scalars, together with $\ft12N(N-1)$ massive W-bosons with
mass $M_W^2=(\phi_a-\phi_b)^2$. In the following we will use the
Weyl symmetry to order $\phi_a<\phi_{a+1}$.

\para
In the previous lecture, instantons arose from the possibility of
winding field configurations non-trivially around the ${\bf
S}^3_\infty$ infinity of Euclidean spacetime. Today we're
interested in particle-like solitons, localized in space rather
than spacetime. These objects are supported by the vev
\eqn{phivev} twisting along its gauge orbit as we circumvent the
spatial boundary $\S^2_\infty$. If we denote the two coordinates
on $\S^2_\infty$ as $\theta$ and $\varphi$, then solitons are
supported by configurations with
$\bra{\phi}\ket=\bra{\phi}(\theta,\varphi)\ket$. Let's classify
the possible windings. A vev of the form \eqn{phivev} is one point
in a space of gauge equivalent vacua, given by $SU(N)/U(1)^{N-1}$
where the stabilizing group in the denominator is the unbroken
symmetry group \eqn{monbreaking} which leaves \eqn{phivev}
untouched. We're therefore left to consider maps:
$\S^2_\infty\rightarrow SU(N)/U(1)^{N-1}$, characterized by
\be \Pi_2\left({SU(N)}/{U(1)^{N-1}}\right)\cong
\Pi_1\left(U(1)^{N-1}\right)\cong \Z^{N-1}\label{monhom}\ee
This classification suggests that we should be looking for $(N-1)$
different types of topological objects. As we shall see, these
objects are monopoles carrying magnetic charge in each of the
$(N-1)$ unbroken abelian gauge fields \eqn{monbreaking}.

\para
Why is winding of the scalar field $\phi$ at infinity associated
with magnetic charge? To see the precise connection is actually a
little tricky
--- details can be found in \cite{tmon,polyakov} and in \cite{gno}
for $SU(N)$ monopoles
--- but there is a simple heuristic argument to see why the two
are related. The important point is that if a configuration is to
have finite energy, the scalar kinetic term $\D_\mu\phi$ must
decay at least as fast as $1/r^2$ as we approach the boundary
$r\rightarrow\infty$. But if $\bra\phi\ket$ varies asymptotically
as we move around $\S^2_\infty$, we have $\partial\phi\sim 1/r$.
To cancel the resulting infrared divergence we must turn on a
corresponding gauge potential $A_\theta\sim 1/r$, leading to a
magnetic field of the form $B\sim 1/r^2$.

\para
Physically, we would expect any long range magnetic field to
propagate through the  massless $U(1)$ photons. This is indeed the
case. If $\D_i\phi\rightarrow 0$ as $r\rightarrow \infty$ then
$[\D_i,\D_j]\phi=-i[F_{ij},\phi]\rightarrow 0$ as
$r\rightarrow\infty$. Combining these two facts, we learn that the
non-abelian magnetic field carried by the soliton is of the form,
\be B_i=\vec{g}\cdot\vec{H}(\theta,\varphi)\ \frac{\hat{r}_i}{4\pi
r^2}\ee
Here the notation $\vec{H}(\theta,\varphi)$ reminds us that the
unbroken Cartan subalgebra twists within the $su(N)$ Lie algebra
as we move around the $\S^2_\infty$.

\subsubsection{Dirac Quantization Condition}

\EPSFIGURE{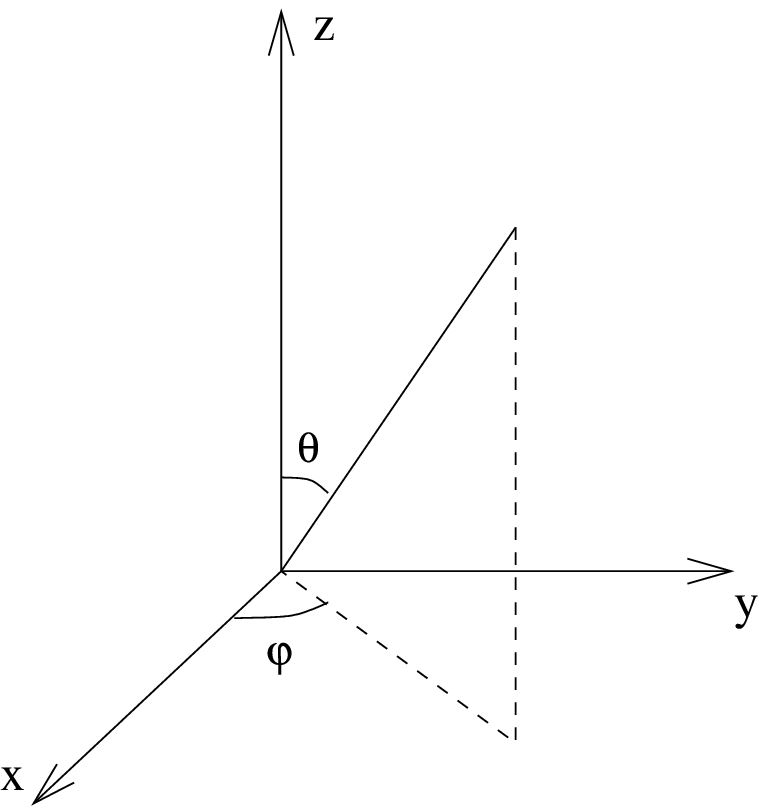,height=110pt}{}
The allowed magnetic charge vectors $\vec{g}$ may be determined by
studying the winding of the scalar field $\phi$ around ${\bf
S}^2_\infty$. However, since the winding is related to the
magnetic charge, and the latter is a characteristic of the long
range behavior of the monopole, it's somewhat easier to neglect
the non-abelian structure completely and study just the $U(1)$
fields. The equivalence between the two methods is reflected in
the equality between first and second homotopy groups in
\eqn{monhom}.

\para
For this purpose, it is notationally simpler to work in unitary,
or singular, gauge in which the vev
$\bra\phi\ket=\vec{\phi}\cdot\vec{H}$ is fixed to be constant at
infinity. This necessarily re-introduces Dirac string-like
singularities for any single-valued gauge potential, but allows us
to globally write the magnetic field in diagonal form,
\be B_i=\diag(g_1,\dots,g_N)\ \frac{\hat{r}_i}{4\pi
r^2}\label{monb}\ee
where $\sum_{a=1}^Ng_a=0$ since the magnetic field lies in $su(N)$
rather than $u(N)$.

\para
What values of $g_a$ are allowed? A variant of Dirac's original
argument, due to Wu and Yang \cite{wy}, derives the magnetic field
\eqn{monb} from two gauge potentials defined respectively on the
northern and southern hemispheres of $\S^2_\infty$:
\be
A_\varphi^N&=&\frac{1-\cos\theta}{4\pi r\sin\theta}\,\vec{g}\cdot\vec{H}\ \nn\\
A^S_\varphi&=& -\frac{1+\cos\theta}{4\pi
r\sin\theta}\,\vec{g}\cdot\vec{H} \ee
where $A^N$ goes bad at the south pole $\theta=\pi$, while $A^S$
sucks at the north pole $\theta=0$. To define a consistent field
strength we require that on the overlap $\theta\neq 0,\pi$, the
two differ by a gauge transformation which, indeed, they do:
\be A^N_i = U(\partial_i+A^S_i)U^{-1}\ee
with $U(\theta,\varphi)=\exp(-i\vec{g}\cdot\vec{H}\varphi/2\pi)$.
Notice that as we've written it, this relationship only holds in
unitary gauge where $\vec{H}$ doesn't depend on $\theta$ or
$\varphi$, requiring that we work in singular gauge. The final
requirement is that our gauge transformation is single valued, so
$U(\varphi)=U(\varphi+2\pi)$ or, in other words, $\exp(i\,
\vec{g}\cdot\vec{H})=1$. This requirement is simply solved by
\be g_a \in 2\pi\Z\ee
This is the Dirac quantization condition \eqn{dirac} in units in
which the electric charge $e=1$, a convention which arises from
scaling the coupling outside the action in \eqn{monact}. In fact,
in our theory the W-bosons have charge 2 under any $U(1)$ while
matter in the fundamental representation would have charge 1.

\para
There's another notation for the magnetic charge vector $\vec{g}$
that will prove useful. We write
\be \vec{g}=2\pi\sum_{a=1}^{N-1}n_a\,\vec{\alpha}_a\label{mong}\ee
where $n_a\in\Z$ by the Dirac quantization condition\footnote{For
monopoles in a general gauge group, the Dirac quantization
condition becomes $\vec{g}=4\pi\sum_a n_a\valpha_a^\star$ where
$\valpha_a^\star$ are simple co-roots.} and $\vec{\alpha}_a$ are
the simple roots of $su(N)$. The choice of simple roots is
determined by defining $\vec{\phi}$ to lie in a positive Weyl
chamber. What this means in practice, with our chosen ordering
$\phi_a<\phi_{a+1}$, is that we can write each root as an
$N$-vector, with
\be \valpha_1&=&(1,-1,0,\ldots, 0)\nn\\
\valpha_2&=&(0,1,-1,\ldots,0)\label{simpleroots}\ee through to \be
\valpha_{N-1}=(0,0,\ldots,1,-1) \ee
Then translating between two different notations for the magnetic
charge vector we have
\be \vec{g}&=&\diag({g_1,\ldots,
g_N})\\&=&2\pi\,\diag(n_1\,,\,n_2-n_1,\ldots,
n_{N-1}-n_{N-2}\,,\,-n_{N-1})\nn\ee
The advantage of working with the integers $n_a$, $a=1,\ldots,N-1$
will become apparent shortly.

\subsubsection{The Monopole Equations}

As in Lecture 1, we've learnt that the space of field
configurations decomposes into different topological sectors, this
time labelled by the vector $\vec{g}$ or, equivalently, the $N-1$
integers $n_a$. We're now presented with the challenge of finding
solutions in the non-trivial sectors. We can again employ a
Bogomoln'yi bound argument (this time actually due to Bogomoln'yi
\cite{bog}) to derive first order equations for the monopoles. We
first set $\partial_0=A_0=0$, so we are looking for time
independent configurations with vanishing electric field. Then the
energy functional of the theory gives us the mass of a magnetic
monopole,
\be M_{\rm mono}&=&\Tr\ \int d^3x\ \frac{1}{e^2}B_i^2
+\frac{1}{e^2} (\D_i\phi)^2 \nn\\ &=& \Tr\ \int d^3x\
\frac{1}{e^2}\left(B_i\mp\D_i\phi\right)^2\pm\frac{2}{e^2}\,B_i\D_i\phi\nn\\
&\geq & \frac{2}{e^2}\int d^3x \ \partial_i\ \Tr (B_i\phi)
\label{monsquare}\ee
where we've used the Bianchi identity $\D_iB_i=0$ when integrating
by parts to get the final line. As in the case of instantons,
we've succeeded in bounding the energy by a surface term which
measures a topological charge. Comparing with the expressions
above we have
\be M_{\rm mono}\geq
\frac{|\vec{g}\cdot\vec{\phi}|}{e^2}=\frac{2\pi}{e^2}\,
\sum_{a=1}^{N-1}n_a\phi_a\ee
with equality if and only if the monopole equations (often called
the Bogomoln'yi equations) are obeyed,
\be B_i=\D_i\phi\ \ \ \ \ \ \ &&\mbox{if\ } \vec{g}\cdot\vec{\phi}
> 0 \nn\\B_i=-\D_i\phi\ \ \ \ \ \ \ &&\mbox{if\ } \vec{g}\cdot\vec{\phi}
< 0\label{monbog}\ee
For the rest of this lecture we'll work with
$\vec{g}\cdot\vec{\phi}>0$ and the first of these equations. Our
path will be the same as in lecture 1: we'll first examine the
simplest solution to these equations and then study its properties
before moving on to the most general solutions. So first:

\subsubsection{Solutions and Collective Coordinates}

The original magnetic monopole described by 't Hooft and Polyakov
occurs in $SU(2)$ theory broken to $U(1)$. We have
$SU(2)/U(1)\cong\S^2$ and $\Pi_2(\S^2)\cong\Z$. Here we'll
describe the simplest such monopole with charge one. To better
reveal the topology supporting this monopole (as well as to
demonstrate explicitly that the solution is smooth) we'll
momentarily revert back to a gauge where the vev winds
asymptotically. The solution to the monopole equation \eqn{monbog}
was found by Prasad and Sommerfield \cite{ps}
\be \phi&=&\frac{\hat{r}_i\sigma^i}{r}\,\left(
vr\coth(vr)-1\right)\nn\\
A_\mu&=&-\epsilon_{i\mu j}\, \frac{\hat{r}^j\sigma^i}{r}
\left(1-\frac{vr}{\sinh vr}\right) \label{monsol}\ee
This solution asymptotes to $\bra\phi\ket = v \sigma^i\hat{r}^i$,
where $\sigma^i$ are the Pauli matrices (i.e. comparing notation
with \eqn{phivev} in, say, the $\hat{r}^3$ direction, we have
$v=-\phi_1=\phi_2$). The $SU(2)$ solution presented above has 4
collective coordinates, although none of them are written
explicitly. Most obviously, there are the three center of mass
coordinates. As with instantons, there is a further collective
coordinate arising from acting on the soliton with the unbroken
gauge symmetry which, in this case, is simply $U(1)$.

\para
For monopoles in $SU(N)$ we can always generate solutions by
embedding the configuration \eqn{monsol} above into a suitable
$SU(2)$ subgroup. Note however that, unlike the situation for
instantons, we can't rotate from one $SU(2)$ embedding to another
since the $SU(N)$ gauge symmetry is not preserved in the vacuum.
Each $SU(2)$ embedding will give rise to a different monopole with
different properties --- for example, they will have magnetic
charges under different $U(1)$ factors.

\para
Of the many inequivalent embeddings of $SU(2)$ into $SU(N)$, there
are $(N-1)$ special ones. These have generators given in the
Cartan-Weyl basis by $\valpha\cdot\vec{H}$ and $E_{\pm\valpha}$
where $\valpha$ is one of the simple roots \eqn{simpleroots}. In a
less sophisticated language, these are simply the $(N-1)$
contiguous $2\times 2$ blocks which lie along the diagonal of an
$N\times N$ matrix. Embedding the monopole in the $a^{\rm th}$
such block gives rise to the magnetic charge $\vec{g}=\valpha_a$.

\subsection{The Moduli Space}

For a monopole with magnetic charge $\vec{g}$, we want to know how
many collective coordinates are contained within the most general
solution. The answer was given by E. Weinberg \cite{erick}. There
are subtleties that don't occur in the instanton calculation, and
a variant of the Atiyah-Singer index theorem due to Callias is
required \cite{callias}. But the result is very simple. Define the
moduli space of monopoles with magnetic charge $\vg$ to be $\mg$.
Then the number of collective coordinates is
\be \dim(\mg)=4\sum_{a=1}^{N-1}n_a\ee
The interpretation of this is as follows. There exist $(N-1)$
"elementary" monopoles, each associated to a simple root
$\valpha_a$, carrying magnetic charge under exactly one of the
$(N-1)$ surviving $U(1)$ factors of \eqn{monbreaking}. Each of
these elementary monopoles has 4 collective coordinates. A
monopole with general charge $\vg$ can be decomposed into $\sum_a
n_a$ elementary monopoles, described by three position coordinates
and a phase arising from $U(1)$ gauge rotations.
\EPSFIGURE{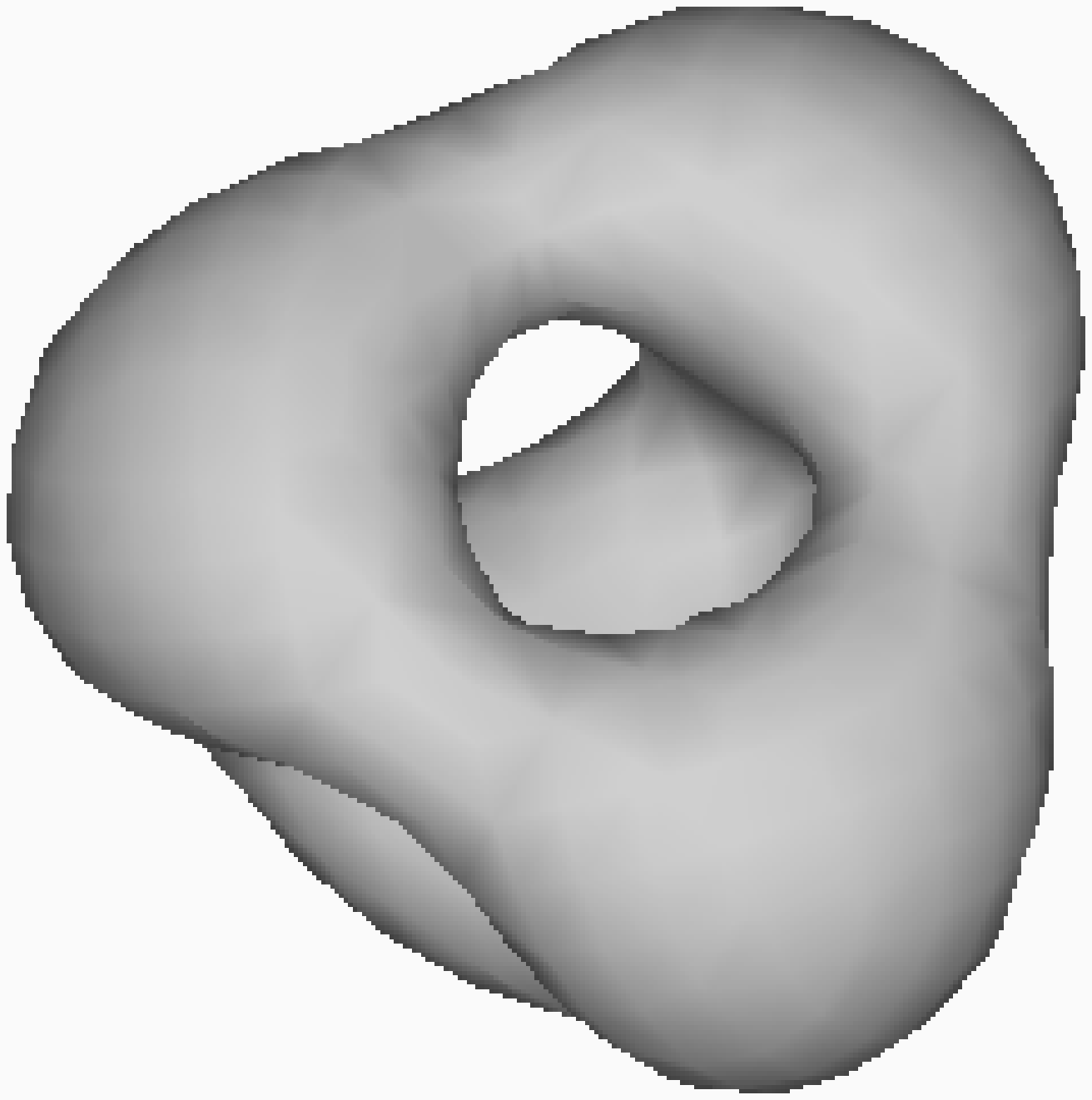,width=105pt}
{}

\para
You should be surprised by the existence of this large class of
solutions since it implies that monopoles can be placed at
arbitrary separation and feel no force. But this doesn't happen
for electrons! Any objects carrying the same charge, whether
electric or magnetic, repel. So what's special about monopoles?
The point is that monopoles also experience a second long range
force due to the massless components of the scalar field $\phi$.
This gives rise to an attraction  between the monopoles that
precisely cancels the electromagnetic repulsion \cite{manforce}.
Such cancellation of forces only occurs when there is no potential
for $\phi$ as in \eqn{monact}.
\EPSFIGURE{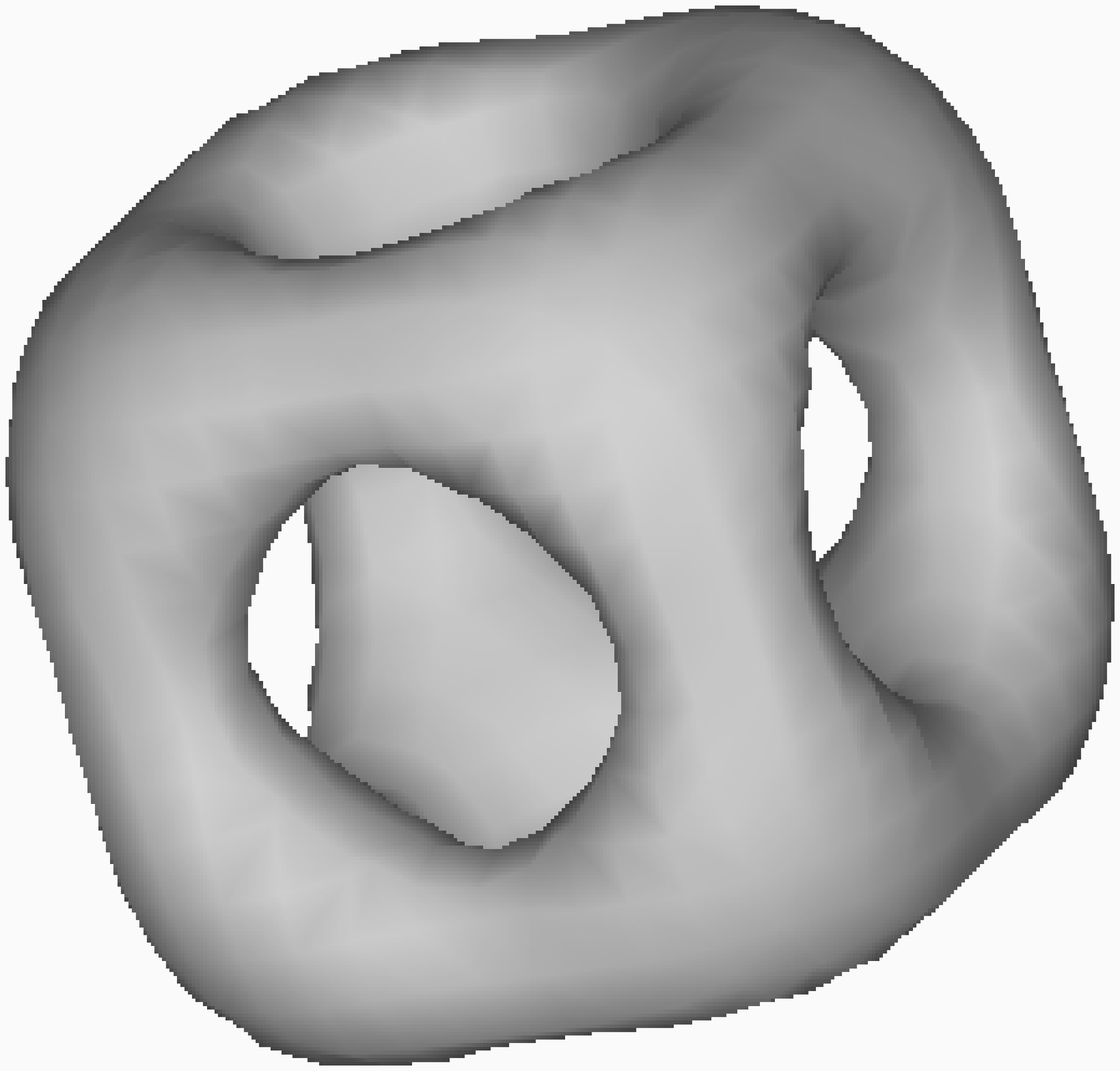,width=95pt}{}
\para
The interpretation of the collective coordinates as positions of
particle-like objects holds only when the monopoles are more
widely separated than their core size. As the monopoles approach,
weird things happen! Two monopoles form a torus. Three monopoles
form a tetrahedron, seemingly splitting into four lumps of energy
as seen in figure 4. Four monopoles form a cube as in figure 5.
(Both of these figures are taken from \cite{conorpaul}). We see
that monopoles really lose their individual identities as the
approach and merge into each other. Higher monopoles form platonic
solids, or buckyball like objects.

\subsubsection{The Moduli Space Metric}

The metric on $\mg$ is defined in a similar fashion to that on the
instanton moduli space $\ikn$. To be more precise, it's defined in
an identical fashion. Literally! The key point is that the
monopole equations $B=\D\phi$ and the instanton equations
$F=\starf$ are really the same: the difference between the two
lies in the boundary conditions. To see this, consider instantons
with $\partial_4=0$ and endow the component of the gauge field
$A_4\equiv\phi$ with a vev $\bra\phi\ket$. We end up with the
monopole equations. So using the notation $\d\phi=\d A_4$, we can
reuse the linearized self-dual equations (1.14) and the gauge
fixing condition (1.17) from the Lecture 1 to define the monopole
zero modes. The metric on the monopole moduli space $\mg$ is again
given by the overlap of zero modes,
\be g_{\alpha\beta}=\frac{1}{e^2}\,\Tr\, \int\,d^3x\ (\d_\alpha
A_i\,\d_\beta A_i
+\d_\alpha\phi\,\d_\beta\phi)\label{monmetric}\ee
The metric on the monopole moduli space has the following
properties:
\begin{itemize}
\item The metric is \hk. \item The metric enjoys an $SO(3)\times
U(1)^{N-1}$ isometry. The former descends from physical rotations
of the monopoles in space. The latter arise from the unbroken
gauge group. The $U(1)^{N-1}$ isometries are tri-holomorphic,
while the $SO(3)$ isometry rotates the three complex structures.

\item The metric is smooth. There are no singular points analogous
to the small instanton singularities of $\ikn$ because, as we have
seen, the scale of the monopole isn't a collective coordinate. It
is fixed to be $L_{\rm mono}\sim 1/M_W$, the Compton wavelength of
the W-bosons.

\item Since the metrics on $\ikn$ and $\mg$ arise from the same
equations, merely endowed with different boundary conditions, one
might wonder if we can interpolate between them. In fact we can.
In the study of instantons on $\R^3\times \S^1$, with a non-zero
Wilson line around the $\S^1$, the $4N$ collective coordinates of
the instanton gain the interpretation of the positions of $N$
"fractional instantons" \cite{fractional,fractional2}. These are
often referred to as calorons and are identified as the monopoles
discussed above. By taking the radius of the circle to zero, and
some calorons to infinity, we can interpolate between the metrics
on $\mg$ and $\ikn$ \cite{leeyiinst}.
\end{itemize}

\subsubsection{The Physical Interpretation of the Metric}

For particles such as monopoles in $d=3+1$ dimensions, the metric
on the moduli space has a beautiful physical interpretation first
described by Manton \cite{manton}. Suppose that the monopoles move
slowly through space. We approximate the motion by assuming that
the field configurations remain close to the static solutions, but
endow the collective coordinates $X^\alpha$ with time dependence:
$X^\alpha\rightarrow X^\alpha(t)$. If monopoles collide at very
high energies this approximation will not be valid. As the
monopoles hit they will spew out massive W-bosons and, on
occasion, even monopole-anti-monopole pairs. The resulting field
configurations will look nothing like the static monopole
solutions. Even for very low-energy scattering it's not completely
clear that the approximation is valid since the theory doesn't
have a mass gap and the monopoles can emit very soft photons.
Nevertheless, there is much evidence that this procedure, known as
the {\it moduli space approximation}, does capture the true
physics of monopole scattering at low energies. The time
dependence of the fields is
\be A_\mu=A_\mu(X^\alpha(t))\ \  \ ,\ \ \ \
\phi=\phi(X^\alpha(t))\ee
which reduces the dynamics of an infinite number of field theory
degrees of freedom to a finite number of collective coordinates.
We must still satisfy Gauss' law,
\be \D_i E_i-i[\phi,\D_0\phi]=0\ee
which can be achieved by setting $A_0=\Omega_\alpha
\dot{X}^\alpha$, where the $\Omega_\alpha$ are the extra gauge
rotations that we introduced in (1.15) to ensure that the zero
modes satisfy the background gauge fixing condition. This means
that the time dependence of the fields is given in terms of the
zero modes,
\be E_{i}=F_{0i}&=&\d_\alpha A_i\ \dot{X}^\alpha\nn\\
\D_0\phi&=&\d_\alpha\phi\ \dot{X}^\alpha \label{physics}\ee
Plugging this into the action \eqn{monact} we find
\be S&=&\Tr\ \int d^4x\
\frac{1}{e^2}\left(E_i^2+B_i^2+(\D_0\phi)^2+(\D_i\phi)^2 \right)
\nn\\ &=& \int dt\ \left( M_{\rm mono} + \ft12
g_{\alpha\beta}\dot{X}^\alpha\dot{X}^\beta\right)
\label{manton}\ee
The upshot of this is analysis is that the low-energy dynamics of
monopoles is given by the $d=0+1$ sigma model on the monopole
moduli space. The equations of motion following from \eqn{manton}
are simply the geodesic equations for the metric
$g_{\alpha\beta}$. We learn that the moduli space metric captures
the velocity-dependent forces felt by monopoles, such that
low-energy scattering is given by geodesic motion.

\para
In fact, this logic can be reversed. In certain circumstances it's
possible to figure out the trajectories followed by several moving
monopoles. From this one can construct a metric on the
configuration space of monopoles such that the geodesics
reconstruct the known motion. This metric agrees with that defined
above in a very different way. This procedure has been carried out
for a number of examples \cite{manmon,gibman,lwy}.

\subsubsection{Examples of Monopole Moduli Spaces}

Let's now give a few examples of monopole moduli spaces. We start
with the simple case of a single monopole where the metric may be
explicitly computed.

\subsubsection*{One Monopole}

Consider the $\vg=\valpha_1$ monopole, which is nothing more that
the charge one  $SU(2)$ solution we saw previously \eqn{monsol}.
In this case we can compute the metric directly. We have two
different types of collective coordinates:
\newcounter{hart}
\begin{list}{\roman{hart})}{\usecounter{hart}}
\item The three translational modes. The linearized monopole
equation and gauge fixing equation are solved by
$\d_{(i)}A_j=-F_{ij}$ and $\d_{(i)}\phi=-\D_i\phi$, so that the
overlap of zero modes is
\be \Tr\ \frac{1}{e^2}\int d^3x\ (\d_{(i)}A_k\, \d_{(j)}A_k
+\d_{(i)}\phi\, \d_{(j)}\phi) = M_{\rm mono}\ \delta_{ij}\ee
\item The gauge mode arises from transformation
$U=\exp(i\phi\chi/v)$, where the normalization has been chosen so
that the collective coordinate $\chi$ has periodicity $2\pi$. This
gauge transformation leaves $\phi$ untouched while the
transformation on the gauge field is $\d A_i=(\D_i\phi)/v$.
\end{list}
Putting these two together, we find that single monopole moduli
space is
\be \M_{\valpha}\cong \R^3\times \S^1\ee
with metric
\be ds^2=M_{\rm mono}\ \left(dX^idX^i+
\frac{1}{v^2}\,d\chi^2\right) \label{onemon}\ee
where $M_{\rm mono}=4\pi v/e^2$ in the notation used in the
solution \eqn{monsol}.

\subsubsection*{Two Monopoles}

Two monopoles in $SU(2)$ have magnetic charge $\vg=2\alpha_1$. The
direct approach to compute the metric that we have just described
becomes impossible since the most general analytic solution for
the two monopole configuration is not available. Nonetheless,
Atiyah and Hitchin were able to determine the two monopole moduli
space using symmetry considerations alone, most notably the
constraints imposed by hyperK\"ahlerity \cite{ah,gibmanah}. It is
\be \M_{2\valpha} \cong \R^3\times\frac{\S^1\times
\M_{AH}}{\Z_2}\label{mah}\ee
where $\R^3$ describes the center of mass of the pair of
monopoles, while $\S^1$ determines the overall phase
$0\leq\chi\leq 2\pi$. The four-dimensional \hk space $\M_{AH}$ is
the famous Atiyah-Hitchin manifold. Its metric can be written as
\be
ds^2=f(r)^2dr^2+a(r)^2\sigma_1^2+b(r)^2\sigma_2^2+c(r)^2\sigma_3^2
\label{ahm}\ee
Here the radial coordinate $r$ measures the separation between the
monopoles in units of the monopole mass. The $\sigma_i$ are the
three left-invariant $SU(2)$ one-forms which, in terms of polar
angles $0\leq\theta\leq \pi$, $0\leq\phi\leq 2\pi$ and $0\leq
\psi\leq 2\pi$, take the form
\be \sigma_1&=&-\sin\psi\ d\theta+ \cos\psi\sin\theta\ d\phi\nn\\
\sigma_2&=&\cos\psi\ d\theta + \sin\psi\sin\theta\ d\phi \nn\\
\sigma_3&=&d\psi + \cos\theta\ d\phi\ee
For far separated monopoles, $\theta$ and $\phi$ determine the
angular separation while $\psi$ is the relative phase. The $\Z_2$
quotient in \eqn{mah} acts as
\be \Z_2: \chi\rightarrow \chi+\pi\ \ \ \ ,\ \ \ \ \psi\rightarrow
\psi+\pi\label{ahz2}\ee
The \hk condition can be shown to relate the four functions
$f,a,b$ and $c$ through the differential equation
\be \frac{2bc}{f}\,\frac{da}{dr}=(b-c)^2-a^2\ee
together with two further equations obtained by cyclically
permuting $a,b$ and $c$. The solutions can be obtained in terms of
elliptic integrals but it will prove more illuminating to present
the asymptotic expansion of these functions. Choosing coordinates
such that $f(r)=-b(r)/r$, we have
\be a^2&=& r^2\left(1-\frac{2}{r}\right)-8r^3 e^{-r} +\ldots\nn\\
b^2&=& r^2\left(1-\frac{2}{r}\right)+8r^3e^{-r} +\ldots\label{abc}\\
c^2 &=& 4\left(1-\frac{2}{r}\right)^{-1}+\ldots\nn\ee
If we suppress the exponential corrections, the metric describes
the velocity dependant forces between two monopoles interacting
through their long range fields. In fact, this asymptotic metric
can be derived by treating the monopoles as point particles and
considering their Li\'{e}nard-Wiechert potentials. Note that in
this limit there is an isometry associated to the relative phase
$\psi$. However, the minus sign before the $2/r$ terms means that
the metric is singular. The exponential corrections to the metric
resolve this singularity and contain
the information about the behavior of the monopoles as their
non-abelian cores overlap.

\para
The Atiyah-Hitchin metric appears in several places in string
theory and supersymmetric gauge theories, including the M-theory
lift of the type IIA O6-plane \cite{o6sen}, the solution of the
quantum dynamics of 3d gauge theories \cite{sw3d}, in intersecting
brane configurations \cite{hw}, the heterotic string compactified
on ALE spaces \cite{hetsen,hetwit} and NS5-branes on orientifold
8-planes \cite{hz}. In each of these places, there is often a
relationship to magnetic monopoles underlying the appearance of
this metric.

\para
For higher charge monopoles of the same type $\vg=n\valpha$, the
leading order terms in the asymptotic expansion of the metric,
associated with the long-range fields of the monopoles,  have been
computed. The result is known as the Gibbons-Manton metric
\cite{gibman}. The full metric on the monopole moduli space
remains an open problem.

\subsubsection*{Two Monopoles of Different Types}

As we have seen, higher rank gauge groups $SU(N)$ for $N\geq 3$
admit monopoles of different types. If a $\vg=\valpha_a$ monopole
and a $\vg=\valpha_b$ monopole live in entirely different places
in the gauge group, so that $\valpha_a\cdot\valpha_b=0$, then they
don't see each other and their moduli space is simply the product
$(\R^3\times \S^1)^2$. However, if they live in neighboring
subgroups so that $\valpha_a\cdot\valpha_b=-1$, then they do
interact non-trivially.

\para
The metric on the moduli space of two neighboring monopoles,
sometimes referred to as the $(1,1)$ monopole, was first computed
by Connell \cite{connell}. But he chose not to publish. It was
rediscovered some years later by two groups when the connection
with electro-magnetic duality made the study of monopoles more
pressing \cite{jeromelowe,lwy1}. It is simplest to describe if the
two monopoles have the same mass, so
$\vec{\phi}\cdot\valpha_a=\vec{\phi}\cdot\valpha_b$. The moduli
space is then
\be \M_{\valpha_1+\valpha_2}\cong \R^3\times\frac{\S^1\times
\M_{TN}}{\Z_2}\ee
where the interpretation of the $\R^3$ factor and $\S^1$ factor
are the same as before. The relative moduli space is the Taub-NUT
manifold, which has metric
\be
ds^2=\left(1+\frac{2}{r}\right)(dr^2+r^2(\sigma_1^2+\sigma_2^2))+
\left(1+\frac{2}{r}\right)^{-1}\sigma_3^2\ee
The $+2/r$ in the metric, rather than the $-2/r$ of
Atiyah-Hitchin, means that the metric is smooth. The apparent
singularity at $r=0$ is merely a coordinate artifact, as you can
check by transforming to the variables $R=\sqrt{r}$. Once again,
the $1/r$ terms capture the long range interactions of the
monopoles, with the minus sign traced to the fact that each sees
the other with opposite magnetic charge (essentially because
$\valpha_1\cdot\valpha_2=-1$). There are no exponential
corrections to this metric. The non-abelian cores of the two
monopoles do not interact.

\para
The exact moduli space metric for a string of neighboring
monopoles, $\vg=\sum_a\valpha_a$ has been determined. Known as the
Lee-Weinberg-Yi metric, it is a higher dimensional generalization
of the Taub-NUT metric \cite{lwy}. It is smooth and has no
exponential corrections.

\subsection{Dyons}

Consider the one-monopole moduli space $\R^3\times\S^1$. Motion in
$\R^3$ is obvious. But what does motion along the $\S^1$
correspond to?

\para
We can answer this by returning to our specific $SU(2)$ solution
\eqn{monsol}. We determined that the zero mode for the $U(1)$
action is $\d A_i=\D_i\phi$ and $\d\phi=0$. Translating to the
time dependence of the fields \eqn{physics}, we find
\be E_i=\frac{(\D_i\phi)}{v}\,\dot{\chi}=\frac{B_i}{v}\
e^2\dot{\chi} \ee
We see that motion along the $\S^1$ induces an electric field for
the monopole, proportional to its magnetic field. In the unbroken
$U(1)$, this gives rise to a long range electric field,
\be \Tr(E_i\phi)=\frac{qve^2\hat{r}_i}{2\pi r^2}\ee
where, comparing with the normalization above, the electric charge
$q$ is given by
\be q=\frac{2\pi\dot{\chi}}{ve^2}\ee
Note that motion in $\R^3$ also gives rise to an electric field,
but this is the dual to the familiar statement that a moving
electric charge produces a magnetic field. Motion in $\S^1$, on
the other hand, only has the effect of producing an electric field
\cite{jz}.

\para
A particle with both electric and magnetic charges is called a
{\it dyon}, a term first coined by Schwinger \cite{schwinger}.
Since we have understood this property from the perspective of the
monopole worldline, can we return to our original theory
\eqn{monact} and find the corresponding solution there? The answer
is yes. We relax the condition $E_i=0$ when completing the
Bogomoln'yi square in \eqn{monsquare} and write
\be M_{\rm dyon}&=&\Tr\ \int d^3x\
\frac{1}{e^2}\left(E_i-\cos\alpha\,\D_i\phi\right)^2 +
\frac{1}{e^2}\left(B_i-\sin\alpha\,\D_i\phi\right)^2 \nn\\
&& + \frac{2}{e^2}\Tr\ \int d^3x\ \partial_i\left(\cos\alpha\,
E_i\phi + \sin\alpha\,B_i\phi\right)\ee
which holds for all $\alpha$. We write the long range magnetic
field as $E_i=\vec{q}\cdot\vec{H}\,e^2\hat{r}^i/4\pi r^2$. Then by
adjusting $\alpha$ to make the bound as tight as possible, we have
\be M_{\rm dyon} \geq
\sqrt{\left(\vec{q}\cdot\vec{\phi}\right)^2+\left(\frac{\vec{g}
\cdot\vec{\phi}}{e^2}\right)^2} \label{dyon}\ee
and, given a solution to the monopole, it is easy to find a
corresponding solution for the dyon for which this bound is
saturated, with the fields satisfying
\be B_i=\sin\alpha\, \D_i\phi\ \ \ \ {\rm and}\ \ \ \
E_i=\cos\alpha\, \D_i\phi\label{dyons}\ee
This method of finding solutions in the worldvolume theory of a
soliton, and subsequently finding corresponding solutions in the
parent 4d theory, will be something we'll see several more times
in later sections.

\para
I have two further comments on dyons.
\begin{itemize}
\item We could add a theta term $\theta F\wedge F$ to the 4d
theory. Careful calculation of the electric Noether charges shows
that this induces an electric charge $\vec{q}=\theta\vec{g}/2\pi$
on the monopole. In the presence of the theta term, monopoles
become dyons. This is known as the Witten effect \cite{witeff}.

\item Both the dyons arising from \eqn{dyons}, and those arising
from the Witten effect, have $\vec{q}\sim \vec{g}$. One can create
dyons whose electric charge vector is not parallel to the magnetic
charge by turning on a vev for a second, adjoint scalar field
\cite{bergman,leeyi14}. These states are $1/4$-BPS in $\N=4$ super
Yang-Mills and correspond to $(p,q)$-string webs stretched between
D3-branes. From the field theory perspective, the dynamics of
these dyons is described by motion on the monopole moduli space
with a potential induced by the second scalar vev
\cite{mepot,blly,gjly}.\end{itemize}

\subsection{Fermi Zero Modes}

As with instantons, when the theory includes fermions they may be
turned on in the background of the monopole without raising the
energy of the configuration. A Dirac fermion $\lambda$ in the
adjoint representation satisfies
\be i\gamma^\mu\D_\mu \lambda - i[\phi,\lambda]=0\ee
Each such fermion carried $4\sum_a{n_a}$ zero modes.

\para
Rather than describing this in detail, we can instead resort again
to supersymmetry. In $\N=4$ super Yang-Mills, the monopoles
preserve one-half the supersymmetry, corresponding to $\N=(4,4)$
on the monopole worldvolume. While, monopoles in $\N=2$
supersymmetric theories preserve $\N=(0,4)$ on their worldvolume.
Monopoles in $\N=1$ theories are not BPS; they preserve no
supersymmetry on their worldvolume.

\para
There is also an interesting story with fermions in the
fundamental representation, leading to the phenomenon of solitons
carrying fractional fermion number \cite{jrewing}. A nice
description of this can be found in \cite{harveymon}.

\subsection{Nahm's Equations}

In the previous section we saw that the ADHM construction gave a
powerful method for understanding instantons, and that it was
useful to view this from the perspective of D-branes in string
theory. You'll be pleased to learn that there exists a related
method for studying monopoles. It's known as the Nahm construction
\cite{nahm}. It was further developed for arbitrary classical
gauge group in \cite{hurtmurray}, while the presentation in terms
of D-branes was given by Diaconescu \cite{diac}.
\begin{figure}[htb]
\begin{center}
\epsfxsize=4.6in\leavevmode\epsfbox{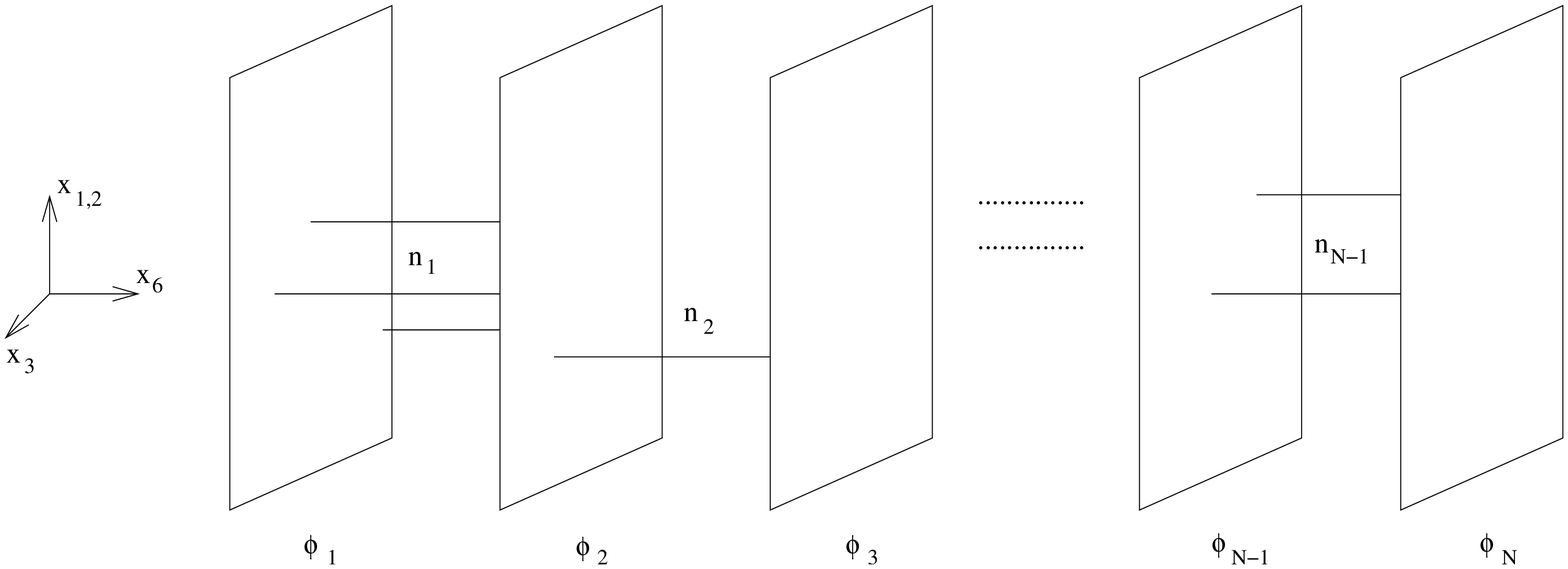}
\end{center}
\caption{The D-brane set-up for monopoles of charge
$\vg=\sum_an_a\valpha_a$.}
\end{figure}
\para
We start with $\N=4$ $U(N)$ super Yang-Mills, realized on the
worldvolume of D3-branes. To reflect the vev
$\bra\phi\ket=\diag(\phi_1\ldots,\phi_N)$, we separate the
D3-branes in a transverse direction, say the $x^6$ direction. The
$a^{\rm th}$ D3-brane is placed at position $x_6=\phi_a$.

\para
As is well known, the W-bosons correspond to fundamental strings
stretched between the D3-branes. The monopoles are their magnetic
duals, the D-strings. At this point our notation for the magnetic
charge vector $\vg=\sum_an_a\valpha_a$ becomes more visual. This
monopole in sector $\vg$ is depicted by stretching $n_a$ D-strings
between the $a^{\rm th}$ and $(a+1)^{\rm th}$ D3-branes.

\para
Our task now is to repeat the analysis of lecture 1 that led to
the ADHM construction: we must read off the theory on the
D1-branes, which we expect give us a new perspective on the
dynamics of magnetic monopoles. From the picture it looks like the
dynamics of the D-strings will  be governed by something like a
$\prod_a U(n_a)$ gauge theory, with each group living on the
interval $\phi_a\leq x_6\leq \phi_{a+1}$. And this is essentially
correct. But what are the relevant equations dictating the
dynamics? And what happens at the boundaries?

\para
To get some insight into this, let's start by considering $n$
infinite D-strings, with worldvolume $x_0,x_6$, and with D3-brane
impurities inserted at particular points $x_6=\phi_a$, as shown
below.
\begin{figure}[htb]
\begin{center}
\epsfxsize=4.3in\leavevmode\epsfbox{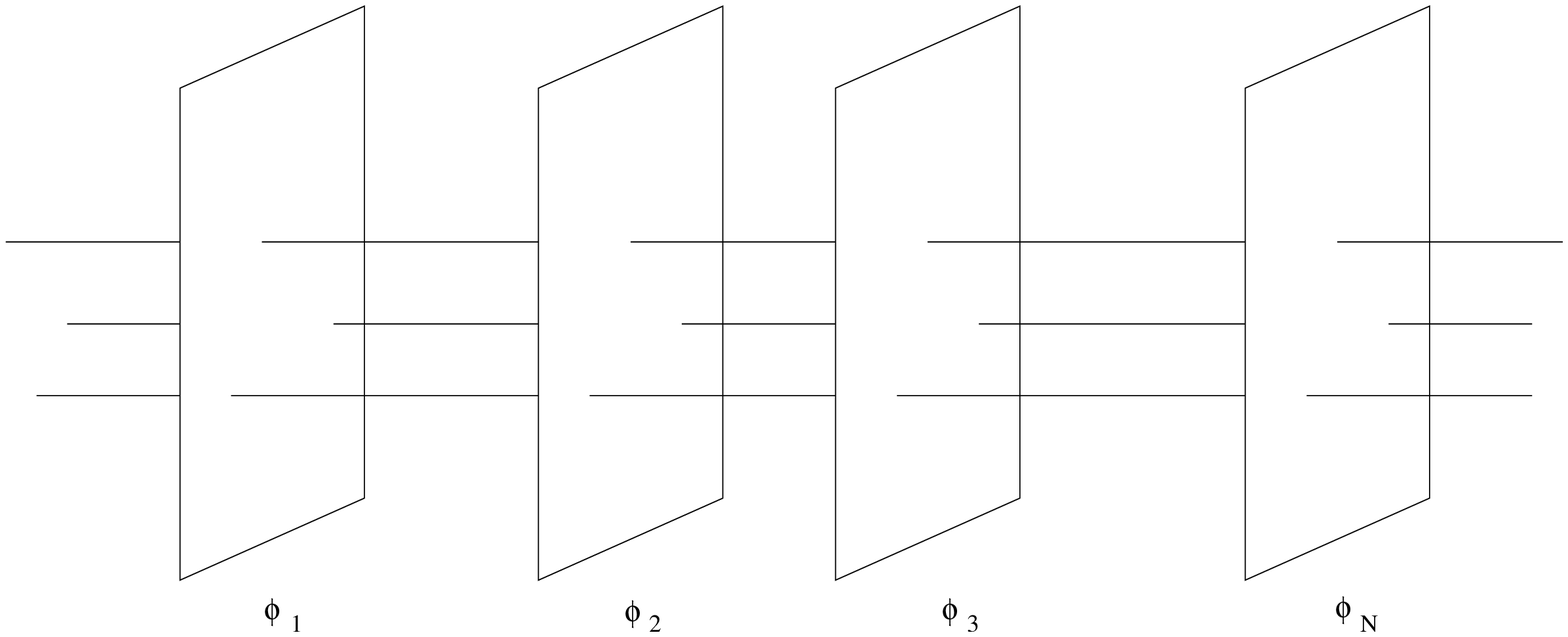}
\end{center}
\caption{The D3-branes give rise to impurities on the worldvolume
of the D1-branes.}
\end{figure}

The theory on the D-strings is a $d=1+1$ $U(n)$ gauge theory with
16 supercharges (known as $\N=(8,8)$). Each D3-brane impurity
donates a hypermultiplet to the theory, breaking supersymmetry by
half to $\N=(4,4)$. As in lecture 1, we write the hypermultiplets
as
\be \omega_a=\left(\begin{array}{c}\psi_a\\
\tilde{\psi}^\dagger_a\end{array}\right) \ \ \ \ a=1,\ldots,N \ee
where $\psi_a$ transforms in the ${\bf n}$ of $U(n)$, while
$\tilde{\psi}_a$ transforms in the $\bar{\bf n}$. The coupling of
these impurities (or defects as they're also known) is uniquely
determined by supersymmetry, and again occurs in a triplet of
D-terms (or, equivalently, a D-term and an F-term). In lecture 1,
I unapologetically quoted the D-term and F-term arising in the
ADHM construction (equation (1.44)) since they can be found in any
supersymmetry text book. However, now we have an impurity theory
which is a little less familiar. Nonetheless, I'm still going to
quote the result, but this time I'll apologize. We could derive
this interaction by examining the supersymmetry in more detail,
but it's easier to simply tell you the answer and then give a
couple of remarks to try and convince you that it's right. It
turns out that the (admittedly rather strange) triplet of D-terms
occurring in the Lagrangian is
\be \Tr\ \left(\frac{\partial X^i}{\partial
x^6}-i[A_6,X^i]-\frac{i}{2}\epsilon_{ijk}[X^j,X^k]+\sum_{a=1}^N
\,\omega^\dagger_a\sigma^i\omega_a\
\delta(x^6-\phi_a)\right)^2\label{nahmd}\ee
In the ground state of the D-strings, this term must vanish. Some
motivating remarks:
\begin{itemize}
\item The configuration shown in figure 7 arises from T-dualizing
the D0-D4 system. This viewpoint makes it clear that $A_6$ is the
right bosonic field to partner $X^i$ in a hypermultiplet.

\item Set $\partial_6=0$. Then, relabelling $A_6=X^4$, this term
is almost the same as the triplet of D-terms appearing in the ADHM
construction. The only difference is the appearance of the
delta-functions.

\item We know that D-strings can end on D3-branes. The
delta-function sources in the D-term are what allow this to
happen. For example, consider a single $n=1$ D-string, so that all
commutators above vanish. We choose $\tilde{\psi}=0$, to find the
triplet of D-terms
\be \partial_6 X^1=0\ \ \ \ ,\ \ \ \ \partial_6X^2=0\ \ \ \ \ , \
\ \ \ \ \partial_6X^3=|\psi|^2\delta(0) \ee
which allows the D-string profile to take the necessary step
(function) to split on the D3-brane as shown below.
\end{itemize}

\EPSFIGURE{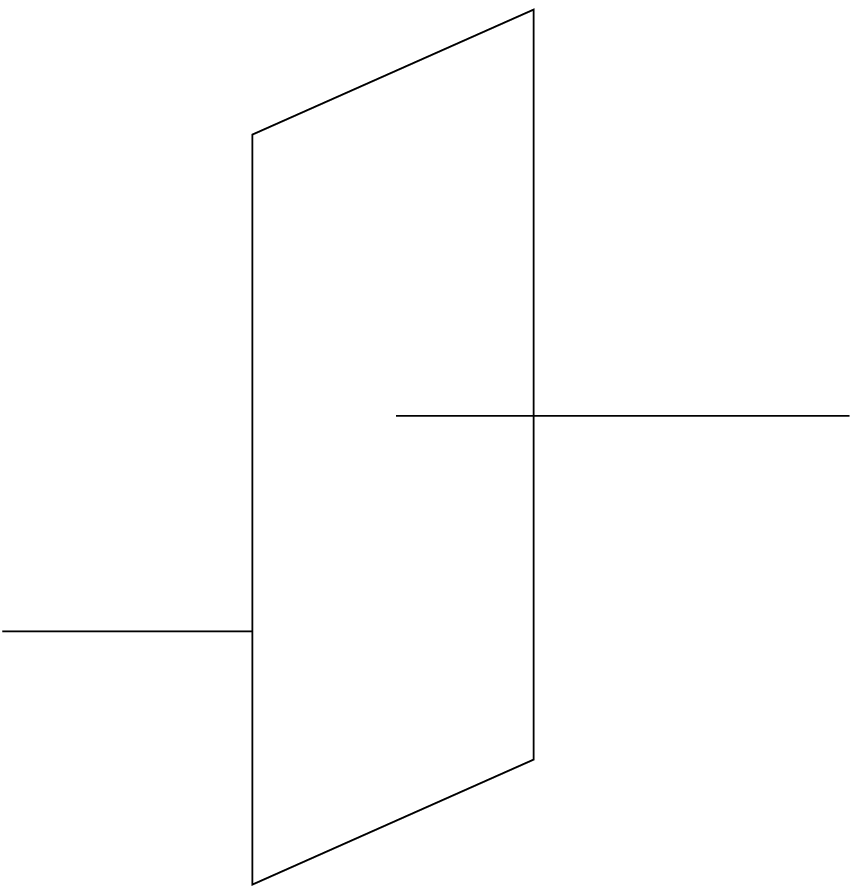,height=110pt}{}
If that wasn't enough motivation, one can find the full
supersymmetry analysis in the original papers \cite{sk,tsimpis}
and, in most detail, in \cite{defectdan}. Accepting \eqn{nahmd} we
can make progress in understanding monopole dynamics by studying
the limit in which several D-string segments, including the
semi-infinite end segments, move off to infinity, leaving us back
with the picture of figure 6.

\para
The upshot of this is that the dynamics of the
$\vec{g}=\sum_an_a\valpha_a$ monopoles are described as follows:
In the interval $\phi_a\leq x_6\leq\phi_{a+1}$, we have a $U(n_a)$
gauge theory, with three adjoint scalars $X^i$, $i=1,2,3$
satisfying
\be \frac{d X^i}{dx^6}-i[A_6,X^i]-\ft12 \epsilon^{ijk}[X^i,X^j]=0
\label{nahm}\ee
These are Nahm's equations.  The boundary conditions imposed at
the end of the interval depend on the number of monopoles in the
neighbouring segment. (Set $n_0=n_N=0$ in what follows)

\para
\underline{$n_a=n_{a+1}$:} The $U(n_a)$ gauge symmetry is extended
to the interval $\phi_a\leq x^6\leq \phi_{a+2}$ and an impurity is
added to the right-hand-side of Nahm's equations
\be \ \frac{d X^i}{dx^6}-i[A_6,X^i]-\ft12
\epsilon^{ijk}[X^i,X^j]=\omega^\dagger_{a+1}\sigma^i\omega_{a+1}\
\delta(x^6-\phi_{a+1})\label{nahmimp}\ee
This, of course, follows immediately from \eqn{nahmd}.

\para
\underline{$n_a=n_{a+1}-1$:} In this case, $X^i\rightarrow
(X_i)_-$, a set of three constant $n_a\times n_a$ matrices as
$x^6\rightarrow(\phi_{a+1})_-$. To the right of the impurity, the
$X^i$ are $(n_a+1)\times (n_a+1)$ matrices. They are required to
satisfy the boundary condition
\be X^i\rightarrow\left(\begin{array}{cc}y^i & a^{i\dagger} \\
a^i & (X^i)_-\end{array}\right)\ \ \ \ \ \ \ {\rm as}\
x_6\rightarrow (\phi_{a+1})_+ \label{bc1}\ee
where $y^i\in\R$ and each $a^i$ is a complex $n_a$-vector. One can
derive this boundary condition without too much trouble by
starting with \eqn{nahmimp} and taking $|\omega|\rightarrow\infty$
to remove one of the monopoles \cite{erickxin}.

\para
\underline{$n_a\leq n_{a+1}-2$} Once again  $X^i\rightarrow
(X^i)_-$ as $x_6\rightarrow (\phi_{a+1})_-$ but, from the other
side, the matrices $X_\mu$ now have a simple pole at the boundary,
\be X^i\rightarrow \left(\begin{array}{cc} J^i/s + Y^i & {\cal
O}(s^\gamma)
\\ {\cal O}(s^\gamma) & (X^i)_- \end{array}\right)\ \ \ \ \ {\rm as\ }
x_6\rightarrow (\phi_{a+1})_+ \label{bc2}\ee
Here $s=(x^6-\phi_{a+1})$ is the distance to the impurity. The
matrices $J^i$ are the irreducible $(n_{a+1}-n_a)\times
(n_{a+1}-n_a)$ representation of $su(2)$, and $Y^i$ are now
constant $(n_{a+1}-n_a)\times (n_{a+1}-n_a)$ matrices. Note that
the simple pole structure is compatible with Nahm's equations,
with both the derivative and the commutator term going like
$1/s^2$. Finally, $\gamma=\ft12(n_{a+1}-n_a-1)$, so the
off-diagonal terms vanish as we approach the boundary. The
boundary condition \eqn{bc2} can also be derived from \eqn{bc1} by
removing a monopole to infinity \cite{erickxin}.

\para
When $n_a>n_{a+1}$, the obvious parity flipped version of the
above conditions holds.

\subsubsection{Constructing the Solutions}

Just as in the case of ADHM construction, Nahm's equations capture
information about both the monopole solutions and the monopole
moduli space. The space of solutions to Nahm's equations
\eqn{nahm}, subject to the boundary conditions detailed above, is
isomorphic to the monopole moduli space $\mg$. The phases of each
monopole arise from the gauge field $A_6$, while $X^i$ carry the
information about the positions of the monopoles. Moreover, there
is a natural metric on the solutions to Nahm's equations which
coincides with the metric on the monopole moduli space. I don't
know if anyone has calculated the Atiyah-Hitchin metric using Nahm
data, but a derivation of the Lee-Weinberg-Yi metric was given in
\cite{murray}.

\para
Given a solution to Nahm's equations, one can explicitly construct
the corresponding solution to the monopole equation. The procedure
is  analogous to the construction of instantons in 1.4.2, although
its a little harder in practice as its not entirely algebraic. We
now explain how to do this. The first step is to build a
Dirac-like operator from the solution to \eqn{nahm}. In the
segment $\phi_a\leq x^6 \leq \phi_{a+1}$, we construct the Dirac
operator
\be \Delta = \frac{d}{d x^6}-iA_6-i(X^i+r^i)\sigma^i \ee
where we've reintroduced the spatial coordinates $r^i$ into the
game. We then look for normalizable zero modes $U$ which are
solutions to the equation
\be \Delta U =0 \ee
One can show that there are $N$ such solutions, and so we consider
$U$ as a $2n_a\times N$-dimensional matrix. Note that the
dimension of $U$ jumps as we move from one interval to the next.
We want to appropriately normalize $U$, and to do so choose to
integrate over all intervals, so that
\be\int_{\phi_1}^{\phi_N}\,dx^6\ U^\dagger U={\bf 1}_N\ee
Once we've figured out the expression for $U$, a Higgs field
$\phi$ and a gauge field $A_i$ which satisfy the monopole equation
are given by,
\be \phi=\int_{\phi_1}^{\phi_N}\,dx^6\ x^6\,U^\dagger U\ \ \ \ ,\
\ \  \ A_i = \int_{\phi_1}^{\phi_N}\, dx^6\ U^\dagger\partial_6
U\ee
The similarity between this procedure and that described in
section 1.4.2 for instantons should be apparent.

\para
In fact, there's a slight complication that I've brushed under the
rug. The above construction only really holds when $n_a\neq
n_{a+1}$. If we're in a situation where $n_a=n_{a+1}$ for some
$a$, then we have to take the hypermultiplets $\omega_a$ into
account, since their value  affects the monopole solution. This
isn't too hard --- it just amounts to adding some extra discrete
pieces to the Dirac operator $\Delta$. Details can be found in
\cite{hurtmurray}.

\para
A string theory derivation of the construction part of the Nahm
construction was recently given in \cite{kojiconst}.

\subsubsection*{An Example: The Single $SU(2)$ Monopole Revisited}

It's very cute to see the single $n=1$ solution \eqn{monsol} for
the $SU(2)$ monopole   drop out of this construction. This is
especially true since the Nahm data is trivial in this case:
$X^i=A_6=0$!

\para
To see how this arises, we look for solutions to
\be \Delta U = \left(\frac{d}{dx^6}-r^i\sigma^i\right)U=0\ee
where $U=U(x^6)$ is a $2\times 2$ matrix. This is trivially solved
by
\be U=\sqrt{\frac{r}{\sinh (2vr)}}\left(\,\cosh(rx^6)\,{\bf
1}_2+\sinh(rx^6) \,\hat{r}^i\sigma^i\,\right)\ee
which has been designed to satisfy the normalizability condition
$\int^{+v}_{-v}U^\dagger U\ dx^6 ={\bf 1}_2$. Armed with this, we
can easily reproduce the monopole solution \eqn{monsol}. For
example, the Higgs field is given by
\be \phi=\int_{-v}^{+v}dx^6\ x^6U^\dagger U =
\frac{\hat{r}^i\sigma^i}{r}\left(vr\coth(vr)-1\right)\ee
as promised. And the gauge field $A_i$ drops out just as easily.
See --- told you it was cute! Monopole solutions with charge of
the type $(1,1,\ldots,1)$ were constructed using this method in
\cite{epj}.

\subsection{What Became of Instantons}

In the last lecture we saw that pure Yang-Mills theory contains
instanton solutions. Now we've added a scalar field, where have
they gone?! The key point to note is that the theory was conformal
before $\phi$ gained its vev. As we saw in Lecture 1, this led to
a collective coordinate $\rho$, the scale size of the instanton.
Now with $\bra\phi\ket\neq 0$ we have introduced a mass scale into
the game and no longer expect $\rho$ to correspond to an exact
collective coordinate. This turns out to be true: in the presence
of a non-zero vev $\bra\phi\ket$, the instanton minimizes its
action by shrinking to zero size $\rho\rightarrow 0$. Although,
strictly speaking, no instanton exists in the theory with
$\bra\phi\ket\neq 0$, they still play a crucial role. For example, the
famed Seiberg-Witten solution can be thought of as summing these
small instanton corrections.

\begin{figure}[htb]
\begin{center}
\epsfxsize=4.2in\leavevmode\epsfbox{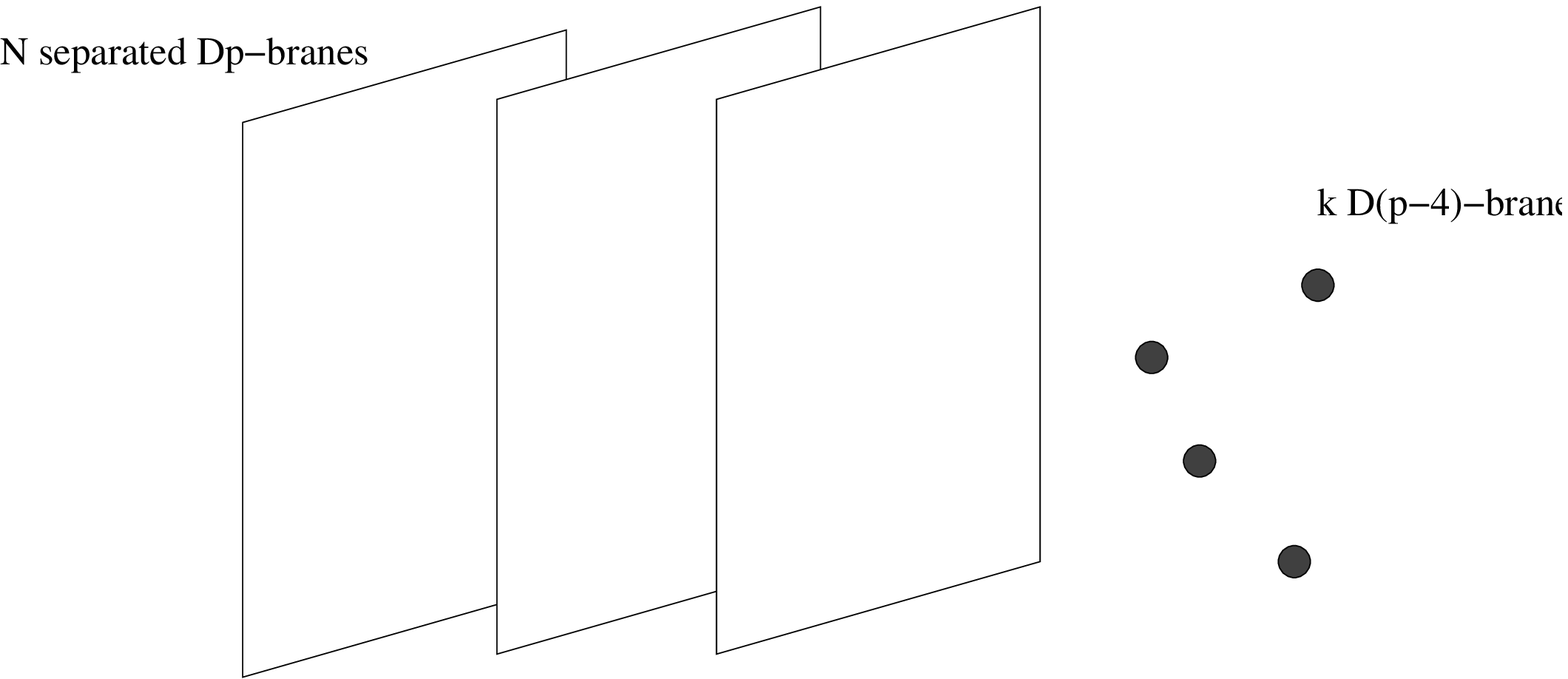}
\end{center}
\caption{Separating the Dp-branes gives rise to a mass for the
hypermultiplets}
\end{figure}
\para
How can we see this behavior from the perspective of the
worldvolume theory? We can return to the D-brane set-up, now with
the D$p$-branes separated in one direction, say $x_6$, to mimic
the vev $\bra\phi\ket$. Each D$p$-D$(p-4)$ string is now stretched
by a different amount, reflecting the fact that each
hypermultiplet has a different mass. The potential on the
worldvolume theory of the D-instantons is now
\be V&=&\frac{1}{g^2} \sum_{m,n=5}^{10}[X_m,X_n]^2
+\sum_{m,\mu}[X_m,X_\mu]^2 +\sum_{a=1}^N \psi^{a\dagger}
(X_m-\phi^m_{\ a})^2\psi_a+\tilde{\psi}^a(X_m-\phi^m_{\
a})^2\tilde{\psi}^\dagger_a\nn
\\&& + g^2\,\Tr\,(\sum_{a=1}^N \psi_a\psi^{a\dagger}-
\tilde{\psi}^\dagger_a\tilde{\psi}^a +[Z,Z^\dagger]
+[W,W^\dagger])^2
+g^2\,\Tr\,|\sum_{a=1}^N\psi_a\tilde{\psi}^a+[Z,W]|^2 \nn\ee
We've actually introduced more new parameters here than we need,
since the D3-branes can be separated in 6 different dimensions, so
we have the corresponding positions $\phi^m_{\ a}$, $m=4,\ldots,9$
and $a=1,\ldots N$. Since we have been dealing with just a single
scalar field $\phi$ in this section, we will set $\phi^m_{\ i}=0$
except for $m=6$ (I know...why 6?!). The parameters $\phi^6_{\
a}=\phi_{a}$ are the components of the vev \eqn{phivev}.

\para
We can now re-examine the vacuum condition for the Higgs branch.
If we wish to try to turn on $\psi$ and $\tilde{\psi}$, we must
first set $X_m=\phi_a$, for some $a$. Then the all $\psi_b$ and
$\tilde{\psi}^b$ must vanish except for $b=a$. But, taking the
trace of the D- and F-term conditions tells us that even $\psi_a$
and $\tilde{\psi}_a$ vanish. We have lost our Higgs branch
completely. The interpretation is that the instantons have shrunk
to zero size. Note that in the case of non-commutativity, the
instantons don't vanish but are pushed to the $U(1)$ instantons
with, schematically, $|\psi|^2\sim\zeta$.

\para
Although the instantons shrink to zero size, there's still
important information to be gleaned from the potential above. One
can continue to  think of the instanton moduli space $\ikn\cong
\M_{\rm Higgs}$ as before, but now with a potential over it. This
potential arises after integrating out the $X_m$ and it is not
hard to show that it is of a very specific form: it is the
length-squared of a triholomorphic Killing vector on $\ikn$
associated with the $SU(N)$ isometry.

\para
This potential on $\ikn$ can be derived directly within  field
theory without recourse to D-branes or the ADHM construction
\cite{dyonicme}. This is the route we follow here. The question we
want to ask is: given an instanton solution, how does the presence
of the $\phi$ vev affect its action? This gives the potential on
the instanton moduli space which is simply
\be V=\int d^4x\ \Tr\,(\D_\mu\phi)^2 \ee
where $\D_\mu$ is evaluated on the background instanton solution.
We are allowed to vary $\phi$ so it minimizes the potential so
that, for each solution to the instanton equations, we want to
find $\phi$ such that
\be \D^2\phi=0\ee
with the boundary condition that
$\phi\rightarrow\langle\phi\rangle$. But we've seen an equation of
this form, evaluated on the instanton background, before. When we
were discussing the instanton zero modes in section 1.2, we saw
that the zero modes arising from the overall $SU(N)$ gauge
orientation were of the form $\delta A_\mu=\D_\mu\Lambda$, where
$\Lambda$ tends to a constant at infinity and satisfies the gauge
fixing condition $\D_\mu\d A_\mu=0$. This means that we can
re-write the potential in terms of the overlap of zero modes
\be V=\int d^4x\ \Tr\ \d A_\mu \d A_\mu \ee
for the particular zero mode $\d A_\mu=\D_\mu\phi$ associated to
the gauge orientation of the instanton. We can give a nicer
geometrical interpretation to this. Consider the action of the
Cartan subalgebra $\vec{H}$ on $\ikn$ and denote the corresponding
Killing vector as $\vec{k}=\vec{k}^\alpha\partial_\alpha$. Then,
since $\phi$ generates the transformation
$\vec{\phi}\cdot\vec{H}$, we can express our zero mode in terms of
the basis $\d A_\mu=(\vec{\phi}\cdot\vec{k}^\alpha)\,\d_\alpha
A_\mu$. Putting this in our potential and performing the integral
over the zero modes, we have the final expression
\be V=
g_{\alpha\beta}\,(\vec{\phi}\cdot\vec{k}^\alpha)\,(\vec{\phi}\cdot
\vec{k}^\beta)\label{kpot}\ee
The potential vanishes at the fixed points of the $U(1)^{N-1}$
action. This is the small instanton singularity (or related points
on the blown-up cycles in the resolved instanton moduli space).
Potentials of the form \eqn{kpot} were first discussed by
Alvarez-Gaume and Freedman who showed that, for tri-holomorphic
Killing vectors $k$, they are the unique form allowed in a
sigma-model preserving eight supercharges \cite{agf}.

\para
The concept of a potential on the instanton moduli space $\ikn$ is
the modern way of viewing what used to known as the "constrained
instanton", that is an approximate instanton-like solution to the
theory with $\bra\phi\ket\neq 0$ \cite{affleck}. These potentials
play an important role in Nekrasov's first-principles computation
of the Seiberg-Witten prepotential \cite{nekr}. Another
application occurs in the five-dimensional theory, where
instantons are particles. Here the motion on the moduli space may
avoid the fate of falling to the zeroes of  \eqn{kpot} by spinning
around the potential like a motorcyclist on the wall of death.
These solutions of the low-energy dynamics are dyonic instantons
which carry electric charge in five dimensions
\cite{dyonicme,eedyonic,kpmzdyonic}.

\subsection{Applications}

Time now for the interesting applications, examining the role that
monopoles play in the quantum dynamics of supersymmetric gauge
theories in various dimensions. We'll look at monopoles in 3, 4, 5
and 6 dimensions in turn.

\subsubsection{Monopoles in Three Dimensions}

In $d=2+1$ dimensions, monopoles are finite action solutions to
the Euclidean equations of motion and the  role they play is the
same as that of instantons in $d=3+1$ dimensions: in a
semi-classical evaluation of the path-integral, one must sum over
these monopole configurations. In 1975, Polyakov famously showed
how a gas of these monopoles leads to linear confinement in
non-supersymmetric Georgi-Glashow model \cite{poly} (that is, an
$SU(2)$ gauge theory broken to $U(1)$ by an adjoint scalar field).

\para
In supersymmetric theories, monopoles give rise to somewhat
different physics. The key point is that they now have fermionic
zero modes, ensuring that they can only contribute to correlation
functions with a suitable number of fermionic insertions to soak
up the integrals over the Grassmannian collective coordinates. In
${\cal N}=1$ and ${\cal N}=2$ theories\footnote{A (foot)note on
nomenclature. In any dimension, the number of supersymmetries
${\cal N}$ counts the number of supersymmetry generators in units
of the minimal spinor. In $d=2+1$ the minimal Majorana spinor has
2 real components. This is in contrast to $d=3+1$ dimensions where
the minimal Majorana (or equivalently Weyl) spinor has 4 real
components. This leads to the unfortunate fact that ${\cal N}=1$
in $d=3+1$ is equivalent to ${\cal N}=2$ in $d=2+1$. It's
annoying. The invariant way to count is in terms of supercharges.
Four supercharges means ${\cal N}=1$ in four dimensions or ${\cal
N}=2$ in three dimensions.} in $d=2+1$ dimensions, instantons
generate superpotentials, lifting moduli spaces of vacua
\cite{ahw}. In ${\cal N}=8$ theories, instantons contribute to
particular 8 fermi correlation functions which have a beautiful
interpretation in terms of membrane scattering in M-theory
\cite{pp,dkm3d}. In this section, I'd like to describe one of the
nicest applications of monopoles in three dimensions which occurs
in theories with ${\cal N}=4$ supersymmetry, or 8 supercharges.

\para
We'll consider ${\cal N}=4$ $SU(2)$ super Yang-Mills. The
superpartners of the gauge field include 3 adjoint scalar fields,
$\phi^\alpha$, $\alpha=1,2,3$ and 2 adjoint Dirac fermions. When
the scalars gain an expectation value
$\langle\phi^\alpha\rangle\neq 0$, the gauge group is broken
$SU(2)\rightarrow U(1)$ and the surviving, massless, bosonic
fields are 3 scalars and a photon. However, in $d=2+1$ dimensions,
the photon has only a single polarization and can be exchanged in
favor of another scalar $\sigma$. We achieve this by a duality
transformation:
\be
F_{ij}=\frac{e^2}{2\pi}\epsilon_{ijk}\partial^k\,\sigma\label{duality}\ee
where we have chosen normalization so that the scalar $\sigma$ is
periodic: $\sigma=\sigma+2\pi$. Since supersymmetry protects these
four scalars against becoming massive, the most general low-energy
effective action we can write down is the sigma-model
\be L_{\rm
low-energy}=\frac{1}{2e^2}\,g_{\alpha\beta}\,\partial_i\phi^\alpha\,\partial^i\phi^\beta
\ee
where $\phi^\alpha=(\phi^1,\phi^2,\phi^3,\sigma)$. Remarkably, as
shown by Seiberg and Witten, the metric $g_{\alpha\beta}$ can be
determined uniquely \cite{sw3d}. It turns out to be an old friend:
it is the Atiyah-Hitchin metric \eqn{ahm}! The dictionary is
$\phi^i=e^2r^i$ and $\sigma=\psi$. Comparing with the functions
$a$, $b$ and $c$ listed in \eqn{abc}, the leading constant term
comes from tree level in our 3d gauge theory, and the $1/r$ terms
arise from a one-loop correction. Most interesting is the $e^{-r}$
term in \eqn{abc}. This comes from a semi-classical monopole
computation in $d=2+1$ which can be computed exactly \cite{dkmvt}.
So we find monopoles arising in two very different ways: firstly
as an instanton-like configuration in the 3d theory, and secondly
in an auxiliary role as the description of the low-energy
dynamics. The underlying reason for this was explained by Hanany
and Witten \cite{hw}, and we shall see a related perspective in
section 2.7.4.

\para
So the low-energy dynamics of $\N=4$ $SU(2)$ gauge theory is
dictated by the two monopole moduli space. It can also be shown
that the low-energy dynamics of the $\N=4$ $SU(N)$ gauge theory in
$d=2+1$ is governed by a sigma-model on the moduli space of $N$
magnetic monopoles in an $SU(2)$ gauge group \cite{ch}. There are
3d quiver gauge theories related to monopoles in higher rank,
simply laced (i.e. ADE) gauge groups \cite{hw,menocite} but, to my
knowledge, there is no such correspondence for monopoles in
non-simply laced groups.

\subsubsection{Monopoles and Duality}

Perhaps the most important application of monopoles is the role
they play in uncovering the web of dualities relating various
theories. Most famous is the S-duality of $\N=4$ super Yang-Mills
in four dimensions. The idea is that we can re write the gauge
theory treating magnetic monopoles as elementary particles rather
than solitons \cite{monolive}. The following is a lightening
review of this large subject. Many more details can be found in
\cite{harveymon}.

\para
The conjecture of S-duality states that we may re-express the
theory, treating monopoles as the fundamental objects, at the
price of inverting the coupling $e\rightarrow 4\pi/e$. Since this
is a strong-weak coupling duality, we need to have some control
over the strong coupling behavior of the theory to test the
conjecture. The window on this regime is provided by the BPS
states \cite{wo}, whose mass is not renormalized in the maximally
supersymmetric $\N=4$ theory which, among other reasons, makes it
a likely place to look for S-duality \cite{osmon}. In fact, this
theory exhibits a more general $SL(2,\Z)$ group of duality
transformations which acts on the complexified coupling
$\tau=\theta/2\pi + 4\pi i /e^2$ by
\be \tau\longrightarrow \frac{a\tau+b}{c\tau + d}\ \ \ \ \ {\rm
with} \ a,b,c,d\in\Z\ \ {\rm and}\ ad-bc=1 \ee
A transformation of this type mixes up what we mean by electric
and magnetic charges. Let's work in the $SU(2)$ gauge theory for
simplicity so that electric and magnetic charges in the unbroken
$U(1)$ are each specified by an integer $(n_e,n_m)$. Then under
the $SL(2,\Z)$ transformation,
\be \left(\begin{array}{c} n_e \\ n_m
\end{array}\right)\longrightarrow \left(\begin{array}{lr} a & -b \\ c
& -d \end{array}\right)\left(\begin{array}{c} n_e \\
n_m\end{array}\right) \ee
The conjecture of S-duality has an important prediction that can
be tested semi-classically: the spectrum must form multiplets
under the $SL(2,Z)$ transformation above. In particular, if
S-duality holds, the existence of the W-boson state
$(n_e,n_m)=(1,0)$ implies the existence of a slew of further
states with quantum numbers $(n_e,n_m)=(a,c)$ where $a$ and $c$
are relatively prime.  The states with magnetic charge $n_m=c=1$
are the dyons that we described in Section 2.3 and can be shown to
exist in the quantum spectrum. But we have to work much harder to
find the states with magnetic charge $n_m=c>1$. To do so  we must
examine the low-energy dynamics of $n_m$ monopoles, described by
supersymmetric quantum mechanics on the monopole moduli space.
Bound states saturating the Bogomoln'yi bound correspond to ground
states of the quantum mechanics. But, as we described in section
1.5.2, this  questions translates into the more geometrical search
for normalizable harmonic forms on the monopole moduli space.

\para
In the $n_m=2$ monopole sector, the bound states were explicitly
demonstrated to exist by Sen \cite{senmon}. S-duality predicts the
existence of a tower of dyon states with charges $(n_e,2)$ for all
$n_e$ odd which  translates into the requirement that there is a
unique harmonic form $\omega$ on the Atiyah-Hitchin manifold. The
electric charge still comes from motion in the ${\S}^1$ factor of
the monopole moduli space \eqn{mah}, but the need for only odd
charges $n_e$ to exist requires that the form $\omega$ is odd
under the $\Z_2$ action \eqn{ahz2}. Uniqueness requires that
$\omega$ is either self-dual or anti-self-dual. In fact, it is the
latter. The ansatz,
\be \omega=F(r)(d\sigma_1-\frac{fa}{bc}dr\wedge\sigma_1)\ee
is harmonic provided that $F(r)$ satisfies
\be \frac{dF}{dr}=-\frac{fa}{bc}F\ee
One can show that this form is normalizable, well behaved at the
center of the moduli space and, moreover, unique. Historically,
the existence of this form was one the compelling pieces of
evidence in favor of S-duality, leading ultimately to an
understanding of strong coupling behavior of many supersymmetric
field theories and string theories.

\para
The discussion above is for $\N=4$ theories. In $\N=2$ theories,
the bound state described above does not exist (a study of the
$\N=(0,4)$ supersymmetric quantum mechanics reveals that the
Hilbert space is identified with holomorphic forms and $\omega$ is
not holomorphic). Nevertheless, there exists a somewhat more
subtle duality between electrically and magnetically charged
states, captured by the Seiberg-Witten solution \cite{sw}. Once
again, there is a semi-classical test of these ideas along the
lines described above \cite{harvgaun}. There is also an interesting
story in this system regarding quantum corrections to the monopole
mass \cite{rebh}.

\subsubsection{Monopole Strings and the $(2,0)$ Theory}

We've seen that the moduli space of a single monopole is ${\cal
M}\cong \R^3\times \S^1$ with metric,
\be ds^2 = M_{\rm
mono}\left(dX^idX^i+\frac{1}{v^2}\,d\chi^2\right)
\label{doesit}\ee
where $\chi\in[0,2\pi)$. It looks as if, at low-energies, the
monopole is moving in a higher dimensional space. Is there any
situation where we can actually interpret this motion in the
$\S^1$ as motion in  an extra, hidden dimension of space?

\para
One problem with interpreting internal degrees of freedom, such as
$\chi$, in terms of  an extra dimension is that there is no
guarantee that motion in these directions will be Lorentz
covariant. For example, Einstein's speed limit tells us that the
motion of the monopole in ${\bf R}^3$ is bounded by the speed of
light: i.e. $\dot{X}\leq 1$. But is there a similar bound on
$\dot{\chi}$\ ? This is an question which goes beyond the moduli
space approximation, which keeps only the lowest velocities, but
is easily answered since we know the exact spectrum of the dyons.
The energy of a relativistically moving dyon is $E^2=M_{\rm
dyon}^2+p_ip_i$, where $p_i$ is the momentum conjugate to the
center of mass $X_i$. Using the mass formula \eqn{dyon}, we have
the full Hamiltonian
\be H_{\rm dyon}=\sqrt{M_{\rm
mono}^2+v^2p_\chi^2+p_ip_i}\label{relham}\ee
where $p_\chi=2q$ is the momentum conjugate to $\chi$. This gives
rise to the Lagrangian,
\be L_{\rm dyon}=-M_{\rm
mono}\sqrt{1-\dot{\chi}^2/v^2-\dot{X}^i\dot{X}^i}\label{rela}\ee
which, at second order in velocities, agrees with the motion on
the moduli space \eqn{doesit}. So, surprisingly, the internal
direction $\chi$ does appear in a Lorentz covariant manner in this
Lagrangian and is therefore a candidate for an extra, hidden,
dimension.

\para
However, looking more closely, our hopes are dashed. From
\eqn{rela} (or, indeed, from \eqn{doesit}), we see that the radius
of the extra dimension is proportional to $1/v$. But the width of
the monopole core is also $1/v$. This makes it a little hard to
convincingly argue that the monopole can happily move in this
putative extra dimension since there's no way the dimension can be
parametrically larger than the monopole itself. It appears that
$\chi$ is stuck in the auxiliary role of endowing monopoles with
electric charge, rather than being promoted to a physical
dimension of space.

\para
Things change somewhat if we consider the monopole as a
string-like object in a $d=4+1$ dimensional gauge theory. Now the
low-energy effective action for a single monopole is simply the
action \eqn{rela} lifted to the two dimensional worldsheet of the
string, yielding the familiar Nambu-Goto action
\be  S_{\rm string}=-T_{\rm mono}\int d^2y\
\sqrt{1-(\partial{\chi})^2/v^2-(\partial{X}^i)^2}\label{rela2}\ee
where $\partial$ denotes derivatives with respect to both
worldsheet coordinates, $\sigma$ and $\tau$. We've rewritten
$M_{\rm mono}=T_{\rm mono}=4\pi v/e^2$ to stress the fact that it
is a tension, with dimension 2 (recall that $e^2$ has dimension
$-1$ in $d=4+1$). As it stands, we're in no better shape. The size
of the circle is still $1/v$, the same as the width of the
monopole string. However, now we have a two dimensional worldsheet
we may apply T-duality. This means exchanging momentum modes
around ${\bf S}^1$ for winding modes  so that
\be\partial\chi={}^\star\partial\tilde{\chi}\ee
We need to be careful with the normalization. A careful study
reveals that,
\be \frac{1}{4\pi}\int\,d^2y\,R^2\,(\partial\chi)^2\ \rightarrow\
\frac{1}{4\pi}\int\,d^2y\,
\frac{1}{R^2}\,(\partial\tilde{\chi})^2\ee
where, up to that important factor of $4\pi$, $R$ is the radius of
the circle measured in string units. Comparing with our
normalization, we have $R^2=8\pi^2/ve^2$, and the dual Lagrangian
is
\be S_{\rm string}=-T_{\rm mono}\int d^2y\
\sqrt{1-(e^2/8\pi^2)^2(\partial{\tilde{\chi}})^2-(\partial{X}^i)^2}\ee
We see that the physical radius of this dual circle is now
$e^2/8\pi^2$. This can be arbitrarily large and, in particular,
much larger than the width of the monopole string. It's a prime
candidate to be interpreted as a real, honest, extra dimension. In
fact, in the maximally supersymmetric Yang-Mills theory in five
dimensions, it is known that this extra dimension is real. It is
precisely the hidden circle that takes us up to the
six-dimensional $(2,0)$ theory that we discussed in section 1.5.2.
The monopole even tells us that the instantons must be the
Kaluza-Klein modes since the inverse radius of the dual circle is
exactly $M_{\rm inst}$. Once again, we see that solitons allow us
to probe important features of the quantum physics where myopic
perturbation theory fails. Note that the derivation above does
rely on supersymmetry since, for the Hamiltonian \eqn{relham} to
be exact, we need the masses of the dyons to saturate the
Bogomoln'yi bound \eqn{dyon}.

\subsubsection{D-Branes in Little String Theory}

Little string theories are strongly interacting string theories
without gravity in $d=5+1$ dimensions. For a review see
\cite{ofer}. The maximally supersymmetric variety can be thought
of as the decoupled theory living on NS5-branes. They come in two
flavors: the type iia little string theory is a $(2,0)$
supersymmetric theory which reduces at low-energies to the
conformal field theory discussed in sections 1.5.2 and 2.7.3. In
contrast, the type iib little string has $(1,1)$ non-chiral
supersymmetry and reduces at low-energies to $d=5+1$ Yang-Mills
theory. When this theory sits on the Coulomb branch it admits
monopole solutions which, in six dimensions, are membranes. Let's
discuss some of the properties of these monopoles in the $SU(2)$
theory.

\para
The low-energy dynamics of a single monopole is the $d=2+1$
dimensional sigma model with target space $\R^3\times \S^1$ and
metric \eqn{doesit}. But, as we already discussed, in $d=2+1$ we
can exchange the periodic scalar $\chi$ for a $U(1)$ gauge field
living on the monopole. Taking care of the normalization, we find
\be F_{mn}=\frac{8\pi^2}{e^2}\epsilon_{mnp}\,
\partial^p\chi\label{3ddual}\ee
with $m,n=0,1,2$ denoting the worldvolume dimensions of the
monopole 2-brane. The low-energy dynamics of this brane can
therefore be written as
\be S_{\rm brane}&=&\int\,d^3x\ \frac{1}{2}T_{\rm
mono}\left((\partial_m
X^i)^2+\frac{1}{v^2}(\partial_m\chi)^2\right) \label{theone}\\
&=&\int\,d^3x\ \frac{1}{2g^2}\left(
(\partial_m\varphi^i)^2+\frac{1}{2}F_{mn}F^{mn}\right)\label{thetwo}\ee
where $g^2=4\pi^2T_{\rm mono}/v^2$ is fixed by the duality
\eqn{3ddual} and insisting that the scalar has canonical kinetic
term dictates $\varphi^i = (8\pi^2/e^2)\,X^i=T_{\rm inst}X^i$.
This normalization will prove important. Including the fermions,
we therefore find  the low-energy dynamics of a monopole membrane
to be free $U(1)$ gauge theory with 8 supercharges (called ${\cal
N}=4$ in three dimensions), containing a photon and three real
scalars.

\para
Six dimensional gauge theories also contain instanton strings.
These are the "little strings" of little string theory. We will
now show that strings can end on the monopole 2-brane. This is
simplest to see from the worldvolume perspective in terms of the
original variable $\chi$. Defining the complex coordinate on the
membrane worldvolume $z=x^4+ix^5$, we have the BPS "BIon" spike
\cite{bion1,bion2} solution of the theory \eqn{theone}
\be X^1+\frac{i}{v}\chi = \frac{1}{v} \log (vz)\ee
Plotting the value of the transverse position $X^1$ as a function
of $|z|$, we see that this solution indeed has the profile of a
string ending on the monopole 2-brane. Since $\chi$ winds once as
we circumvent the origin $z=0$, after the duality transformation
we see that this string sources a radial electric field. In other
words, the end of the string is charged under the $U(1)$ gauge
field on the brane \eqn{3ddual}. We have found a D-brane in the
six-dimensional little string theory.

\para
Having found the string solution from the perspective of the
monopole worldvolume theory, we can ask whether we can find a
solution in the full $d=5+1$ dimensional theory. In fact, as far
as I know, no one has done this. But it is possible to write down
the first order equations that this solution must solve
\cite{unpub}. They are the dimensional reduction of equations
found in \cite{cdfm} and read
\be & F_{23}+F_{45}=\D_1\phi\ \ \ ,\ \ \ F_{35}=-F_{42}\ \ \ ,\ \
\ F_{34}=-F_{25}& \nn\\ &F_{31}=\D_2\phi\ \ \ \ ,\ \ \ \
F_{12}=\D_3\phi\ \ \ \ ,\ \ \ \ F_{51}=\D_4\phi\ \ \ \ ,\ \ \ \
F_{14}=\D_5\phi&\ee
Notice that among the solutions to these equations are instanton
strings stretched in the $x^1$ directions, and monopole 2-branes
with spatial worldvolume $(x^4,x^5)$. It would be interesting to
find an explicit solution describing the instanton string ending
on the monopole brane.

\EPSFIGURE{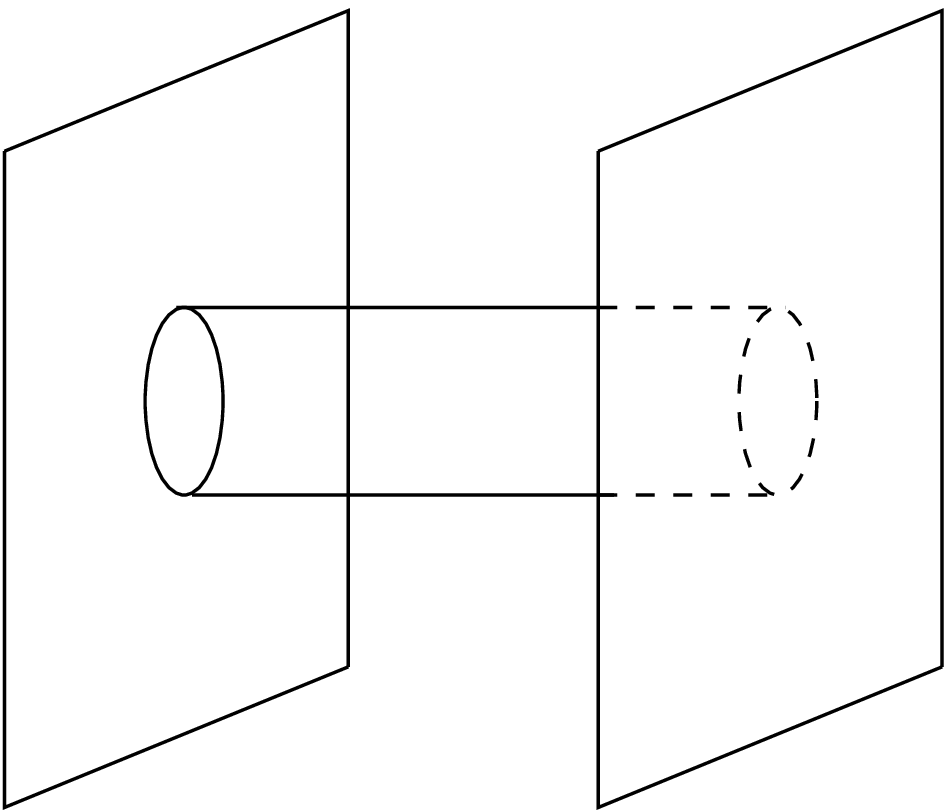,height=100pt}{}
\para
We find ourselves in a rather familiar situation. We have
string-like objects which can terminate on D-brane objects, where
their end is electrically charged. Yet all this is within the
context of a gauge theory, with no reference to string theory or
gravity. Let's remind ourselves about some further properties of
D-branes in string theory to see if the analogy can be pushed
further. For example, there are two methods to understand the
dynamics of D-branes in string theory, using either closed or open
strings. The first method --- the closed string description ---
uses the supergravity solution for D-branes to compute their
scattering. In contrast, in the second method --- the open string
description --- the back-reaction on the bulk is ignored. Instead
the strings stretched between two branes are integrated in, giving
rise to new, light fields of the worldvolume theory as the branes
approach. In flat space, this enhances $U(1)^n$ worldvolume gauge
symmetry to $U(n)$ \cite{wittna}. The quantum effects from these
non-abelian fields capture the scattering of the D-branes. The
equivalence of these two methods is assured by open-closed string
duality, where the diagram drawn in figure 10 can be interpreted
as tree-level closed string or one-loop open string exchange.
Generically the two methods have different regimes of validity.

\para
Is there an analogous treatment for our monopole D-branes? The
analogy of the supergravity description is simply the Manton
moduli space approximation described in section 2.2. What about
the open string description? Can we integrate in the light states
arising from instanton strings stretched between two D-branes?
They have charge $(+1,-1)$ under the two branes and, by the
normalization described above, mass $T_{\rm
inst}|X^i_1-X^i_2|=|\varphi_1^i-\varphi^i_2|$. Let's make the
simplest assumption that quantization of these strings gives rise
to W-bosons, enhancing the worldvolume symmetry of $n$ branes to
$U(n)$. Do the quantum effects of these open strings mimic the
classical scattering of monopoles? Of course they do! This is
precisely the calculation we described in section 2.7.1: the
Coulomb branch of the $U(n)$ ${\cal N}=4$ super Yang-Mills in
$d=2+1$ dimensions is the $n$ monopole moduli space.

\para
The above discussion is not really new. It is nothing more than
the "Hanany-Witten" story \cite{hw}, with attention focussed on
the NS5-brane worldvolume rather than the usual 10-dimensional
perspective. Nevertheless, it's interesting that one can formulate
the story without reference to 10-dimensional string theory. In
particular, if we interpret our results in terms of open-closed
string duality summarized in figure 10, it strongly suggests that
the bulk six-dimensional Yang-Mills fields can be thought of as
quantized loops of instanton strings.

\para
To finish, let me confess that, as one might expect, the closed
and open string descriptions have different regimes of validity.
The bulk calculation is valid in the full quantum theory only if
we can ignore higher derivative corrections to the six-dimensional
action. These scale as $e^{2n}\partial^{2n}$. Since the size of
the monopole is $\partial\sim v^{-1}$, we have the requirement
$v^2e^2\ll 1$ for the "closed string" description to be valid.
What about the open string description? We integrate in an object
of energy $E=T_{\rm inst}\Delta X$, where $\Delta X$ is the
separation between branes. We do not want to include higher
excitations of the string which scale as $v$. So we have $E\ll v$.
At the same time, we want $\Delta X>1/v$, the width of the branes,
in order to make sense of the discussion. These two requirements
tell us that $v^2e^2\gg 1$. The reason that the two calculations
yield the same result, despite their different regimes of
validity, is due to a non-renormalization theorem, which
essentially boils down the restrictions imposed by the
hyperK\"ahler nature of the metric.

\newpage
\section{Vortices}

In this lecture, we're going to discuss vortices. The motivation
for studying vortices should be obvious: they are one of the most
ubiquitous objects in physics. On table-tops, vortices appear as
magnetic flux tubes in superconductors and fractionally charged
quasi-excitations in quantum Hall fluids. In the sky, vortices in
the guise of cosmic strings  have been one one of the most
enduring themes in cosmology research. With new gravitational wave
detectors coming on line, there is hope that we may be able to see
the distinctive signatures of these strings as the twist and whip.
Finally, and more formally, vortices play a crucial role in
determining the phases of low-dimensional quantum systems: from
the phase-slip of superconducting wires, to the physics of strings
propagating on Calabi-Yau manifolds, the vortex is key.

\para
As we shall see in detail below, in four dimensional theories
vortices are string like objects, carrying magnetic flux threaded
through their core. They are the semi-classical cousins of the
more elusive QCD flux tubes. In what follows we will primarily be
interested in the dynamics of infinitely long, parallel vortex
strings and the long-wavelength modes they support. There are a
number of reviews on the dynamics of vortices in four dimensions,
mostly in the context of cosmic strings
\cite{vilshel,kibhind,polcos}.

\subsection{The Basics}

In order for our theory to support vortices, we must add a further
field to our Lagrangian. In fact we must make two deformations
\begin{itemize}
\item We increase the gauge group from $SU(N)$ to $U(N)$. We could
have done this before now, but as we have considered only fields
in the adjoint representation the central $U(1)$ would have simply
decoupled.

\item We add matter in the fundamental representation of $U(N)$.
We'll add $N_f$ scalar fields $q_i$, $i=1\ldots,N_f$.
\end{itemize}
The action that we'll work with throughout this lecture  is
\be S&=&\int d^4x\
\Tr\left(\frac{1}{2e^2}F^{\mu\nu}F_{\mu\nu}+\frac{1}{e^2}({\cal
D}_\mu\phi)^2\right) +\sum_{i=1}^{N_f}|{\cal D}_\mu q_i|^2 \nn\\
&& \ \ \ \ \ \ \ \ \ \ \ \ \ -\sum_{i=1}^{N_f}q_i^\dagger\phi^2
q_i -\frac{e^2}{4}\,\Tr\,(\sum_{i=1}^{N_f}q_iq_i^\dagger-v^2\,
1_N)^2\label{vlag}\ee
The potential is of the type admitting a completion to $\N=1$ or
$\N=2$ supersymmetry. In this context, the final term is called
the D-term. Note that everything in the bracket of the D-term is
an $N\times N$ matrix.  Note also that the couplings in front of
the potential are not arbitrary: they have been tuned to critical
values.

\para
We've included a new parameter, $v^2$, in the potential. Obviously
this will induce a vev for $q$. In the context of supersymmetric
gauge theories, this parameter is known as a Fayet-Iliopoulos
term.

\para
We are interested in ground states of the theory with vanishing
potential. For $N_f<N$, one can't set the D-term to zero since the
first term is, at most, rank $N_f$, while the $v^2$ term is rank
$N$. In the context of supersymmetric theories, this leads to
spontaneous supersymmetry breaking. In what follows we'll only
consider $N_f\geq N$. In fact, for the first half of this section
we'll restrict ourselves to the simplest case:
\be N_f=N\ee
With this choice, we can view $q$ as an $N\times N$ matrix $q^a_{\
i}$, where $a$ is the color index and $i$ the flavor index. Up to
gauge transformations, there is a unique ground state of the
theory,
\be \phi=0  \ \  \ \ \ ,\ \ \ \ \ q^a_{\ i}=v\delta^a_{\ i}
\label{qvev}\ee
Studying small fluctuations around this vacuum, we find that all
gauge fields and scalars are massive, and all have the same mass
$M^2=e^2v^2$. The fact that all masses are equal is a consequence
of tuning the coefficients of the potential.

\para
The theory has a $U(N)_G\times SU(N)_F$ gauge and flavor symmetry.
On the quark fields $q$ this acts as
\be q\rightarrow UqV\ \ \ \ \ \ U\in U(N)_G,\ \ \ V\in SU(N)_F\ee
The vacuum expectation value \eqn{qvev} is preserved only for
transformations of the form $U=V$, meaning that we have the
pattern of spontaneous symmetry breaking
\be U(N)_G\times SU(N)_F\rightarrow SU(N)_{\rm diag} \ee
\EPSFIGURE{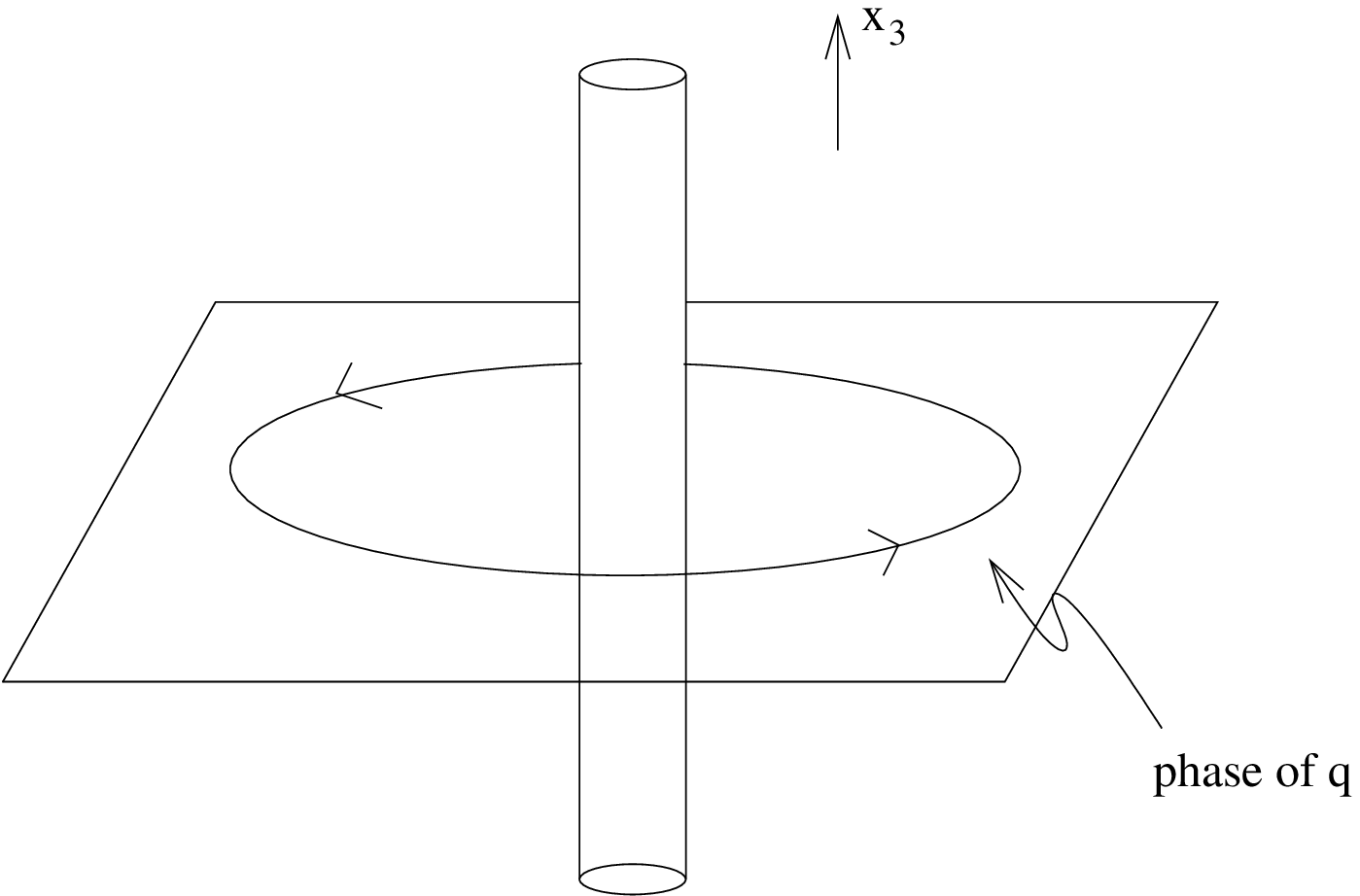,height=110pt}{}
\noindent This is known as the color-flavor-locked phase in the
high-density QCD literature \cite{cfl}.

\para
When $N=1$, our theory is the well-studied abelian Higgs model,
which has been known for many years to support vortex strings
\cite{abrikosov,no}. These vortex strings also exist in the
non-abelian theory and enjoy rather rich properties, as we shall
now see. Let's choose the strings to lie in the $x^3$ direction.
To support such objects, the scalar fields $q$ must wind around
$\S^1_\infty$ at spatial infinity in the $(x^1,x^2)$ plane,
transverse to the string. As we're used to by now, such winding is
characterized by the homotopy group, this time
\be \Pi_1\left(U(N)\times SU(N)/SU(N)_{\rm diag}\right)\cong \Z\ee
Which means that we can expect vortex strings supported by a
single winding number $k\in \Z$. To see that this winding of the
scalar is associated with magnetic flux, we use the same trick as
for monopoles. Finiteness of the quark kinetic term requires that
$\D q\sim 1/r^2$ as $r\rightarrow\infty$. But a winding around
$\S^1_\infty$ necessarily means that $\partial q\sim 1/r$. To
cancel this, we must turn on $A \rightarrow i\partial  q\,q^{-1}$
asymptotically. The winding of the scalar at infinity is
determined by an integer $k$, defined by
\be 2\pi k =\Tr\oint_{\S^1_\infty}\ i\partial_\theta q\ q^{-1} =
\Tr\oint_{\S^1_\infty} A_\theta = \Tr\int dx^1dx^2\ B_3
\label{andrew}\ee
This time however, in contrast to the case of magnetic monopoles,
there is no long range magnetic flux. Physically this is because
the theory has a mass gap, ensuring any excitations die
exponentially. The result, as we shall, is that the magnetic flux
is confined in the center of the vortex string.

\para
The Lagrangian of equation \eqn{vlag} is very special, and far
from the only theory admitting vortex solutions. Indeed, the
vortex zoo is well populated with different objects, many
exhibiting curious properties. Particularly interesting examples
include Alice strings \cite{alice,alford}, and  vortices in
Chern-Simons theories \cite{dunne}. In this lecture we shall stick
with the vortices arising from \eqn{vlag} since, as we shall see,
they are closely related to the instantons and monopoles described
in the previous lectures. To my knowledge, the properties of
non-abelian vortices in this model were studied only quite
recently in \cite{vib} (a related model,  sharing similar
properties, appeared at the same time \cite{auzzi}).

\subsection{The Vortex Equations}

To derive the vortex equations we once again perform the
Bogomoln'yi completing the square trick (due, once again, to
Bogomoln'yi \cite{bog}). We look for static strings in the $x^3$
direction, so make the ansatz $\partial_0=\partial_3=0$ and
$A_0=A_3=0$. We also set $\phi=0$. In fact $\phi$ will not play a
role for the remainder of this lecture, although it will be
resurrected in the following lecture. The tension (energy per unit
length) of the string is
\be T_{\rm vortex} &=& \int dx^1 dx^2\
\Tr\,\left(\frac{1}{e^2}B_3^2+\frac{e^2}{4}(\sum_{i=1}^N
q_iq_i^\dagger-v^2\,1_N)^2\right) + \sum_{i=1}^N|\D_1q_i|^2
+|\D_2q_i|^2\nn\\ &=& \int dx^1dx^2\ \frac{1}{e^2}\Tr\,\left(B_3
\mp \frac{e^2}{2}(\sum_{i=1}^Nq_iq_i^\dagger - v^2\,1_N)\right)^2
+\sum_{i=1}^N|\D_1q_i\mp i\D_2q_i|^2\nn\\ &&\mp v^2\int dx^1dx^2 \
\Tr\,B_3 \label{vbog}\ee
To get from the first line to the second, we need to use the fact
that $[D_1,D_2]=-iB_3$, to cancel the cross terms from the two
squares. Using \eqn{andrew}, we find that the tension of the
charge $|k|$ vortex is bounded by
\be T_{\rm vortex}\geq 2\pi v^2\,|k| \ee
In what follows we focus on vortex solutions with winding $k<0$.
(These are mapped into $k>0$ vortices by a parity transformation,
so there is no loss of generality). The inequality is then
saturated for configurations obeying the vortex equations
\be B_3=\frac{e^2}{2}(\sum_i q_iq_i^\dagger -v^2\,1_N)\ \ \ \ ,\ \
\ \ \D_zq_i=0\label{vortex}\ee
where we've introduced the complex coordinate $z=x^1+ix^2$ on the
plane transverse to the vortex string, so
$\partial_z=\ft12(\partial_1-i\partial_2)$. If we choose $N=1$,
then the Lagrangian \eqn{vlag} reduces to the abelian-Higgs model
and, until recently, attention mostly focussed on this abelian
variety of the equations \eqn{vortex}. However, as we shall see
below, when the vortex equations are non-abelian, so each side of
the first equation \eqn{vortex} is an $N\times N$ matrix, they
have a much more interesting structure.

\para
Unlike monopoles and instantons, no analytic solution to the
vortex equations is known. This is true even for a single $k=1$
vortex in the $U(1)$ theory. There's nothing sinister about this.
It's just that differential equations are hard and no one has
decided to call the vortex solution a special function and give it
a name! However, it's not difficult to plot the solution
numerically and the profile of the fields is sketched below. The
energy density is localized within a core of the vortex of size
$L=1/ev$, outside of which all fields return exponentially to
their vacuum.
\begin{figure}[htb]
\begin{center}
\epsfxsize=4.2in\leavevmode\epsfbox{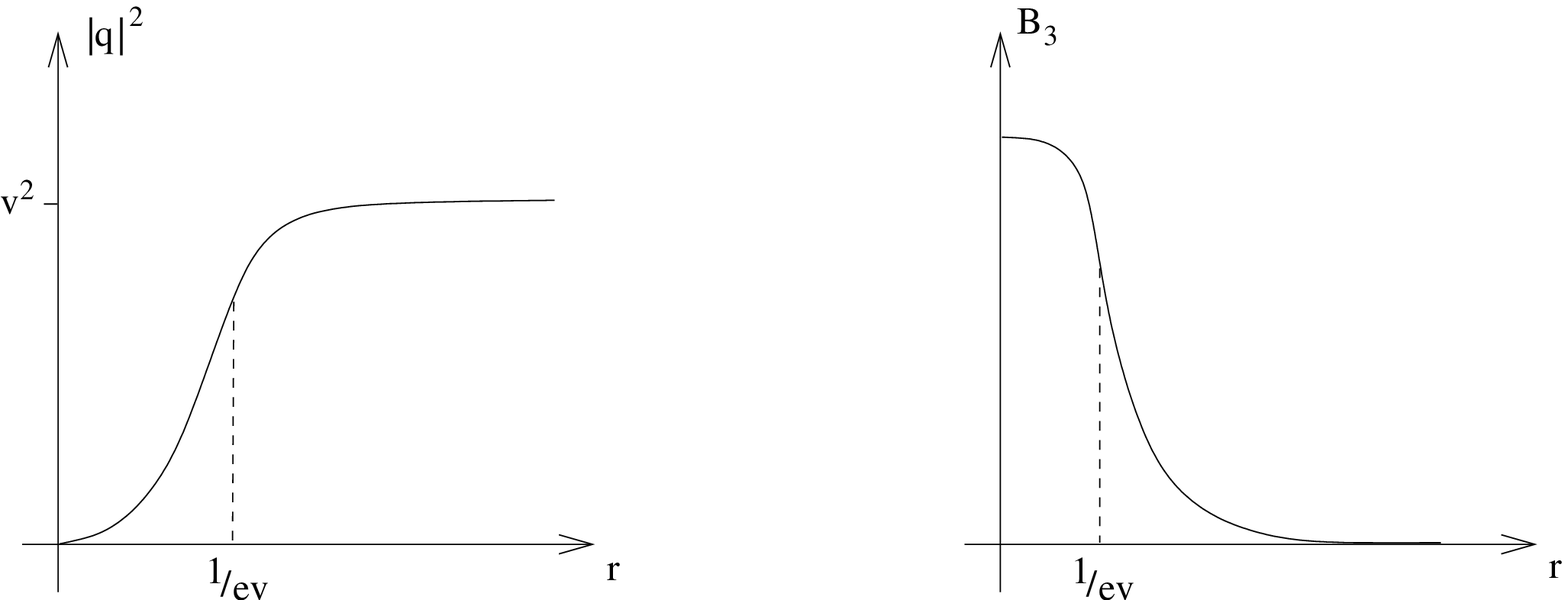}
\end{center}
\caption{A sketch of the vortex profile.}
\end{figure}

\para
The simplest $k=1$ vortex in the abelian $N=1$ theory has just two
collective coordinates, corresponding to its position on the
$z$-plane. But what are the collective coordinates of a vortex in
$U(N)$. We can use the same idea we saw in the instanton lecture,
and embed the abelian vortex --- let's denote it $q^\star$ and
$A_z^\star$ --- in the $N\times N$ matrices of the non-abelian
theory. We have
\be A_z=\left(\begin{array}{cccc} A^\star_z \ & & & \\ & 0\  & & \\
& & \ddots & \\ & & & \ 0\end{array}\right) \ \ \ \ ,\ \ \ \ \
q=\left(\begin{array}{cccc} q^\star \ & & & \\ & v\  & & \\ & &
\ddots & \\ & & & \ v\end{array}\right)\ee
where the columns of the $q$ matrix carry the color charge, while
the rows carry the flavor charge. We have chosen the embedding
above to lie in the upper left-hand corner but this isn't unique.
We can rotate into other embeddings by acting with the $SU(N)_{\rm
diag}$ symmetry preserved in the vacuum. Dividing by the
stabilizer, we find the internal moduli space of the single
non-abelian vortex to be
\be SU(N)_{\rm diag}/S[U(N-1)\times U(1)] \cong \CP^{N-1} \ee
The appearance of $\CP^{N-1}$ as the internal space of the vortex
is interesting: it tells us that the low-energy dynamics of a
vortex string is the much studied quantum $\CP^{N-1}$ sigma model.
We'll see the significance of this in the following lecture. For
now, let's look more closely at the moduli of the vortices.

\subsection{The Moduli Space}

We've seen that a single vortex has $2N$ collective coordinates: 2
translations, and $2(N-1)$ internal modes, dictating the
orientation of the vortex in color and flavor space. We denote the
moduli space of charge $k$ vortices in the $U(N)$ gauge theory as
$\vkn$. We've learnt above that
\be {\cal V}_{1,N}\cong \C\times\CP^{N-1}\ee
What about higher $k$? An index theorem \cite{erickvort,vib} tells
us that the number of collective coordinates is
\be \dim (\vkn)=2kN \label{vortindex}\ee
Look familiar? Remember the result for $k$ instantons in $U(N)$
that we found in lecture 1: $\dim({\ikn})=4kN$. We'll see more of
this similarity between instantons and vortices in the following.

\para
As for previous solitons, the counting \eqn{vortindex} has a
natural interpretation: $k$ parallel vortex strings may be placed
at arbitrary position, each carrying $2(N-1)$ independent
orientational modes. Thinking physically in terms of forces
between vortices, this is a consequence of tuning the coefficient
$e^2/4$ in front of the D-term in \eqn{vlag} so that the mass of
the gauge bosons equals the mass of the $q$ scalars. If this
coupling is turned up, the scalar mass increases and so mediates a
force with shorter range than the gauge bosons, causing the
vortices to repel. (Recall the general rule: spin 0 particles give
rise to attractive forces; spin 1 repulsive). This is a type II
non-abelian superconductor. If the coupling decreases, the mass of
the scalar decreases and the vortices attract. This is a
non-abelian type I superconductor. In the following, we keep with
the critically coupled case \eqn{vlag} for which the first order
equations \eqn{vortex} yield solutions with vortices at arbitrary
position.

\subsubsection{The Moduli Space Metric}

There is again a natural metric on $\vkn$ arising from taking the
overlap of zero modes. These zero modes must solve the linearized
vortex equations together with a suitable background gauge fixing
condition. The linearized vortex equations read
\be \D_z\d A_{\bar{z}}-\D_{\bar{z}}\d A_z=\frac{ie^2}{4}(\d q\,
q^\dagger + q\,\d q^\dagger)\ \ \ \ {\rm and}\ \ \ \ \D_z\d q=i\d
A_z q\label{fiirst}\ee
where $q$ is to be viewed as an $N\times N$ matrix in these
equations. The gauge fixing condition is
\be \D_z\d A_{\bar{z}}+\D_{\bar{z}}\d A_z=-\frac{ie^2}{4}(\d
q\,q^\dagger-q\,\d q^\dagger) \ee
which combines with the first equation in \eqn{fiirst} to give
\be \D_{\bar{z}}\d A_z=-\frac{ie^2}{4}\d q\,q^\dagger\ee
Then, from the index theorem, we know that there are $2kN$ zero
modes $(\d_\alpha A_z,\d_\alpha q)$, $\alpha,\beta=1,\ldots,2kN$
solving these equations, providing a metric on $\vkn$ defined by
\be g_{\alpha\beta} =\Tr\int dx^1dx^2\ \frac{1}{e^2}\d_\alpha
A_a\d_\beta A_{\bar{z}}+\frac{1}{2}\d_\alpha q\d_\beta q^\dagger\
+\ {\rm h.c.}\ee
The metric has the following properties \cite{taubes,samols}
\begin{itemize}
\item The metric is K\"ahler. This follows from similar arguments
to those given for hyperK\"ahlerity of the instanton moduli space,
the complex structure now descending from that on the plane
$\R^2$, together with the obvious complex structure on $q$.

\item The metric is smooth. It has no singularities as the
vortices approach each other. Strictly speaking this statement has
been proven only for abelian vortices. For non-abelian vortices,
we shall show this using branes in the following section.

\item The metric inherits a $U(1)\times SU(N)$ holomorphic
isometry from the rotational and internal symmetry of the
Lagrangian.

\item The metric is unknown for $k\geq 2$. The leading order,
exponentially suppressed, corrections to the flat metric were
computed recently \cite{manspeight}.
\end{itemize}

\subsubsection{Examples of Vortex Moduli Spaces}

\subsubsection*{A Single $U(N)$ Vortex}

We've already seen above that the moduli space for a single $k=1$
vortex in $U(N)$ is
\be \V_{1,N}\cong \C\times \CP^{N-1}\ee
where the isometry group $SU(N)$ ensures that $\CP^{N-1}$ is
endowed with the round, Fubini-Study metric. The only question
remaining is the size, or K\"ahler class, of the $\CP^{N-1}$. This
can be computed either from a D-brane construction \cite{vib} or,
more conventionally, from the overlap of zero modes \cite{sy}.
We'll see the former in the following section. Here let's sketch
the latter. The orientational zero modes of the vortex take the
form
\be \d A_z=\D_z\Omega\ \ \ \ ,\ \ \ \ \ \d q=i(\Omega
q-q\Omega_0)\ee
where the gauge transformation asymptotes to $\Omega\rightarrow
\Omega_0$, and $\Omega_0$ is the flavor transformation. The gauge
fixing condition requires
\be\D^2\Omega =
\frac{e^2}{2}\{\Omega,qq^\dagger\}-2qq^\dagger\Omega_0 \ee
By explicitly computing the overlap of these zero modes, it can be
shown that the size of the $\CP^{N-1}$ is
\be r=\frac{4\pi}{e^2}\label{ree}\ee
This important equation will play a crucial role in the
correspondence between 2d sigma models and 4d gauge theories that
we'll meet in the following lecture.

\subsubsection*{Two $U(1)$ Vortices}

The moduli space of two vortices in a $U(1)$ gauge theory is
topologically
\be \V_{k=2,N=1}\cong\C\times \C/\Z_2\label{twovort}\ee
where the $\Z_2$ reflects the fact that the two solitons are
indistinguishable. Note that the notation we used above actually
describes more than the topology of the manifold because,
topologically, $\C^k/\Z_k\cong \C^k$ (as any polynomial will tell
you). So when I write $\C/\Z_2$ in \eqn{twovort}, I  mean that
asymptotically the space is endowed with the flat metric on
$\C/\Z_2$. Of course, this can't be true closer to the origin
since we know the vortex moduli space is complete. The cone must
be smooth at the tip, as shown in figure 13. The metric on the
cone has been computed numerically \cite{rushellard}, no analytic
form is known. The deviations from the flat, singular, metric on
the cone are exponentially suppressed and parameterized by the
size of the vortex $L\sim 1/ev$.
\begin{figure}[htb]
\begin{center}
\epsfxsize=4.2in\leavevmode\epsfbox{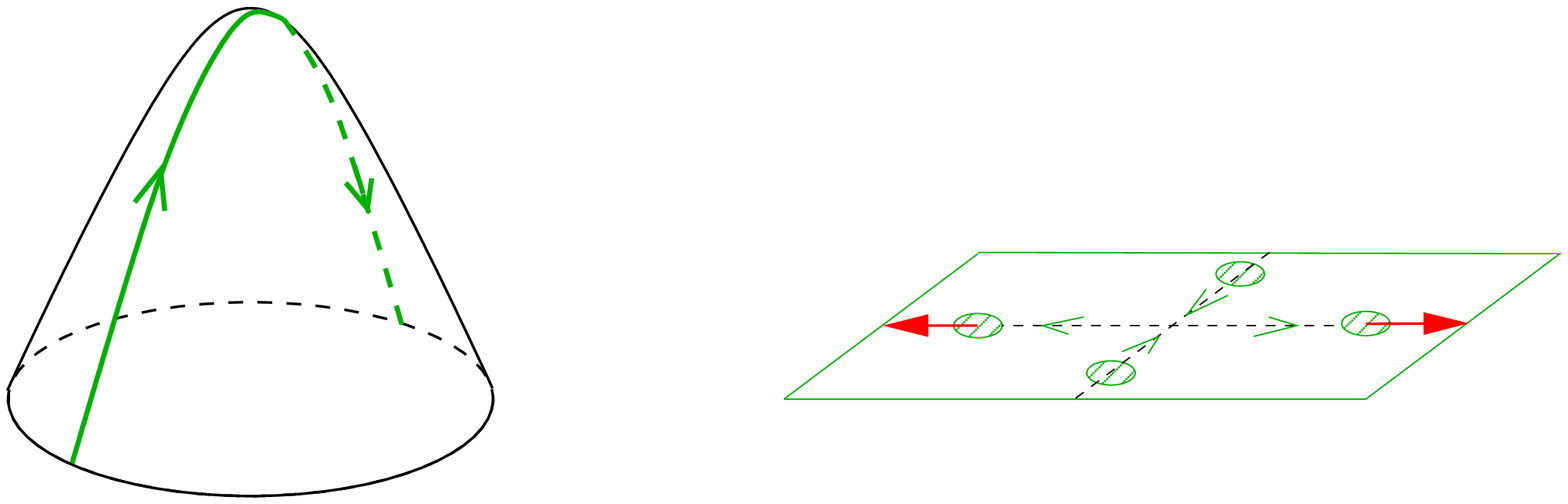}
\end{center}
\caption{Right-angle scattering from the moduli space of two
vortices.}
\end{figure}
%
\para
Even without the exact form of the metric, we learn something very
important about vortices. Consider two vortices colliding head on.
This corresponds to the trajectory in moduli space that goes up
and over the tip of the cone, as shown in the figure. What does
this correspond to in real space? One might think that the
vortices collide and rebound. But that's wrong: it would
correspond to the trajectory going to the tip of the cone, and
returning down the same side. Instead, the trajectory corresponds
to vortices scattering at right angles \cite{ruback}. The key
point is that the $\Z_2$ action in \eqn{twovort}, arising because
the vortices are identical, means that the single valued
coordinate on the moduli space is $z^2$ rather than $z$, the
separation between the vortices. The collision sends
$z^2\rightarrow -z^2$ or $z\rightarrow iz$. This result doesn't
depend on the details of the metric on the vortex moduli space,
but follows simply from the fact that, near the origin, the space
is smooth. Right-angle scattering of this type is characteristic
of soliton collisions, occurring also for magnetic monopoles.

\para
For $k\geq 3$ $U(1)$ vortices, the moduli space is topologically
and asymptotically $\C^k/\Z_k$. The leading order exponential
corrections to the flat metric on this space are known, although
the full metric is not \cite{manspeight}.

\subsection{Brane Construction}

For both instantons and monopoles, it was fruitful to examine the
solitons from the perspective of D-branes. This allowed us to
re-derive the ADHM and Nahm constructions respectively. What about
for vortices? Here we present a D-brane constuction of vortices
\cite{vib} that will reveal interesting information about the
moduli space of solutions although, ultimately, won't be as
powerful as the ADHM and Nahm constructions described in previous
sections.

\EPSFIGURE{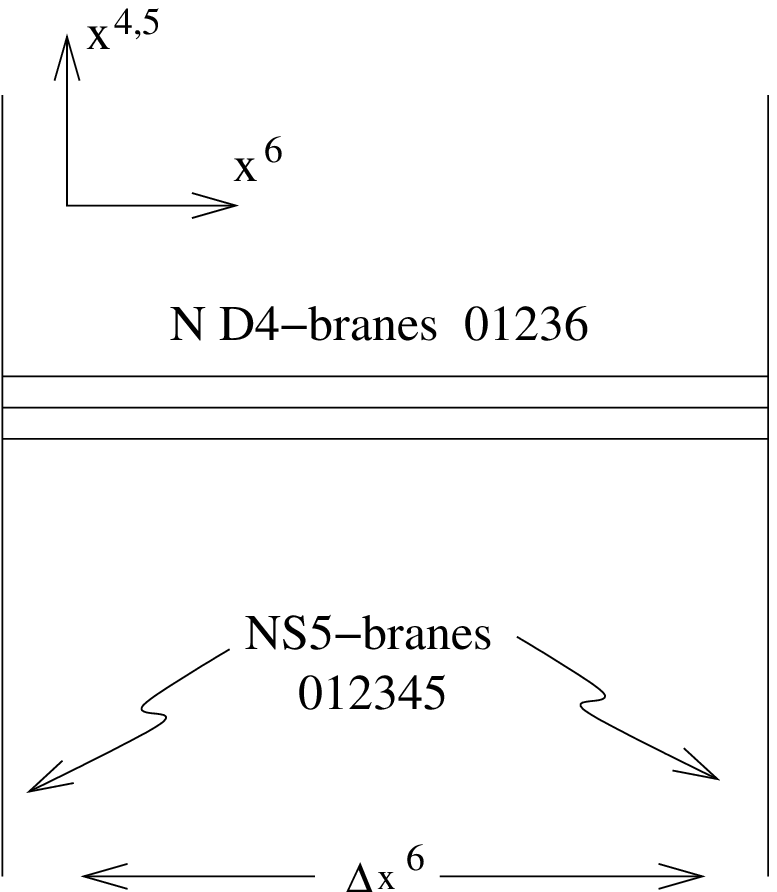,height=150pt}{}
\para
We use the brane set-ups of Hanany and Witten \cite{hw},
consisting of D-branes suspended between a pair of NS5-branes. We
work in type IIA string theory, and build the $d=3+1$, $U(N)$
gauge theory\footnote{In fact, for four-dimensional theories the
overall $U(1)$ decouples in the brane set-up, and we have only
$SU(N)$ gauge theory \cite{witten4d}. This doesn't affect our
study of the vortex moduli space; if you're bothered by this,
simply T-dualize the problem to type IIB where you can study
vortices in $d=2+1$ dimensions.} with $\N=2$ supersymmetry. The
D-brane set-up is shown in figure 14, and consists of $N$
D4-branes with worldvolume $01236$, stretched between two
NS5-branes, each with worldvolume $012345$, and separated in the
$x^6$ direction. The gauge coupling $e^2$ is determined by the
separation between the NS5-branes,
\be \frac{1}{e^2}=\frac{\Delta x^6\,l_s}{2g_s} \ee
where $l_s$ is the string length, and $g_s$ the string coupling.
The D4-branes may slide up and down between the NS5-branes in the
$x^4$ and $x^5$ direction. This  corresponds to turning on a  vev
for the complex adjoint scalar in the $\N=2$ vector multiplet.
Since we consider only  a real adjoint scalar $\phi$ in our
theory, we have
\be \phi_a=\left.\frac{x^4}{l_s^2}\right|_{D4_a} \ee
and we'll take all D4-branes to lie coincident in the $x^5$
direction.

\EPSFIGURE{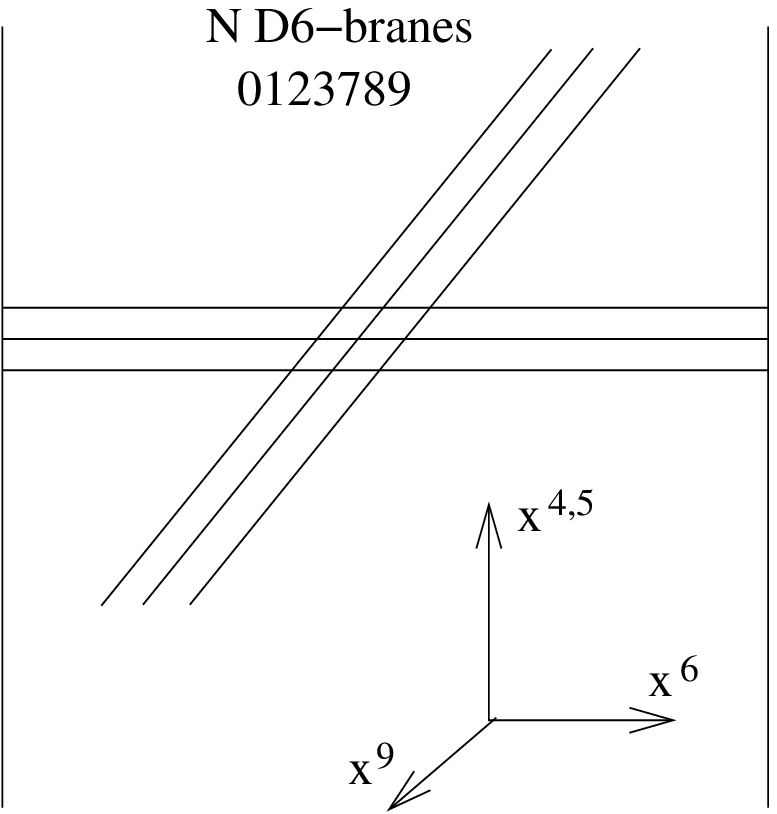,height=120pt}{}
\para
The hypermultiplets arise in the form of $N$ D6-branes with
worldvolume $0123789$. The positions of the D6-branes in the
$x^4+ix^5$ directions will correspond to complex masses for the
hypermultiplets. We shall consider these in the following section,
but for now we set all  D6-branes to lie at the origin of the
$x^4$ and $x^5$ plane.

\para
We also need to turn on the FI parameter $v^2$. This is achieved
by taking the right-hand NS5-brane and pulling it out of the page
in the $x^9$ direction. In order to remain in the ground state,
the D4-branes are not allowed to tilt into the $x^9$ direction:
this would break supersymmetry and increase their length,
reflecting a corresponding increase in the ground state energy of
the theory. Instead, they must split on the D6-branes. Something
known as the S-rule \cite{hw,srule} tells us that only one
D4-brane can end on a given D6-brane while preserving
supersymmetry, ensuring that we need at least $N$ D6-branes to
find a zero-energy ground state. The final configuration is drawn
in the figure 16, with the field theory dictionary given by
\be v^2=\frac{\Delta x^9}{(2\pi)^3g_sl_s^3}\ee
\EPSFIGURE{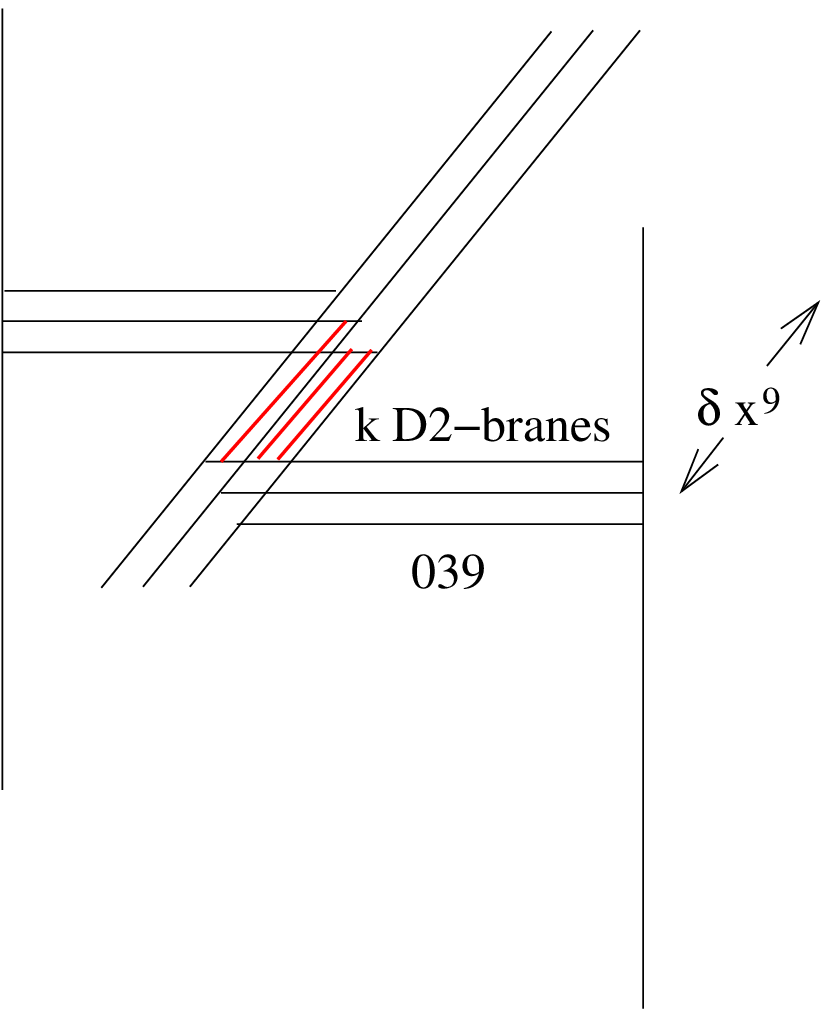,height=150pt}{}
\noindent Now we've built our theory, we can look to find the
vortices. We expect them to appear as other D-branes in the
configuration. There is a unique BPS D-brane with the correct
mass: it is a D2-brane, lying coincident with the D6-branes, with
worldvolume 039, as shown in figure 16 \cite{nohta}. The $x^3$ direction here
is the direction of the vortex string.
%
%
%

\para
The problem is: what is the worldvolume theory on the D2-branes.
It's hard to read off the theory directly because of the boundary
conditions where the D2-branes end on the D4-branes. But, already
by inspection, we might expect that it's related to the
D$p$-D$(p-4)$ system described in Lecture 1 in the context of
instantons. To make progress we play some brane games. Move the
D6-branes to the right. As they pass the NS5-brane, the
Hanany-Witten transition occurs and the right-hand D4-branes
disappear \cite{hw}. We get the configuration shown in figure 17.

\DOUBLEFIGURE{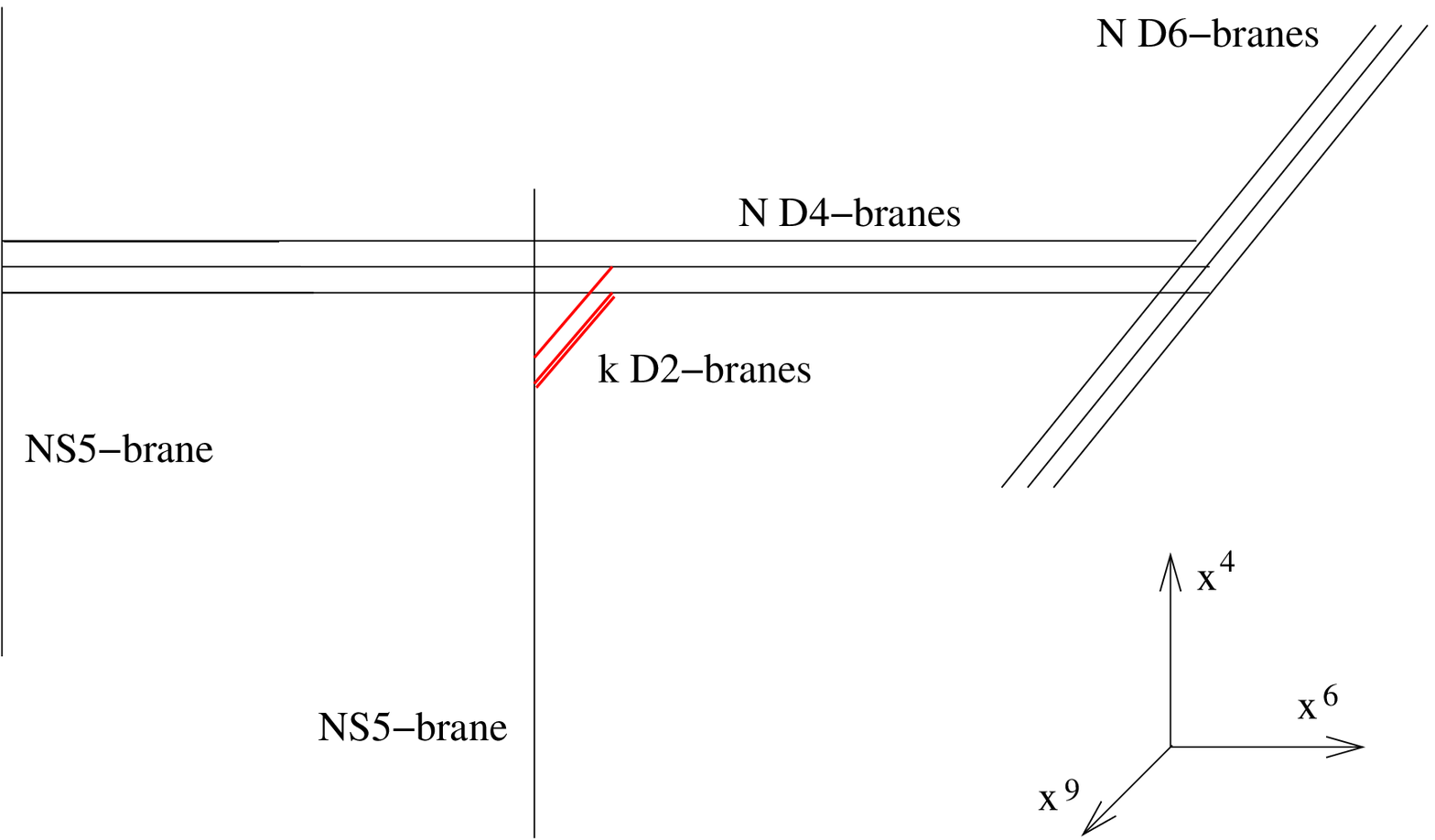,width=230pt}{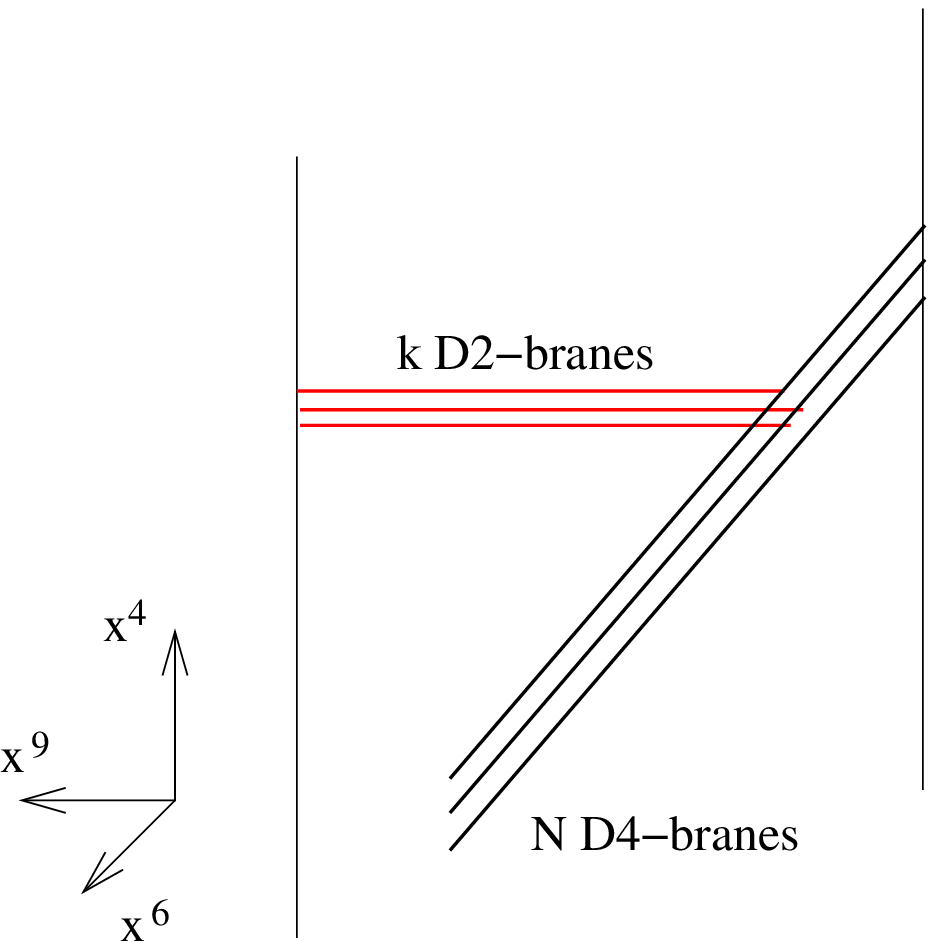,width=140pt}
{Moving the D6-branes}{Rotating our viewpoint}
\para
Let's keep the D6-branes moving. Off to infinity. Finally, we
rotate our perspective a little, viewing the D-branes from a
different angle, shown in figure 18. This is our final D-brane
configuration and we can now read off the dynamics.

\para
We want to determine the theory on the D2-branes in figure 18.
Let's start with the easier problem in figure 19. Here the
D4-branes extend to infinity in both $x^6\rightarrow \pm\infty$
directions, and the D2-branes end on the other NS5. The theory on
the D2-branes is simple to determine: it is a $U(k)$ gauge theory
with 4 real adjoint scalars, or two complex scalars
\be \sigma=X^4+iX^5\ \ \ ,\ \ \ \ Z=X^1+iX^2\ee
\EPSFIGURE{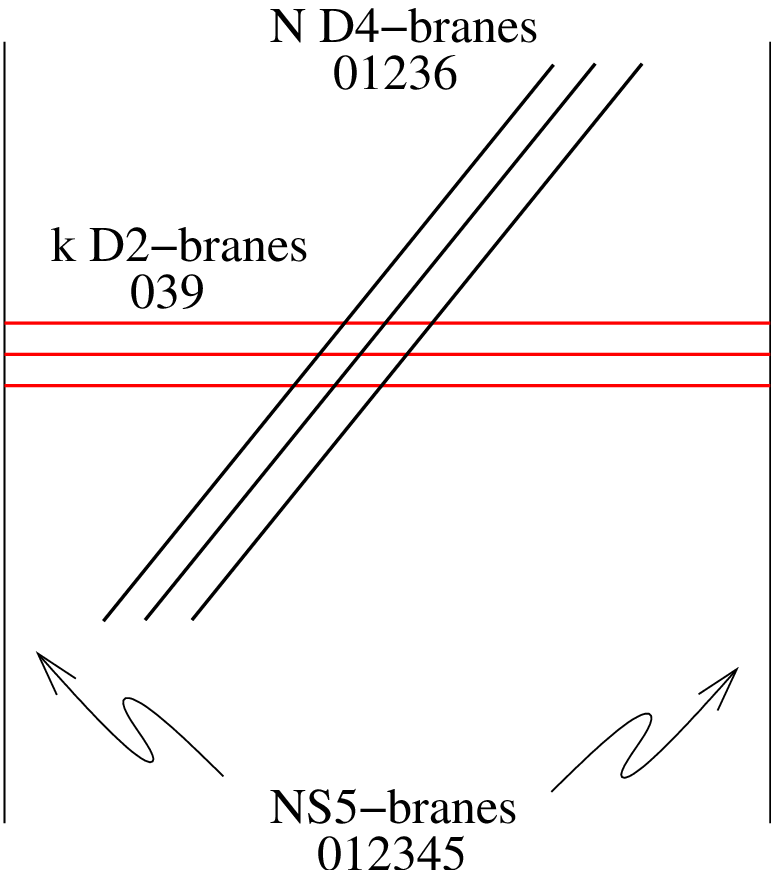,height=130pt}{}
\noindent which combine to give the $\N=(4,4)$ theory in $d=1+1$.
The D4-branes contribute hypermultiplets $(\psi_a,\tilde{\psi}_a)$
with $a=1,\ldots, N$. These hypermultiplets get a mass only when
the D2-branes and D6-branes are separated in the $X^4$ and $X^5$
directions. This means we have a coupling like
\be \sum_{a=1}^N\psi^\dagger_a\,\{\sigma^\dagger,\sigma\}\,\psi_a+
\tilde{\psi}_a\,\{\sigma^\dagger,\sigma\}\,\tilde{\psi}^\dagger_a\label{coupling}\ee
But there is  no such coupling between the hypermultiplets and
$Z$. The coupling \eqn{coupling} breaks supersymmetry to
$\N=(2,2)$. So we now understand the D2-brane theory of figure 19.
However, the D2-brane theory that we're really interested in,
shown in figure 18, differs from this in two ways
\begin{itemize}
\item The right-hand NS5-brane is moved out of the page. But we
already saw in the manoeuvres around figure 16 that this induces a
FI parameter on brane theory. Except this this time the FI
parameter is for the D2-brane theory. It's given by
\be r=\frac{\Delta x^6}{2\pi g_sl_s} =
\frac{4\pi}{e^2}\label{fir}\ee
\item We only have half of the D4-branes, not all of them. If a
full D4-brane gives rise to a hypermultiplet, one might guess that
half a D4-brane should give rise to half a hypermultiplet,
otherwise known as a chiral multiplet. Although the argument is a
little glib, it turns out that this is the correct answer
\cite{hh}.
\end{itemize}

We end up with the gauge theory in $d=1+1$ dimensions with
$\N=(2,2)$ supersymmetry
\be \mbox{$U(k)$ Gauge Theory\ } &+& \mbox{\ Adjoint Chiral Multiplet $Z$} \nn\\
&+& \mbox{\ $N$ Fundamental Chiral Multiplets $\psi_a$}\nn\ee
This theory has a FI parameter $r=4\pi/e^2$. Now this should be
looking very familiar --- it's very similar to the instanton
theory we described in Lecture 1. We'll return to this shortly.
For now let's keep examining our vortex theory. The potential for
the various scalars is dictated by supersymmetry and is given by
\be
V&=&\frac{1}{g^2}\Tr\,|[\sigma,\sigma^\dagger]|^2+\Tr\,|[\sigma,Z]|^2+
\Tr\,|[\sigma,Z^\dagger]|^2
+\sum_{a=1}^N\,\psi^\dagger_a\sigma^\dagger\sigma\psi_a \nn\\ && +
\frac{g^2}{2}\,\Tr\,\left(\sum_a\psi_a\psi_a^\dagger +
[Z,Z^\dagger]-r\,1_k\right)^2\ee
Here $g^2$ is an auxiliary gauge coupling which we take to
infinity $g^2\rightarrow\infty$ to restrict us to the Higgs
branch, the vacuum moduli space defined by
\be \M_{\rm Higgs}\cong\{\sigma=0,V=0\}/U(k) \ee
Counting the various degrees of freedom, the Higgs branch has real
dimension $2kN$. From the analogy with the instanton case, it is
natural to conjecture that this is the vortex moduli space
\cite{vib}
\be \vkn\cong\M_{\rm Higgs} \ee
While the ADHM construction has a field theoretic underpinning, I
know of no field theory derivation of the above result for
vortices. So what evidence do we have that the Higgs branch indeed
coincides with the vortex moduli space? Because of the FI
parameter, $\M_{\rm Higgs}$ is a smooth manifold, as is $\vkn$
and, obviously the dimensions work out. Both spaces have a
$SU(N)\times U(1)$ isometry which, in the above construction, act
upon $\psi$ and $Z$ respectively. Finally, in all cases we can
check, the two spaces agree (as, indeed, do their K\"ahler
classes). Let's look at some examples.

\subsubsection{Examples of Vortex Moduli Spaces Revisited}

\subsubsection*{One Vortex in $U(N)$}

The gauge theory for a single $k=1$ vortex in $U(N)$ is a $U(1)$
gauge theory. The adjoint scalar $Z$ decouples, parameterizing the
complex plane $\C$, leaving us with the $N$ charged scalars
satisfying
\be \sum_{a=1}^N|\psi_a|^2=r\ee
modulo the $U(1)$ action $\psi_a\rightarrow e^{i\alpha}\psi_a$.
This gives us the moduli space
\be \V_{1,N}\cong \C\times \CP^{N-1}\label{iso}\ee
where the $\CP^{N-1}$ has the correct K\"ahler class $r=4\pi/e^2$
in agreement with \eqn{ree}. The metric on $\CP^{N-1}$ is, again,
the round Fubini-Study metric.

\subsubsection*{$k$ Vortices in $U(1)$}

The Higgs branch corresponding to the $k$ vortex moduli space is
\be \{\psi\psi^\dagger+[Z,Z^\dagger]=r\,1_k\}/U(k) \ee
which is asymptotic to the cone $\C^k/\Z_k$, with the
singularities resolved. This is in agreement with the vortex
moduli space. {\it However}, the metric on $\M_{\rm Higgs}$
differs by power law corrections from the flat metric on the
orbifold $\C^k/\Z_k$. But, as we've discussed, $\vkn$ differs from
the flat metric by exponential corrections.

\para
More recently, the moduli space of two vortices in $U(N)$ was
studied in some detail and shown to possess interesting and
non-trivial topology \cite{hashtong}, with certain expected
features of ${\cal V}_{2,N}$ reproduced by the Higgs branch.

\para
In summary, it is conjectured that the vortex moduli space $\vkn$
is isomorphic to the Higgs branch \eqn{iso}. But, except for the
case $k=1$ where the metric is determined by the isometry, the
metrics do not agree. A direct field theory proof of this
correspondence remains to be found.

\subsubsection{The Relationship to Instantons}

As we've mentioned a few times, the vortex theory bears a striking
resemblance to the ADHM instanton theory we met in Lecture 1. In
fact, the gauge theoretic construction of vortex moduli space
$\vkn$ involves exactly half the fields of the ADHM construction.
Or, put another way, the vortex moduli space is half of the
instanton moduli space. We can state this more  precisely: $\vkn$
is a complex, middle dimensional submanifold of $\ikn$. It can be
defined by looking at the action of the isometry rotating the
instantons in the $x^3-x^4$ plane. Denote the corresponding
Killing vector as $h$. Then
\be \vkn\cong\left.\ikn\right|_{h=0}\ee
where $\ikn$ is the resolved instanton moduli space with
non-commutativity parameter $\theta_{\mu\nu}=r
\bar{\eta}^3_{\mu\nu}$. We'll see a physical reason for this
relationship shortly.

\para
An open question: The ADHM construction is constructive. As we
have seen, it allows us to build solutions to $F=\starf$ from the
variables of the Higgs branch. Does a similar construction exist
for vortices?

\para
Relationships between the instanton and vortex equations have been
noted in the past. In particular, a twisted reduction of
instantons in $SU(2)$ Yang-Mills on ${\bf R}^2\times {\bf S}^2$
gives rise to the $U(1)$ vortex equations \cite{fm}. While this
relationship appears to share several characteristics to the
correspondence described above, it differs in many important
details. It don't understand the relationship between the two
approaches.

\subsection{Adding Flavors}

Let's now look at vortices in a $U(N_c)$ gauge theory with
$N_f\geq N_c$ flavors. Note that we've added subscripts to denote
color and flavor. In theories with $N_c=1$ and $N_f >1$, these
were called semi-local vortices \cite{av,avrev,markv,marksemi}. The name
derives from the fact the theory has both a gauge (local) group
and a flavor (global) group. But for us, it's not a great name as
all our theories have both types of symmetries, but it's only when
$N_f>N_c$ that the extra properties of "semi-local" vortices
become apparent.

\para
The Lagrangian \eqn{vlag} remains but, unlike before, the theory
no longer has a mass gap in vacuum. Instead there are $N_c^2$
massive scalar fields and scalars, and $2N_c(N_f-N_c)$ massless
scalars. At low-energies, the theory reduces to a $\sigma$-model
on the Higgs branch of the gauge theory \eqn{vlag},
\be \M_{\rm Higgs}\cong
\{\sum_{i=1}^{N_f}q_iq_i^\dagger=v^2\,1_{N_c}\}/U(N_c)\cong
G(N_c,N_f)\ee
When we have an abelian $N_c=1$ theory, this Higgs branch is the
projective space $G(1,N_f)\cong \CP^{N_f-1}$. For non-abelian
theories, the Higgs branch is the Grassmannian $G(N_c,N_f)$, the
space of $\C^{N_c}$ planes in $\C^{N_f}$. In a given vacuum, the
symmetry breaking pattern is $U(N_c)\times SU(N_f)\rightarrow
S[U(N_c)\times U(N_f-N_c)]$.

\para
The first order vortex equations \eqn{vortex} still give solutions
to the full Lagrangian, now with the flavor index running over
values $i=1\ldots, N_f$. Let's denote the corresponding vortex
moduli space as $\hat{\V}_{k,N_c,N_f}$, so our previous notation
becomes $\vkn\cong \hat{\V}_{k,N,N}$. The index theorem now tells
us the dimension of the vortex moduli space
\be \dim(\hat{\V}_{k,N_c,N_f})=2kN_f \ee
The dimension depends only on the number of flavors, and the
semi-local vortices inherit new modes. These modes are related to
scaling modes of the vortex --- the size of the vortex becomes a
parameter, just as it was for instantons \cite{leese}.

\para
These vortices arising in the theory with extra flavors are
related to other solitons, known as a sigma-model lumps. (These
solitons have other names, depending on the context, sometimes
referred to as "textures", "Skyrmions" or, in the context of
string theory, "worldsheet instantons"). Let's see how this works.
At low-energies (or, equivalently, in the strong coupling limit
$e^2\rightarrow\infty$) our gauge theory flows to the sigma-model
on the Higgs branch $\M_{\rm Higgs}\cong G(N_c,N_f)$. In this
limit our vortices descend to lumps, objects which gain their
topological support once we compactify the $(x^1-x^2)$-plane at
infinity, and wrap this sphere around $\M_{\rm Higgs}\cong
G(N_c,N_f)$ \cite{phases,schroers}
\be \Pi_2(G(N_c,N_f))\cong \Z\ee
When $N_f=N_c$ there is no Higgs branch, the vortices have size
$L=1/ev$ and become singular as $e^2\rightarrow\infty$. In
contrast, when $N_f>N_c$, the vortices may have arbitrary size and
survive the strong coupling limit. However, while the vortex
moduli space is smooth, the lump moduli space has singularities,
akin to the small instanton singularities we saw in Lecture 1. We
see that the gauge coupling  $1/e^2$ plays the same role for lumps
as $\theta$ plays for Yang-Mills instantons.

\para
The brane construction for these vortices is much like the
previous section - we just need more $D6$ branes. By performing
the same series of manoeuvres, we can deduce the worldvolume
theory. It is again a $d=1+1$ dimensional, $\N=(2,2)$ theory with
\be \mbox{$U(k)$ Gauge Theory\ } &+& \mbox{\ Adjoint Chiral
Multiplet $Z$} \nn\\ &+& \mbox{\ $N_c$ Fundamental Chiral
Multiplets $\psi_a$} \nn\\ &+& \mbox{\ $(N_f-N_c)$
Anti-Fundamental Chiral Multiplets $\tilde{\psi}_a$}\nn\ee
Once more, the FI parameter is $r=4\pi/e^2$. The D-term constraint
of this theory is
\be \sum_{a=1}^{N_c}\psi_a\psi^\dagger_a-\sum_{b=1}^{N_f-N_c}
\tilde{\psi}_b^\dagger \tilde{\psi}_b+[Z,Z^\dagger]=r\,1_k
\label{mored}\ee
A few comments
\begin{itemize}
\item Unlike the moduli space $\vkn$, the presence of the
$\tilde{\psi}$ means that this space doesn't collapse as we send
$r\rightarrow 0$. Instead, in this limit it develops singularities
at $\psi=\tilde{\psi}=0$ where the $U(k)$ gauge group doesn't act
freely. This is the manifestation of the discussion above.

\item The metric inherited from the D-term \eqn{mored} again
doesn't coincide with the metric on the vortex moduli space
$\hat{\V}_{k,N_c,N_f}$. In fact, here the discrepancy is more
pronounced, since the metric on $\hat{\V}_{k,N_c,N_f}$ has
non-normalizable modes: the directions in moduli space
corresponding to the scaling the solution are suffer an infra-red
logarithmic divergence \cite{ward,leese}. The vortex theory
arising from branes doesn't capture this.
\end{itemize}

\subsubsection{Non-Commutative Vortices}

As for instantons, we can consider vortices on the non-commutative
plane
\be [x^1,x^2]=i\vartheta \ee
These objects were first studied in \cite{jak}. How does this
affect the moduli space? In the ADHM construction for instantons,
we saw that non-commutivity added a FI parameter to the D-term
constraints. But, for vortices, we already have a FI parameter:
$r=4\pi/e^2$. It's not hard to show using D-branes \cite{vib},
that the effect of non-commutivity is to deform,
\be r=\frac{4\pi}{e^2}+2\pi v^2\vartheta\ee
This has some interesting consequences.  Note that for $N_f=N_c$,
there is a critical FI parameter $\vartheta_c=-v^2/e^2$ for which
$r=0$. At this point the vortex moduli space becomes singular. For
$\vartheta<\vartheta_c$, no solutions to the D-term equations
exist. Indeed, it can be shown that in this region, no solutions
to the vortex equations exist either \cite{bak}. We see that the
Higgs branch correctly captures the physics of the vortices.

\para
For $N_f>N_c$, the Higgs branch makes an interesting prediction:
the vortex moduli space should undergo a topology changing
transition as $\vartheta \rightarrow\vartheta_c$. For example, in
the case of a single $k=1$ vortex in $U(2)$ with $N_f=4$, this is
the well-known flop transition of the conifold. To my knowledge,
no one has confirmed this behavior of the vortex moduli space from
field theory. Nor has anyone found a use for it!

\subsection{What Became Of.......}

Let's now look at what became of the other solitons we studied in
the past two lectures.

\subsubsection{Monopoles}

Well, we've set $\phi=0$ throughout this lecture and, as we saw,
the monopoles live on the vev of $\phi$. So we shouldn't be
surprised if they don't exist in our theory \eqn{vlag}. We'll see
them reappear in the following section.

\subsubsection{Instantons}

These are more interesting. Firstly the vev $q\neq 0$ breaks
conformal invariance, causing the instantons to collapse. This is
the same behavior that we saw in Section 2.6. But recall that in
the middle of the vortex string, $q\rightarrow 0$. So maybe it's
possible for the instanton to live inside the vortex string, where
the non-abelian gauge symmetry is restored. To see that this can
indeed occur, we can look at the worldsheet of the vortex string.
As we've seen, the low-energy dynamics for a single string is
\be \mbox{$U(1)$ with N charged chiral multiplets and FI parameter
$r=4\pi/e^2$} \nn\ee
But this falls into the class of theories we discussed in section
3.5. So if the worldsheet is Euclidean, the theory on the vortex
string itself admits a vortex solution: a vortex in a vortex. The
action of this vortex is \cite{vstring}
\be S_{\rm vortex\ in\ vortex} = 2\pi r =\frac{8\pi^2}{e^2}=S_{\rm
inst} \ee
which is precisely the action of the Yang-Mills instanton. Such a
vortex has $2N$ zero modes which include scaling modes but, as we
mentioned previously, not all are normalizable.

\para
There is also a 4d story for these instantons buried in the vortex
string. This arises by completing the square in the Lagrangian in
a different way to  \eqn{vbog}. We still set $\phi=0$, but now
allow for all fields to vary in all four dimensions
\cite{vstring}. We write $z=x^1+ix^2$ and $w=x^3-ix^4$,
\be S&=& \int d^4x\ \frac{1}{2e^2}\Tr\
F_{\mu\nu}F^{\mu\nu}+\sum_{i=1}^{N_f}|\D_\mu
q_i|^2+\frac{e^2}{4}\Tr(\sum_{i=1}^{N_f}q_iq_i^\dagger -v^2\,{\bf
1}_{N_c})^2\nn\\ &=& \int d^4x\
\frac{1}{2e^2}\Tr\,\left(F_{12}-F_{34}-\frac{e^2}{2}(\sum_{i=1}^{N_f}
q_iq_i^\dagger-v^2\,{\bf 1}_{N_c})\right)^2 \nn\\ &&
+\sum_{i=1}^{N_f}|\D_zq_i|^2+|\D_\omega q_i|^2
+\frac{1}{e^2}\Tr\left((F_{14}-F_{23})^2+(F_{13}+F_{24})^2\right)
\nn\\ && + \frac{1}{e^2}\Tr\,F_{\mu\nu}{}^\star
F^{\mu\nu}+F_{12}v^2+F_{34}v^2 \nn\\
&\geq& \int d^4x\ \frac{1}{e^2}\Tr\,F_{\mu\nu}{}^\star
F^{\mu\nu}+\Tr\,(F_{12}v^2+F_{34}v^2)\ee
The last line includes three topological charges, corresponding to
instantons, vortex strings in the $(x^1-x^2)$ plane, and further
vortex strings in the $(x^3-x^4)$ plane. The Bogomoln'yi equations
describing these composite solutions are
\be F_{14}=F_{23}\ \ ,\ \ F_{13}=F_{24}\ \   ,\  \
F_{12}-F_{34}=\frac{e^2}{2}(\sum_{i=1}^{N_f}q_iq_i^\dagger-v^2{\bf
1}_{N_c})\ \ ,\ \  \D_zq_i=\D_wq_i=0 \nn\ee
It is not known if solutions exist, but the previous argument
strongly suggests that there should be solutions describing an
instanton trapped inside a vortex string. Some properties of this
configuration were studied in \cite{inhiggs}.

\para
The observation that a vortex in the vortex string is a Yang-Mills
instanton gives some rationale to the fact that $\vkn\subset\ikn$.

\subsection{Fermi Zero Modes}

In this section, I'd like to describe an important feature of
fermionic zero modes on the vortex string: they are chiral. This
means that a Weyl fermion in four dimensions will give rise to a
purely left-moving (or right-moving) mode on the (anti-) vortex
worldsheet. In fact, a similar behavior occurs for instantons and
monopoles, but since this is the first lecture where the solitons
are string-like in four-dimensions, it makes sense to discuss this
phenomenon here.

\para
The exact nature of the fermionic zero modes depends on the
fermion content in four dimensions. Let's stick with the
supersymmetric generalization of the Lagrangian \eqn{vlag}. Then
we have the gaugino $\lambda$, an adjoint valued Weyl fermion
which is the superpartner of the gauge field. We also have
fermions in the fundamental representation, $\chi_i$ with
$i=1,\ldots, N$, which are the superpartners of the scalars $q_i$.
These two fermions mix through Yukawa couplings of the form
$q^\dagger_i{\lambda}\chi_i$, and the Dirac equations read
\be -i\Dbarslash\lambda+i\sqrt{2}\sum_{i=1}^Nq_i\bar{\chi}_i=0\ \
\ &{\rm and}&\ \ \ \
-i\Dslash\bar{\chi}_i-i\sqrt{2}q_i^\dagger\lambda=0
\label{lambdap}\ee
where the Dirac operators take the form,
\be \Dslash\equiv\sigma^\mu{\cal D}_\mu=\left(\begin{array}{cc}
{\cal D}_+ & {\cal D}_z \\ {\cal D}_{\bar{z}} & {\cal
D}_-\end{array}\right)\ \ \  \ {\rm and}\ \ \ \ \
\Dbarslash\equiv\bar{\sigma}^\mu{\cal
D}_\mu=\left(\begin{array}{cc} {\cal D}_- & -{\cal D}_z \\ -{\cal
D}_{\bar{z}} & {\cal D}_+\end{array}\right)\label{ds}\ee
which, as we can see, nicely split into ${\cal D}_\pm={\cal
D}_0\pm{\cal D}_3$ and ${\cal D}_z={\cal D}_1-i{\cal D}_2$ and
${\cal D}_{\bar{z}}={\cal D}_1+i{\cal D}_2$. The bosonic fields in
\eqn{lambdap} are evaluated on the vortex solution which,
crucially, includes $\D_zq_i=0$ for the vortex (or
$\D_{\bar{z}}q_i=0$ for the anti-vortex). We see the importance of
this if we take the first equation in \eqn{lambdap} and hit it
with $\Dslash$, while hitting the second equation with
$\Dbarslash$. In each equation terms of the form ${\cal D}_zq_i$
will appear, and subsequently vanish as we evaluate them on the
vortex background. Let's do the calculation. We split up the
spinors into their components $\lambda_\alpha$ and
$(\chi_\alpha)_i$ with $\alpha=1,2$ and, for now, look for zero
modes that don't propagate along the string, so
$\partial_+=\partial_-=0$. Then the Dirac equations in component
form become
\be (-{\cal D}_z{\cal D}_{\bar{z}}+2q_iq_i^\dagger)\lambda_1=0\ \
&{\rm and}&\ \  (-{\cal D}_{\bar{z}}{\cal
D}_z+2q_iq_i^\dagger)\lambda_2-\sqrt{2}({\cal
D}_{\bar{z}}q_i)\bar{\chi}_{1i}=0 \nn\\ (-{\cal D}_{\bar{z}}{\cal
D}_z\delta^j_{\ i}+2q_i^\dagger q_j)\bar{\chi}_{2j}=0 \ \  &{\rm
and}&\ \ (-{\cal D}_z{\cal D}_{\bar{z}}\delta^j_{\
i}+2q_iq_j^\dagger)\bar{\chi}_{1j}-\sqrt{2} ({\cal D}_z
q_i^\dagger) \lambda_2 =0  \nn \ee
The key point is that the operators appearing in the first column
are positive definite, ensuring that $\lambda_1$ and $\chi_{2i}$
have no zero modes. In contrast, the equations for $\lambda_2$ and
$\bar{\chi}_{1i}$ do have zero modes, guaranteed by the index. We
therefore know that any zero modes of the vortex are of the form,
\be \lambda=\left(\begin{array}{c} 0 \\
\lambda\end{array}\right)\ \ \ \ {\rm and}\ \ \ \
\bar{\chi}_i=\left(\begin{array}{c} \bar{\chi}_{i} \\ 0
\end{array} \right)\ee
If we repeat the analysis for the anti-vortex, we find that the
other components turn on. To see the relationship to the chirality
on the worldsheet, we now allow the zero modes to propagate along
the string, so that $\lambda=\lambda(x^0,x^3)$ and
$\bar{\chi}_i=\bar{\chi}_i(x^0,x^3)$. Plugging this ansatz back
into the Dirac equation, now taking into account the derivatives
${\cal D}_\pm$ in \eqn{ds}, we find the equations of motion
\be {\partial}_+\lambda=0 \ \ \ \ {\rm and}\ \ \ \
{\partial}_+\bar{\chi}_i=0\ee
Or, in other words, $\lambda=\lambda(x_-)$ and
$\bar{\chi}=\bar{\chi}(x_-)$: both are right movers.

\para
In fact, the four-dimensional theory with only fundamental
fermions $\chi_i$ is anomalous. Happily, so is the $\CP^{N-1}$
theory on the string with only right-moving fermions, suffering
from the sigma-model anomaly \cite{sigmaanom}. To rectify this,
one may add four dimensional Weyl fermions $\tilde{\chi}_i$ in the
anti-fundamental representation, which provide left movers on the
worldsheet. If the four-dimensional theory has ${\cal N}=2$
supersymmetry, the worldsheet theory preserves ${\cal N}=(2,2)$
\cite{edel}. Alternatively, one may work with a chiral,
non-anomalous ${\cal N}=1$ theory in four-dimensions, resulting in
a chiral non-anomalous ${\cal N}=(0,2)$ theory on the worldsheet.

\subsection{Applications}

Let's now turn to discussion of applications of vortices in
various field theoretic contexts. We review some of  the roles
vortices play as finite action, instanton-like, objects in two
dimensions, as particles in three dimensions, and as strings in
four dimensions.

\subsubsection{Vortices and Mirror Symmetry}

Perhaps the most important application of vortices in string
theory is in the context of the $d=1+1$ dimensional theory on the
string itself. You might protest that the string worldsheet theory
doesn't involve a gauge field, so why would it contain vortices?!
The trick, as described by Witten \cite{phases}, is to view
sigma-models in terms of an auxiliary gauge theory known as a {\it
gauged linear sigma model}. We've already met this trick several
times in these lectures: the sigma-model target space is the Higgs
branch of the gauge theory. Witten showed how to construct gauge
theories that have compact Calabi-Yau manifolds as their Higgs
branch.

\para
In $d=1+1$ dimensions, vortices are finite action solutions to the
Euclidean equations of motion. In other words, they play the role
of instantons in the theory. As we explained Section 3.5 above,
the vortices are related to worldsheet instantons wrapping the
2-cycles of the Calabi-Yau Higgs branch. It turns out that it is
much easier to deal with vortices than directly with worldsheet
instantons (essentially because their moduli space is free from
singularities). Indeed, in a beautiful paper, Morrison and Plesser
succeeded in summing the contribution of all vortices in the
topological A-model on certain Calabi-Yau manifolds, showing that
it agreed with the classical prepotential derived from the B-model
on the mirror Calabi-Yau \cite{mp}.

\para
More recently, Hori and Vafa used vortices to give a proof of
${\cal N}=(2,2)$ mirror symmetry for all Calabi-Yau which can be
realized as complete intersections in toric varieties \cite{hv}.
Hori and Vafa work with dual variables, performing the so-called
Rocek-Verlinde transformation to twisted chiral superfields
\cite{rocver}. They show that vortices contribute to a two fermi
correlation function which, in terms of these dual variables, is
cooked up by a superpotential. This superpotential then captures
the relevant quantum information about the original theory.
Similar methods can be used in ${\cal N}=(4,4)$ theories to derive
the T-duality between NS5-branes and ALE spaces
\cite{ghm,mens5,next1,next2}, with the instantons providing the
necessary ingredient to break translational symmetry after
T-duality, leading to localized, rather than smeared, NS5-branes.

\subsubsection{Swapping Vortices and Electrons}

In lecture 2, we saw that it was possible to rephrase
four-dimensional field theories, treating the monopoles as
elementary particles instead of solitons. This trick, called
electric-magnetic duality, gives key insight into the strong
coupling behavior of four-dimensional field theories. In three
dimensions, vortices are particle like objects and one can ask the
same question: is it possible to rewrite a quantum field theory,
treating the vortices as fundamental degrees of freedom?

\para
The answer is yes. In fact, condensed matter theorists have been
using this trick for a number of years (see for example
\cite{wenzee}). Things can be put on  a much more precise footing
in the supersymmetric context, with the first examples given by
Intriligator and Seiberg \cite{is}. They called this phenomenon
"mirror symmetry" in three dimensions as it had some connection to
the mirror symmetry of Calabi-Yau manifolds described above.

\para
Let's describe the basic idea. Following Intriligator and Seiberg,
we'll work with a theory with eight supercharges (which is ${\cal
N}=4$ supersymmetry in three dimensions). Each gauge field comes
with three real scalars and four Majorana fermions. The charged
matter, which we'll refer to as "electrons", lives in a
hypermultiplet, containing two complex scalars together with two
Dirac fermions. The theory we start with is:
\be \mbox{Theory A: } U(1) \mbox{ with $N$ charged
hypermultiplets}\nn\ee
The vortices in this theory fall into the class described in
Section 3.5. Each vortex has $2N$ zero modes but, as we discussed,
not all of these zero modes are normalizable. The overall center
of mass is, of course, normalizable (the vortex has mass $M=2\pi
v^2$) but the remaining $2(N-1)$ modes of a single vortex are
logarithmically divergent.

\para
We now wish to rewrite this theory, treating the vortices as
fundamental objects. What properties must the theory have in order
to mimic the behavior of the vortex? It will prove useful to think
of each vortex as containing $N$ individual "fractional vortices".
We postulate that these fractional vortices suffer a logarithmic
confining potential, so that any number $n<N$ have a
logarithmically divergent mass, but $N$ together form a state with
finite mass. Such a system would exhibit the properties of the
vortex zero modes described above: the $2N$ zero modes correspond
to the positions of the $N$ fractional vortices. They can move
happily as a whole, but one pays a logarithmically divergent cost
to move these objects individually. (Note: a logarithmically
divergent cost isn't really that much!)

\EPSFIGURE{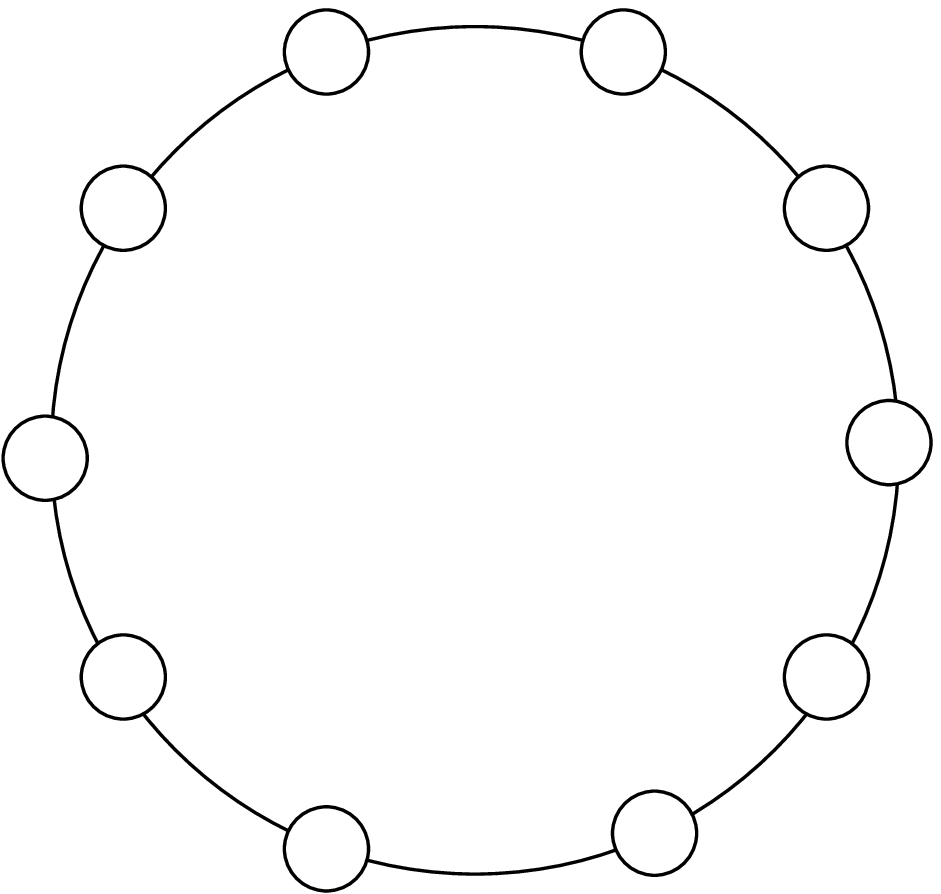,height=110pt}{}
\para
In fact, it's very easy to cook up a theory with these properties.
In $d=2+1$, an electron experiences logarithmic confinement, since
its electric field goes as $E\sim 1/r$ so its energy $\int d^2x\,
E^2$ suffers a logarithmic infra-red divergence. These electrons
will be our "fractional vortices". We will introduce $N$ different
types of electrons and, in order to assure that only bound states
of all $N$ are gauge singlets, we introduce $N-1$ gauge fields
with couplings dictated by the quiver diagram shown in the figure.
Recall that quiver diagrams are read in the following way: the
nodes of the quiver are gauge groups, each giving a $U(1)$ factor
in this case. Meanwhile, the links denote hypermultiplets with
charge $(+1,-1)$ under the gauge groups to which it is attached.
Although there are $N$ nodes in the quiver, the overall $U(1)$
decouples, leaving us with the theory
\be \mbox{Theory B: } U(1)^{N-1} \mbox{ with $N$
hypermultiplets}\nn \ee
This is the Seiberg-Intriligator mirror theory, capturing the same
physics as Theory A.  The duality also works the other way, with
the electrons of Theory A mapping to the vortices of Theory B. It
can be shown that the low-energy dynamics of these two theories
exactly agree. This statement can be made precise at the
two-derivative level. The Higgs branch of Theory A coincides with
the Coulomb branch of Theory B: both are $T^\star(\CP^{N-1})$.
Similarly, the Coulomb branch of Theory A coincides with the Higgs
branch of Theory A: both are the $A_{N-1}$ ALE space.

\para
There are now many mirror pairs of theories known in three
dimensions. In particular, it's possible to tinker with the mirror
theories so that they actually coincide at all length scales,
rather than simply at low-energies \cite{kapstrass}. Mirror pairs
for non-abelian gauge theories are known, but are somewhat more
complicated to due to presence of instanton corrections (which,
recall, are monopoles in three dimensions)
\cite{nonabmirror,nonabmirror2,nonabmirror3,nonabmirror4,ncmir}.
Finally, one can find mirror pairs with less supersymmetry
\cite{ahiss,n2mirror}, including mirrors for interesting
Chern-Simons theories \cite{kitaohta,csmirrors,csmirrors1}. Finite quantum
correction to the vortex mass in ${\cal N}=2$ theories was described in
\cite{alwaysmore}. The
Chern-Simons mirrors reduce to Hori-Vafa duality under
compactification to two dimensions \cite{ahkt}.

\subsubsection{Vortex Strings}

In $d=3+1$ dimensions, vortices are string like objects. There is
a very interesting story to be told about how we quantize vortex
worldsheet theory, which is a sigma-model on $\vkn$. But this will
have to wait for the next lecture.

\para
Here let me mention an application of vortices in the context of
cosmic strings which shows that reconnection of vortices in gauge
theories is inevitable at low-energies. Reconnection of strings
means that they swap partners as they intersect as shown in the
figure. In general, it's a difficult problem to determine whether
reconnection occurs and requires numerical study. However, at
low-energies we may reliably employ the techniques of the moduli
space approximation that we learnt above
\cite{edturok,eps,hanhash}.

\para
The first step is to reduce the dynamics of cosmic strings to that
of particles by considering one of two spatial slices shown in the
figure. The vertical slice cuts the strings to reveal a
vortex-anti-vortex pair. After reconnection, this slice no longer
intersects the strings, implying the annihilation of this pair.
Alternatively, one can slice horizontally to reveal two vortices.
Here the smoking gun for reconnection is the right-angle
scattering of the vortices at (or near) the interaction point.
Such $90^{\rm o}$ degree scattering is a requirement since, as is
clear from the figure, the two ends of each string are travelling
in opposite directions after the collision. By varying the slicing
along the string, one can reconstruct the entire dynamics of the
two strings in this manner and show the inevitability of
reconnection at low-energies.

\begin{figure}[htb]
\begin{center}
\epsfxsize=4.2in\leavevmode\epsfbox{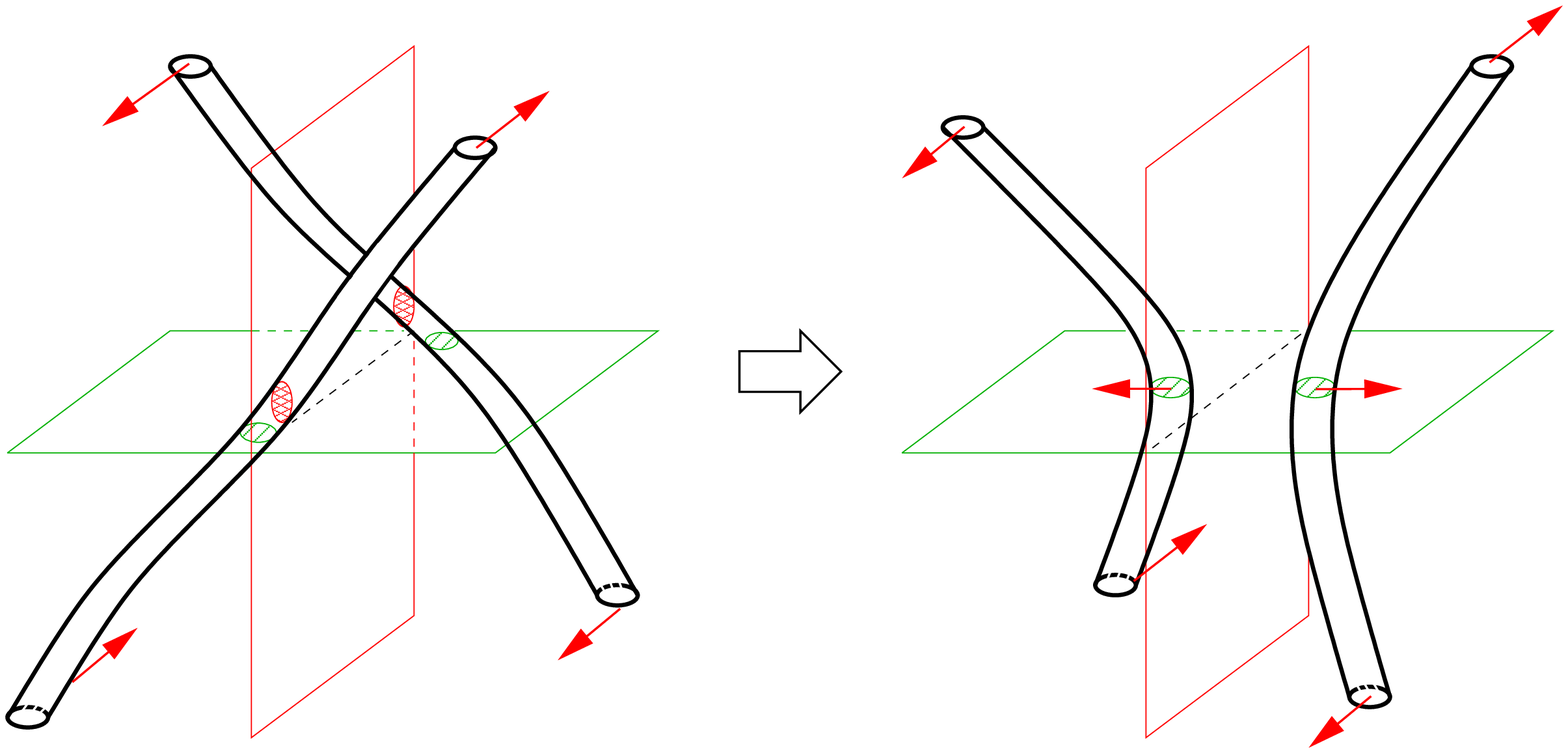}
\end{center}
\caption{The reconnection cosmic strings. Slicing vertically, one
sees a vortex-anti-vortex pair annihilate. Slicing horizontally,
one sees two vortices scattering at right angles.}
\end{figure}

\para
Hence, reconnection of cosmic strings requires both the
annihilation of vortex-anti-vortex pairs and the right-angle
scattering of two vortices. The former is expected (at least for
suitably slow collisions). And we saw in Section 3.3.2 that the
latter occurs for abelian vortices in the moduli space
approximation. We conclude that abelian cosmic strings do
reconnect at low energies. Numerical simulations reveal that these
results are robust, holding for very high energy collisions
\cite{matzner}.

\para
For cosmic strings in non-abelian theories this result continues
to hold, with strings reconnecting except for very finely tuned
initial conditions \cite{hashtong}. However, in this case there
exist mechanisms to push the strings to these finely tuned
conditions, resulting in a probability for reconnection less than
1.

\para
Recently, there has been renewed interest in the reconnection of
cosmic strings, with the realization that cosmic strings may be
fundamental strings, stretched across the sky \cite{witcos}. These
objects differ from abelian cosmic strings as they have a reduced
probability of reconnection, proportional to the string coupling
$g^2_s$ \cite{pol,jjp}. If cosmic strings are ever discovered, it
may be possible to determine their probability of reconnection,
giving a vital clue to their microscopic origin. The recent
developments of this story have been nicely summarized in the
review \cite{polcos}.

\newpage
\section{Domain Walls}

So far we've considered co-dimension 4 instantons, co-dimension 3
monopoles and co-dimension 2 vortices. We now come to co-dimension
1 domain walls, or kinks as they're also known. While BPS domain
walls exist in many supersymmetric theories (for example, in
Wess-Zumino models \cite{cecvaf}), there exists a special class of
domain walls that live in gauge theories with 8 supercharges. They
were first studied by Abraham and Townsend \cite{at} and  have
rather special properties. These will be the focus of this
lecture. As we shall explain below, the features of these domain
walls are inherited from the other solitons we've met, most
notably the monopoles.

\subsection{The Basics}

To find domain walls, we need to deform our theory one last time.
We add masses $m_i$ for the fundamental scalars $q_i$. Our
Lagrangian is that of a $U(N_c)$ gauge theory, coupled to a real
adjoint scalar field $\phi$ and $N_f$ fundamental scalars $q_i$
\be S&=&\int d^4x\
\Tr\left(\frac{1}{2e^2}F^{\mu\nu}F_{\mu\nu}+\frac{1}{e^2}({\cal
D}_\mu\phi)^2\right) +\sum_{i=1}^{N_f}|{\cal D}_\mu q_i|^2 \nn\\
&& \ \ \ \ \ \ \ \ \ \ \ \ \
-\sum_{i=1}^{N_f}q_i^\dagger(\phi-m_i)^2 q_i
-\frac{e^2}{4}\,\Tr\,(\sum_{i=1}^{N_f}q_iq_i^\dagger-v^2\,
1_N)^2\label{dwlag}\ee
Notice the way the masses mix with $\phi$, so that the true mass
of each scalar is $|\phi-m_i|$. Adding masses in this way is
consistent with $\N=2$ supersymmetry. We'll pick all masses to be
distinct and, without loss of generality, choose
\be m_i<m_{i+1} \ee
As in Lecture 3, there are vacua with $V=0$ only if $N_f\geq N_c$.
The novelty here is that, for $N_f>N_c$, we have multiple isolated
vacua. Each vacuum is determined by a choice of $N_c$ distinct
elements from a set of $N_f$
\be \Xi=\{\xi(a):\ \xi(a)\neq\xi(b)\ {\mbox {\rm for}\ } a\neq
b\}\ee
where $a=1,\ldots,N_c$ runs over the color index, and $\xi(a)\in
\{1,\ldots,N_f\}$. Let's set $\xi(a)<\xi(a+1)$. Then, up to a Weyl
transformation, we can set the first term in the potential to
vanish by
\be \phi={\rm diag}(m_{\xi(1)},\ldots,m_{\xi(N_c)})\ee
This allows us to turn on the particular components $q^a_{\ i}\sim
\delta^a_{\ i=\xi(a)}$ without increasing the energy. To cancel
the second term in the potential, we require
\be q^a_{\ i}=v\delta^a_{\ i=\xi(a)}\ee
The number of vacua of this type is
\be N_{\rm vac}=\left(\begin{array}{c}N_f \\
N_c\end{array}\right)=\frac{N_f!}{N_c!(N_f-N_c)!}\label{nvac}\ee
Each vacuum has a mass gap in which there are $N_c^2$ gauge bosons
with $M_\gamma^2=e^2v^2+|m_{\xi(a)}-m_{\xi(b)}|^2$, and
$N_c(N_f-N_c)$ quark fields with mass $M_q^2=|m_{\xi(a)}-m_i|^2$
with $i\notin\Xi$.

\para
Turning on the masses has explicitly broken the $SU(N_f)$ flavor
symmetry to
\be SU(N_f)\rightarrow U(1)^{N_f-1}_F\label{dwflavor}\ee
while the $U(N_c)$ gauge group is also broken completely in the
vacuum. (Strictly speaking it is a combination of the $U(N_c)$
gauge group and $U(1)_F^{N_f-1}$ that survives in the vacuum).

\subsection{Domain Wall Equations}

The existence of isolated vacua implies the existence of a domain
wall, a configuration that interpolates from a given vacuum
$\Xi_-$ at $x^3\rightarrow -\infty$ to a distinct vacuum $\Xi_+$
at $x^3\rightarrow +\infty$. As in each previous lecture, we can
derive the first order equations satisfied by the domain wall
using the Bogomoln'yi trick. We'll chose $x^3$ to be the direction
transverse to the wall, and set
$\partial_0=\partial_1=\partial_3=0$ as well as $A_0=A_1=A_2=0$.
The tension of the domain wall can be written as \cite{kinky}
\be T_{\rm wall}&=&\int dx^3\
\frac{1}{e^2}\Tr\,\left(\D_3\phi+\frac{e^2}{2}(\sum_{i=1}^{N_f}
q_iq_i^\dagger-v^2) \right)^2
-\D_3\phi\,(\sum_{i=1}^{N_f}q_iq_i^\dagger-v^2)\nn\\
&&\ \
+\sum_{i=1}^{N_f}\left(|\D_3q_i+(\phi-m_i)q_i|^2-q_i^\dagger(\phi-m_i)
\D_3q_i-\D_3q_i^\dagger (\phi-m_i)q_i\right)\nn\\ &\geq &
v^2\,\left[\Tr\phi\right]^{+\infty}_{-\infty}\ee
With our vacua $\Xi_-$ and $\Xi_+$ at left and right infinity, we
have the tension of the domain wall bounded by
\be T_{\rm wall}\geq
v^2[\Tr\phi]^{+\infty}_{-\infty}=v^2\sum_{i\in\Xi_+}m_i-v^2\sum_{i\in\Xi_-}m_i\ee
and the minus signs have been chosen so that this quantity is
positive (if this isn't the case we must swap left and right
infinity and consider the anti-wall). The bound is saturated when
the domain wall equations are satisfied,
\be \D_3\phi=-\frac{e^2}{2}(\sum_{i=1}^{N_f}q_iq_i^\dagger-v^2)\ \
\ \ ,\ \ \ \ \D_3q_i=-(\phi-m_i)q_i\label{dw}\ee
Just as the monopole equations $\D\phi=B$ arise as the dimensional
reduction of the instanton equations $F=\starf$, so the domain
wall equations \eqn{dw} arise from the dimensional reduction of
the vortex equations. To see this, we look for solutions to the
vortex equations with $\partial_2=0$ and relabel $x^1\rightarrow
x^3$ and $(A_1,A_2)\rightarrow (A_3,\phi)$. Finally, the analogue
of turning on the vev in going from the instanton to the monopole,
is to turn on the masses $m_i$ in going from the vortex to the
domain wall. These can be thought of as a "vev" for $SU(N_f)$ the
flavor symmetry.

\subsubsection{An Example}

The simplest theory admitting a domain wall is $U(1)$ with $N_f=2$
scalars $q_i$. The domain wall equations are
\be \partial_3\phi=-\frac{e^2}{2}(|q_1|^2+|q_2|^2-v^2)\ \ \ \ ,\ \
\ \ \D_3q_i=-(\phi-m_i)q_i\label{domestic}\ee
We'll chose $m_2=-m_1=m$. The general solution to these equations
is not known. The profile of the wall depends on the value of the
dimensionless constant $\gamma=e^2v^2/m^2$. For $\gamma\ll 1$, the
wall can be shown to have a three layer structure, in which the
$q_i$ fields decrease to zero in the outer layers, while $\phi$
interpolates between its two expectation values at a more
leisurely pace \cite{sydbrane}. The result is a domain wall with
width $L_{\rm wall}\sim m/e^2v^2$. Outside of the wall, the fields
asymptote exponentially to their vacuum values.

\begin{figure}[htb]
\begin{center}
\epsfxsize=2.8in\leavevmode\epsfbox{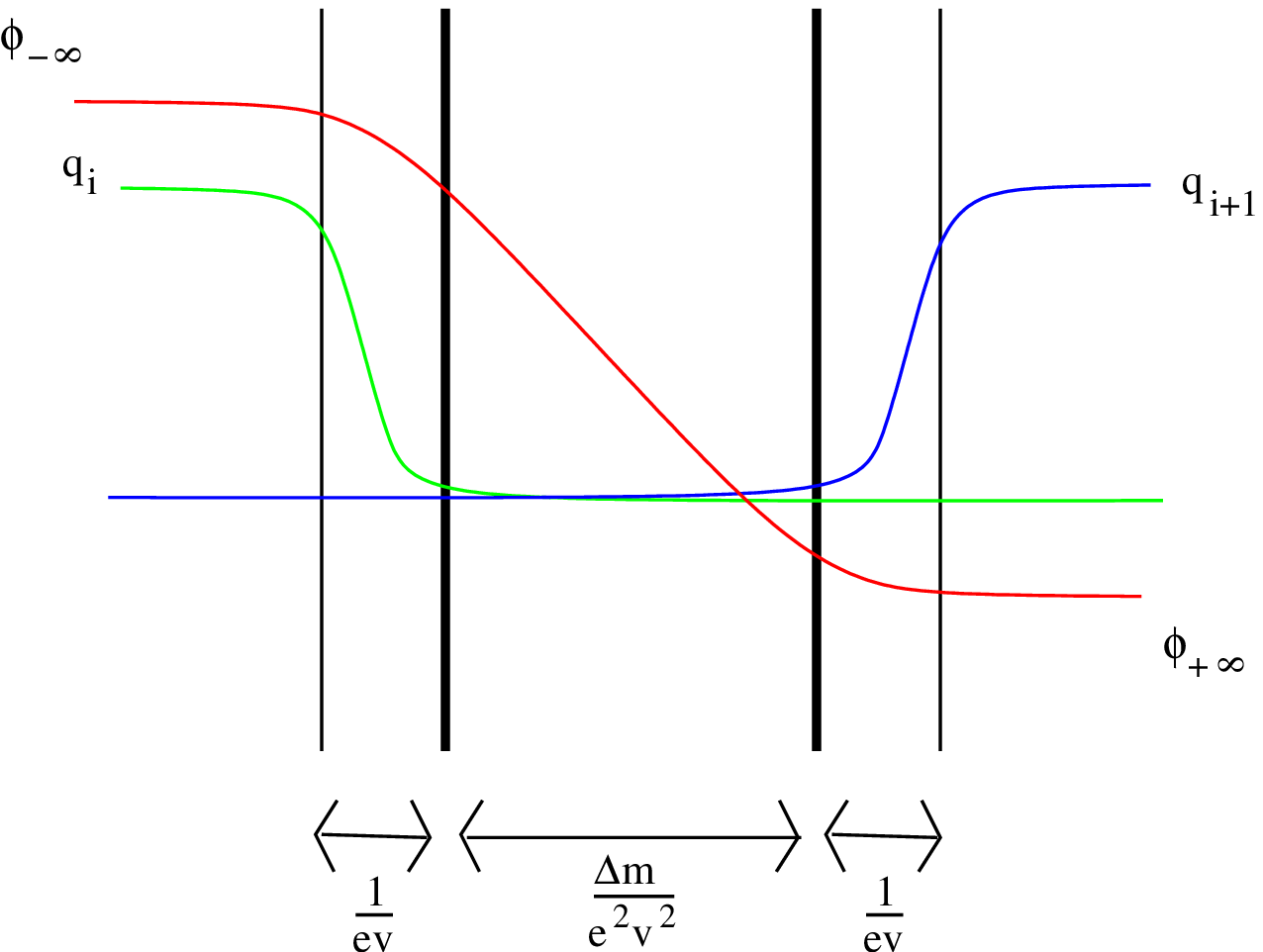}
\end{center}
\caption{The three layer structure of the domain wall when
$e^2v^2\ll m^2$.}
\end{figure}

In the opposite limit $\gamma\gg 1$, the inner segment collapses
and the two outer layers coalesce, leaving us with a domain wall
of width $L_{\rm wall}\sim 1/m$. In fact, if we take the limit
$e^2\rightarrow \infty$, the first equation \eqn{domestic} becomes
algebraic while the second is trivially solved. We find the
profile of the domain wall to be \cite{tong}
\be q_1=\frac{v}{A}\,e^{-m(x_3-X)+i\theta}\ \ \ ,\ \ \ \
q_2=\frac{v}{A}\,e^{+m(x_3-X)-i\theta}\label{dwsol}\ee
where $A^2=e^{-2m(x_3-X)}+e^{+2m(x_3-X)}$.

\para
The solution \eqn{dwsol} that we've found in the
$e^2\rightarrow\infty$ limit has two collective coordinates, $X$
and $\theta$. The former is simply the position of the domain wall
in the transverse $x^3$ direction. The latter is also easy to see:
it arises from acting on the domain wall with the $U(1)_F$ flavor
symmetry of the theory \cite{at}:
\be U(1)_F: q_1\rightarrow e^{i\theta}q_1\ \ ,\ \ q_2\rightarrow
e^{-i\theta}q_2\ee
In each vacuum, this coincides with the $U(1)$ gauge symmetry.
However, in the interior of the domain wall, it acts
non-trivially, giving rise to a phase collective coordinate
$\theta$ for the solution. It can be shown that $X$ and $\theta$
remain the only two collective coordinates of the domain wall when
we return to finite $e^2$ \cite{sakaisol}.

\subsubsection{Classification of Domain Walls}

So we see above that the simplest domain wall has two collective
coordinates. What about the most general domain wall,
characterized by the choice of vauca $\Xi_-$ and $\Xi_+$ at left
and right infinity. At first sight it appears a little daunting to
classify these objects. After all, a strict classification of the
topological charge requires a statement of the vacuum at left and
right infinity, and the number of vacua increases exponentially
with $N_f$. To ameliorate this sense of confusion, it will help to
introduce a coarser classification of domain walls which will
capture some information about the topological sector, without
specifying the vacua completely. This classification, introduced
in \cite{boojum}, will prove most useful when relating our domain
walls to the other solitons we've met previously. To this end,
define the $N_f$-vector
\be \vec{m}=(m_1,\ldots,m_{N_f}) \ee
We can then write the tension of the domain wall as
\be T_{\rm wall}=v^2\vec{g}\cdot\vec{m}\ee
which defines a vector $\vec{g}$ that contains entries $0$ and
$\pm 1$ only. Following the classification of monopoles in Lecture
2, let's decompose  this vector as
\be \vec{g}=\sum_{i=1}^{N_f}n_i\valpha_i\ee
with $n_i\in\Z$ and the $\valpha_i$ the simple roots of $su(N_f)$,
\be \valpha_1&=&(1,-1,0,\ldots,0)\nn\\
\valpha_2&=&(0,1,-1,\ldots,0)\nn\\\valpha_{N_f-1}&=&(0,\ldots,0,1,-1)\nn\ee
Since the vector $\vec{g}$ can only contain $0$'s, $1$'s and
$-1$'s, the integers $n_i$ cannot be arbitrary. It's not hard to
see that this restriction means that neighboring $n_i$'s are
either equal or differ by one: $n_i=n_{i+1}$ or $n_i=n_{i+1}\pm
1$.

\subsection{The Moduli Space}

A choice of $\vec{g}$ does not uniquely determine a choice of
vacua at left and right infinity. Nevertheless, domain wall
configurations which share the same $\vec{g}$ share certain
characteristics, including the number of collective coordinates.
The collective coordinates carried by a given domain wall was
calculated in a number of situations in \cite{multi,keith,sakai}.
Using our classification, the index theorem tells us that there
are solutions to the domain wall equations \eqn{dw} only if
$n_i\geq0$ for all $i$. Then the number of collective coordinates
is given by \cite{boojum},
\be \dim{\cal W}_{\vec{g}} = 2\sum_{i=1}^{N_f-1}n_i\label{dwcc}\ee
where ${\cal W}_{\vg}$ denotes the moduli space of any set of
domain walls with charge $\vg$. Again, this should be looking
familiar! Recall the result for monopoles with charge $\vg$ was
$\dim({\cal M}_{\vec{g}})=4\sum_an_a$. The interpretation of the
result \eqn{dwcc} is, as for monopoles, that there are $N_f-1$
elementary types of domain walls associated to the simple roots
$\vg=\valpha_i$. A domain wall sector in sector $\vg$ then splits
up into $\sum_in_i$ elementary domain walls, each with its own
position and phase collective coordinate.

%
%

\subsubsection{The Moduli Space Metric}

The low-energy dynamics of multiple, parallel, domain walls is
described, in the usual fashion, by a sigma-model from the domain
wall worldvolume to the  target space is $\W_{\vg}$. As with other
solitons, the domain walls moduli space $\W_{\vg}$ inherits a
metric from the zero modes of the solution. In notation such that
$q=q^a_{\ i}$ is an $N_c\times N_f$ matrix, the linearized domain
wall equations \eqn{dw}
\be \D_3\d\phi-i[\d A_3,\phi]&=&-\frac{e^2}{2}(\d q\,q^\dagger+q\d
q^\dagger)\nn\\\D_3\d q-i\d A_3\,q&=&-(\phi\d q+\d \phi\, q-\d
q\,m) \ee
where $m=\diag(m_1,\ldots,m_{N_f})$ is an $N_f\times N_f$ matrix.
Again, these are to be supplemented by a background gauge fixing
condition,
\be \D_3\d A_3-i[\phi,\d\phi]=i\frac{e^2}{2}(q\d q^\dagger-\d q\,
q^\dagger)\ee
and the metric on the moduli space $\W_{\vg}$ is defined by the
overlap of these zero modes,
\be g_{\alpha\beta}=\int dx^3\ \Tr\left(\frac{1}{e^2}\left[
\d_\alpha A_3\,\d_\beta
A_3+\d_\alpha\phi\,\d_\beta\phi\right]+\d_\alpha q\,\d_\beta
q^\dagger+\d_\beta q\,\d_\alpha q^\dagger\right)\ee
By this stage, the properties of the metric on the soliton moduli
space should be familiar. They include.
\begin{itemize}
\item The metric is K\"ahler.

\item The metric is smooth. There is no singularity as two domain
walls approach each other.

\item The metric inherits a $U(1)^{N-1}$ isometry from the action
of the unbroken flavor symmetry \eqn{dwflavor} acting on the
domain wall.
\end{itemize}

\subsubsection{Examples of Domain Wall Moduli Spaces}

Let's give some simple examples of domain wall moduli spaces.

\subsubsection*{One Domain Wall}

We've seen that a  single elementary domain wall $\vg=\valpha_1$
(for example, the domain wall described above in the theory with
$N_c=1$ and $N_f=2$) has two collective coordinates: its center of
mass $X$ and a phase $\theta$. The moduli space is
\be \W_{\alpha}\cong \R\times \S^1\ee
The metric on this space is simple to calculate. It is
\be ds^2= (v^2\vec{m}\cdot\vg)\ dX^2+v^2\,d\theta^2 \ee
with the phase collective coordinate living in $\theta\in
[0,2\pi)$.

\subsubsection*{Two Domain Walls}

We can't have two domain walls of the same type, say
$\vg=2\valpha_1$, since there is no choice of vacua that leads to
this charge. Two elementary domain walls must necessarily be of
different types, $\vg=\valpha_i+\valpha_j$ for $i\neq j$. Let's
consider $\vg=\valpha_1+\valpha_2$.

\para
The moduli space is simplest to describe if the two domain walls
have the same mass, so
$\vec{m}\cdot\valpha_a=\vec{m}\cdot\valpha_b$. The moduli space is
\be \W_{\valpha_1+\valpha_2}\cong \R\times\frac{\S^1\times \M_{\rm
cigar}}{\Z_2}\ee
where the interpretation of the $\R$ factor and $\S^1$ factor are
the same as before. The relative moduli space has the topology and
asymptotic form of a cigar,
\begin{figure}[htb]
\begin{center}
\epsfxsize=3.2in\leavevmode\epsfbox{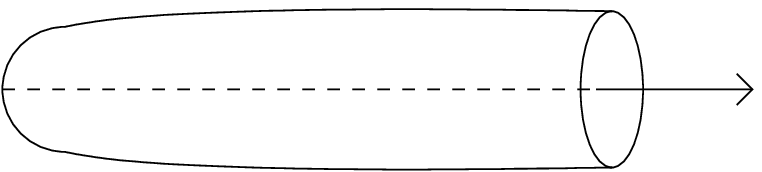}
\end{center}
\caption{The relative moduli space of two domain walls is a
cigar.}
\end{figure}
The relative separation between domain walls is denoted by $R$.
The tip of the cigar, $R=0$, corresponds to the two domain walls
sitting on top of each other. At this point the relative phase of
the two domain walls degenerates, resulting in a smooth manifold.
The metric on this space has been computed in the
$e^2\rightarrow\infty$ limit, although it's not particularly
illuminating \cite{tong} and gives a good approximation to the
metric at large finite $e^2$ \cite{finite}. Asymptotically, it
deviates from the flat metric on the cylinder by exponentially
suppressed corrections $e^{-R}$, as one might expect since the
profile of the domain walls is exponentially localized.

\subsection{Dyonic Domain Walls}

You will have noticed that, rather like monopoles, the domain wall
moduli space includes a phase collective coordinate $\S^1$ for
each domain wall. For the monopole, excitations along this $\S^1$
give rise to dyons, objects with both magnetic and electric
charges. For domain walls, excitations along this $\S^1$ also give
rise to dyonic objects, now carrying both topological (kink)
charge and flavor charge. Abraham and Townsend called these
objects "Q-kinks" \cite{at}.

\para
First order equations of motion  for these dyonic domain walls may
be obtained by completing the square in the Lagrangian
\eqn{dwlag}, now looking for configurations that depend on both
$x^0$ and $x^3$, allowing for a non-zero electric field $F_{03}$.
We have
\be T_{\rm wall}&=&\int dx^3\ \frac{1}{e^2}\Tr\left(\cos\alpha\,
\D_3\phi +
\frac{e^2}{2}(\sum_{i=1}^{N_f}|q_iq_i^\dagger-v^2)\right)^2
-\cos\alpha\,\D_3\phi\,(\sum_{i=1}^{N_f}q_iq_i^\dagger-v^2)\nn\\
&&
+\sum_{i=1}^{N_f}\left(|\D_3q_i+\cos\alpha\,(\phi-m_i)q_i|^2-\cos\alpha\,(
q_i^\dagger(\phi-m_i)\D_3q_i+\  {\rm h.c.})\right)\nn\\
&&+ \frac{1}{e^2}\Tr(F_{03}-\sin\alpha\,\D_3\phi)^2 +
\frac{1}{e^2}\sin\alpha\,F_{03}\D_3\phi\nn\\ && +
\sum_{i=1}^{N_f}\left(|{\cal D}_0q_i+ i\sin\alpha(\phi-m_i)q_i|^2
-\sin\alpha(iq_i^\dagger(\phi-m_i)\D_0q_i +\ {\rm
h.c})\right)\nn\ee
As usual, insisting upon the vanishing of the total squares yields
the Bogomoln'yi equations. These are now to augmented with Gauss'
law,
\be \D_3 F_{03}=ie^2\sum_{i=1}^{N_f}(q_i{\cal D}_0{q}^\dagger_i
-({\cal D}_0q_i)q_i^\dagger) \ee
Using this, we may re-write the cross terms in the energy-density
to find the Bogomoln'yi bound,
\be T_{\rm wall}\geq \pm
v^2[\Tr\phi]^{+\infty}_{-\infty}\,\cos\alpha +
(\vec{m}\cdot\vec{S})\sin\alpha \ee
\EPSFIGURE{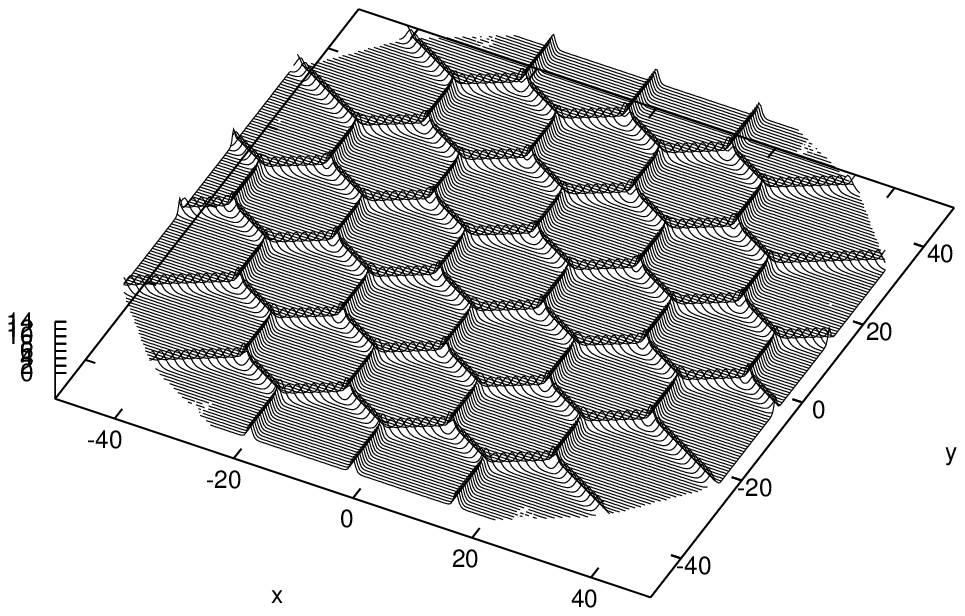,height=130pt}{}
\noindent where $\vec{S}$ is the Noether charge associated to the
surviving $U(1)^{N_f-1}$ flavor symmetry, an $N_f$-vector with
$i^{\rm th}$ component given by
\be S_i=i(q_i\D_0q_i^\dagger-({\cal D}_0{q}_i)q_i^\dagger) \ee
Maximizing with respect to $\alpha$ results in the Bogomoln'yi
bound for dyonic domain walls,
\be {\cal
H}\geq\sqrt{v^4(\vec{m}\cdot\vec{g})^2+(\vec{m}\cdot\vec{S})^2}
\label{dwdyonspec}\ee
This square-root form is familiar from the spectrum of dyonic
monopoles that we saw in Lecture 2. More on this soon. For now,
some further comments, highlighting the some similarities between
dyonic domain walls and monopoles.

\begin{itemize}
\item There is an analog of the Witten effect. In two dimensions,
where the domain walls are particle-like objects, one may add a
theta term of the form $\theta F_{01}$. This induces a flavor
charge on the domain wall, proportional to its topological charge,
$\vec{S}\sim\vec{g}$ \cite{nick}.

\item  One can construct dyonic domain walls with $\vec{g}$ and
$\vec{S}$ not parallel if we turn on complex masses and,
correspondingly, consider a complex adjoint scalar $\phi$
\cite{18korea,18japan}. The resulting 1/4 and 1/8-BPS states are
the analogs of the 1/4-BPS monopoles we briefly mentioned in
Lecture 2.

\item The theory with complex masses also admits interesting
domain wall junction configurations \cite{intersect,junction}.
Most notably, Eto et al. have recently found beautiful webs of
domain walls, reminiscent of $(p,q)$5-brane webs of IIB string
theory, with complicated moduli as the strands of the web shift,
causing cycles to collapse and grow \cite{web,naweb}. Examples
include the intricate honeycomb structure shown in figure 24
(taken from \cite{web}).

\end{itemize}

Other aspects of these domain walls were discussed in
\cite{arai,losev,ivanovask,allquarter,sakaimore}. A detailed
discussed of the mass renormalization of supersymmetric kinks
in $d=1+1$ dimensions can be found in \cite{kinkren1,kinkren2}.

\subsection{The Ordering of Domain Walls}

The cigar moduli space for two domain walls illustrates an
important point: domain walls cannot pass each other. In contrast
to other solitons, they must satisfy a particular ordering on the
line. This is apparent in the moduli space of two domain walls
since the relative separation takes values in $R\in \R^+$ rather
than $\R$. The picture in spacetime shown in figure 25.
\begin{figure}[htb]
\begin{center}
\epsfxsize=4.7in\leavevmode\epsfbox{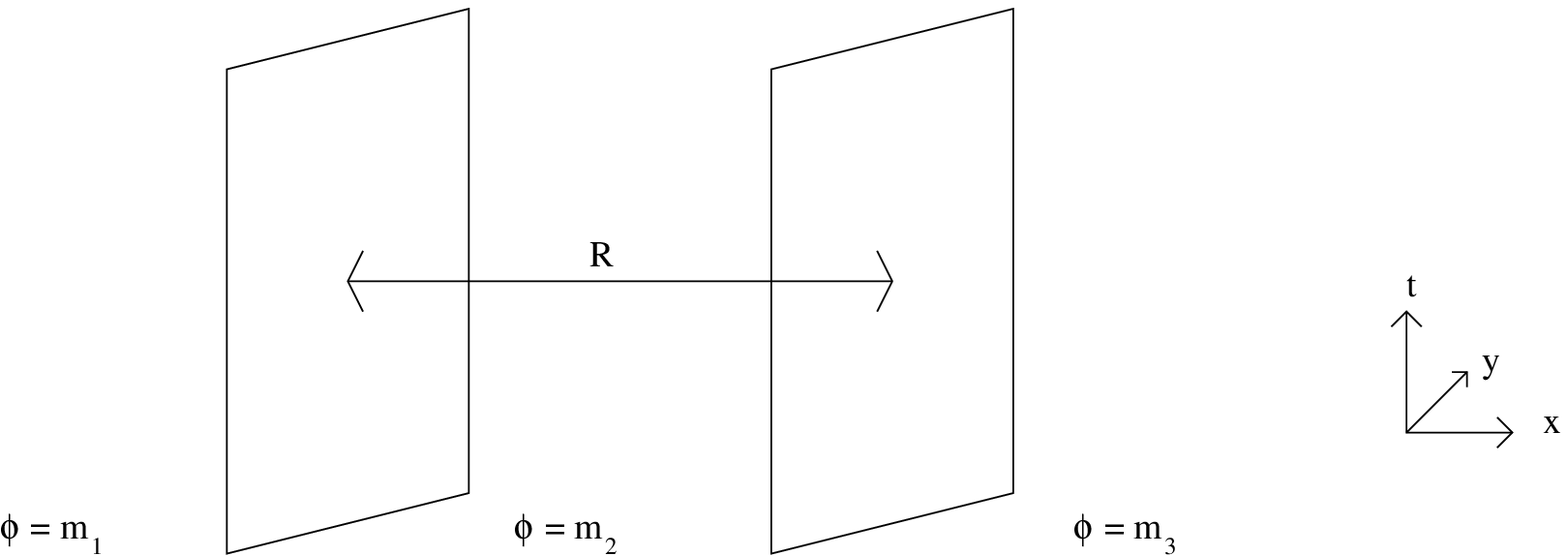}
\end{center}
\caption{Two interacting domain walls cannot pass through each
other. The $\valpha_1$ domain wall is always to the left of the
$\valpha_2$ domain wall.}
\end{figure}

\para
However, it's not always true that domain walls cannot pass
through each other. Domain walls which live in different parts of
the flavor group, so have $\valpha_i\cdot\valpha_j=0$, do not
interact so can happily move through each other. When these domain
walls are two of many in a topological sector $\vg$, an
interesting pattern of interlaced walls arises, determined by
which walls bump into each other, and which pass through each
other. This pattern was first explored in  \cite{sakai}. Let's see
how the ordering emerges. Start at left infinity in a particular
vacuum $\Xi_-$. Then each elementary domain wall shifts the vacuum
by increasing a single element $\xi(a)\in\Xi$ by one. The
restriction that the $N_c$ elements are distinct means that only
certain domain walls can occur. This point is one that is best
illustrated by a simple example:

\subsubsection*{An Example: $N_c=2$, $N_f=4$}

Consider the domain walls in the $U(2)$ theory with $N_f=4$
flavors. We'll start at left infinity in the vacuum
$\Xi_-=\{1,2\}$ and end at right infinity in the vacuum
$\Xi_+=\{3,4\}$. There are two different possibilities for the
intermediate vacua. They are:
\be &&\Xi_-=\{1,2\}\longrightarrow\{1,3\}\longrightarrow\{1,4\}
\longrightarrow\{2,4\}\longrightarrow\{3,4\}=\Xi_+\nn\\
&&\Xi_-=\{1,2\}\longrightarrow\{1,3\}\longrightarrow\{2,3\}
\longrightarrow\{2,4\}\longrightarrow\{3,4\}=\Xi_+ \nn\ee
In terms of domain walls, these two ordering become,
\be && \valpha_2\longrightarrow \valpha_3\longrightarrow
\valpha_1\longrightarrow \valpha_2  \nn\\
&&\valpha_2\longrightarrow\valpha_1\longrightarrow
\valpha_3\longrightarrow \valpha_2 \label{expos}\ee
%
%
We see that the two $\valpha_2$ domain walls must play bookends to
the $\valpha_1$ and $\valpha_3$ domain walls. However, one expects
that these middle two walls are able to pass through each other.

\subsubsection*{The General Ordering of Domain Walls}

We may generalize the discussion above to deduce the rule for
ordering of general domain walls \cite{sakai}. One finds that the
$n_i$ elementary $\valpha_i$ domain walls must be interlaced
between the $\valpha_{i-1}$ and $\valpha_{i+1}$ domain walls.
(Recall that $n_i=n_{i+1}$ or $n_i=n_{i+1}\pm 1$ so the concept of
interlacing is well defined). The final pattern of domain walls is
captured in figure 26, where $x^3$ is now plotted horizontally and
the vertical position of the domain wall simply denotes its type.
We shall see this vertical position take on life in the D-brane
set-up we describe shortly.
\begin{figure}[htb]
\begin{center}
\epsfxsize=4.2in\leavevmode\epsfbox{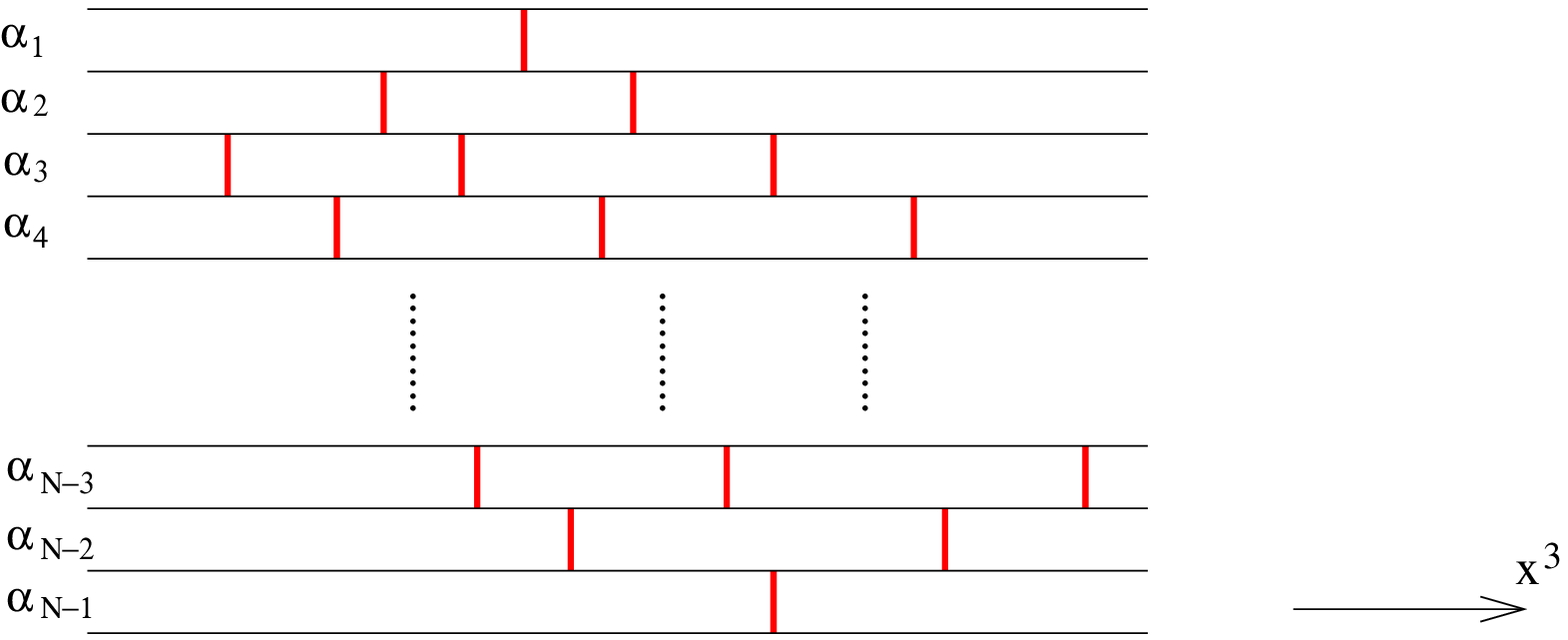}
\end{center}
\caption{The ordering of many domain walls. The horizontal
direction is their position, while the vertical denotes the type
of domain wall.}
\end{figure}

\para
Notice that the $\valpha_1$ domain wall is trapped between the two
$\valpha_2$ domain walls. These in turn are trapped between the
three $\valpha_3$ domain walls. However, the relative positions of
the $\valpha_1$ and middle $\valpha_3$ domain walls are not fixed:
these objects can pass through each other.

\subsection{What Became Of......}

Now let's play our favorite game and ask what happened to the
other solitons now that we've turned on the masses. We start
with.....

\subsubsection{Vortices}

The vortices described in the previous lecture enjoyed zero modes
arising from their embedding in $SU(N)_{\rm diag}\subset
U(N_c)\times SU(N_f)$. Let go back to the situation with
$N_f=N_c=N$, but with the extra terms from \eqn{dwlag} added to
the Lagrangian,
\be V =
\frac{1}{e^2}\Tr\,(\D_\mu\phi)^2+\sum_{i=1}^Nq_i^\dagger(\phi-m_i)^2q_i
\label{vv}\ee
As we've seen, this mass term breaks $SU(N)_{\rm diag}\rightarrow
U(1)^{N-1}_{\rm diag}$, which means we can no longer rotate the
orientation of the vortices within the gauge and flavor groups. We
learn that the masses are likely to lift some zero modes of the
vortex moduli space \cite{monflux,sy,vstring}.

\para
The vortex solutions that survive are those whose energy isn't
increased by the extra terms in $V$ above. Or, in other words,
those vortex configurations which vanish when evaluated on $V$
above. If we don't want the vortex to pick up extra energy from
the kinetic terms $\D\phi^2$, we need to keep $\phi$ in its
vacuum,
\be \phi={\rm diag}(\phi_1,\ldots,\phi_N) \ee
which means that only the components $q^a_{\ i}\sim \delta^a_{\
i}$ can turn on keeping $V=0$.

\para
For the single vortex $k=1$ in $U(N)$, this means that the
internal moduli space $\CP^{N-1}$ is lifted, leaving behind $N$
different vortex strings, each with magnetic field in a different
diagonal component of the gauge group,
\be B_3&=&{\rm diag}(0,\ldots,B^\star_3,\ldots,0)\nn\\q&=&{\rm
diag}(v,\ldots,q^\star,\ldots,v)\ee
In summary, rather than having a moduli space of vortex strings,
we are left with $N$ different vortex strings, each carrying
magnetic  flux in a different $U(1)\subset U(N)$.

\para
How do we see this from the perspective of the vortex worldsheet?
We can re-derive the vortex theory using the brane construction of
the previous lecture, but now with the D6-branes separated in the
$x^4$ direction, providing masses $m_i$ for the hypermultiplets
$q_i$. After performing the relevant brane-game manipulations, we
find that these translate into masses $m_i$ for the chiral
multiplets in the vortex theory. The potential for the vortex
theory (3.30) is replaced by,
\be
V&=&\frac{1}{g^2}\Tr\,|[\sigma,\sigma^\dagger]|^2+\Tr\,|[\sigma,Z]|^2+
\Tr\,|[\sigma,Z^\dagger]|^2\nn\\&&
+\sum_{a=1}^N\,\psi^\dagger_a(\sigma-m_a)^2\psi_a  +
\frac{g^2}{2}\,\Tr\,\left(\sum_a\psi_a\psi_a^\dagger +
[Z,Z^\dagger]-r\,1_k\right)^2\ee
where $r=2\pi/e^2$ as before. The masses $m_i$ of the
four-dimensional theory have descended to masses $m_a$ on the
vortex worldsheet.

\para
To see the implications of this, consider the theory on a single
$k=1$ vortex. The potential is simply,
\be V_{k=1}=\sum_{a=1}^N(\sigma-m_a)^2|\psi_a|^2
+\frac{g^2}{2}(\sum_{a=1}^N|\psi_a|^2-r)^2\ee
Whereas before we could set $\sigma=0$, leaving $\psi_a$ to
parameterize $\CP^{N-1}$, now the Higgs branch is lifted. We have
instead $N$ isolated vacua,
\be \sigma=m_a\ \ \ \ ,\ \ \ \ \
|\psi_b|^2=r\delta_{ab}\label{isvac}\ee
These correspond to the $N$ different vortex strings we saw above.

\subsubsection*{A Potential on the Vortex Moduli Space}

We can view the masses $m_i$ as inducing a potential on the Higgs
branch of the vortex theory after integrating out $\sigma$. This
potential is equal to the length of Killing vectors on the Higgs
branch associated to the $U(1)^{N-1}\subset SU(N)_{\rm diag}$
isometry. This is the same story we saw in Lecture 2.6, where the
a vev for $\phi$ induced a potential on the instanton moduli
space.

\para
In fact, just as we saw for instantons, this result can also be
derived directly within the field theory itself \cite{vstring}.
Suppose we fix a vortex configuration $(A_z,q)$ that solves the
vortex equations before we introduce masses. We want to determine
how much the new terms \eqn{vv} lift the energy of this vortex. We
minimize $V$ by solving the equation of motion for $\phi$ in the
background of the vortex,
\be
\D^2\phi=\frac{e^2}{2}\sum_{i=1}^{N}\{\phi,q_iq_i^\dagger\}-2q_iq_i^\dagger
m_i\ee
subject to the vev boundary condition $\phi\rightarrow {\rm
diag}(m_1,\ldots,m_N)$ as $r\rightarrow \infty$. But we have seen
this equation before! It is precisely the equation (3.21) that an
orientational zero mode of the vortex must satisfy. This means
that we can write the excess energy of the vortex in terms of the
relevant orientational zero mode
\be V=\int d^2x \ \frac{2}{e^2}\Tr\,\d A_z\,\d
A_{\bar{z}}+\frac{1}{2}\sum_{i=1}^N\d q_i\,\d q_i^\dagger\ee
for the particular orientation zero mode $\d A_z=\D_z\phi$ and $\d
q_i=i(\phi q_i-q_im_i)$. We can give a nicer geometrical
interpretation to this following the discussion in Section 2.6.
Denote the Cartan subalgebra of $SU(N)_{\rm diag}$ as $\vec{H}$,
and the associated Killing vectors on $\vkn$ as $\vec{k}_\alpha$.
Then, since $\phi$ generates the transformation
$\vec{m}\cdot\vec{H}$, we can express our zero mode in terms of
the basis $\d A_z=(\vec{m}\cdot\vec{k}^\alpha)\,\d_\alpha A_z$ and
$\d q_i=(\vec{m}\cdot\vec{k}^\alpha)\,\d_\alpha q_i$. Putting this
in our potential and performing the integral over the zero modes,
we have the final expression
\be V=
g_{\alpha\beta}\,(\vec{m}\cdot\vec{k}^\alpha)\,(\vec{m}\cdot
\vec{k}^\beta)\label{vkpot}\ee
This potential vanishes at the fixed points of the $U(1)^{N-1}$
action. For the one-vortex moduli space $\CP^{N-1}$, it's not hard
to see that this gives rise to the $N$ vacuum states described
above \eqn{isvac}.

\subsubsection{Monopoles}

To see where the monopoles have gone, it's best if we first look
at the vortex worldsheet theory \cite{monflux}. This is now have a
$d=1+1$ dimensional theory with isolated vacua, guaranteeing the
existence of domain wall, or kink, in the worldsheet. In fact, for
a single $k=1$ vortex, the theory on the worldsheet is precisely
of the form \eqn{dwlag} that we started with at the beginning of
this lecture. (For $k>1$, the presence of the adjoint scalar $Z$
means that isn't precisely the same action, but is closely
related). The equations describing kinks on the worldsheet are the
same as \eqn{dw},
\be \partial_3\sigma=g^2(\sum_{a=1}^N|\psi_a|^2-r)\ \ \ \ ,\ \ \ \
\D_3\psi_a=(\sigma-m_a)\psi_a\ee
where we should take the limit $g^2\rightarrow\infty$, in which
the first equation becomes algebraic. What's the interpretation of
this kink on the worldsheet? We can start by examining its mass,
\be M_{\rm
kink}=(\vec{m}\cdot\vec{g})\,r=\frac{2\pi}{e^2}\,(\vec{\phi}\cdot\vec{g})=M_{\rm
mono}\label{kmequal}\ee
So the kink has the same mass as the monopole! In fact, it also
has the same quantum numbers. To see this, recall that the
different vacua on the vortex string correspond to flux tubes
lying in different $U(1)\subset U(N)$ subgroups. For example, for
$N=2$, the kink must take the form shown in figure 27.
\begin{figure}[htb]
\begin{center}
\epsfxsize=3.7in\leavevmode\epsfbox{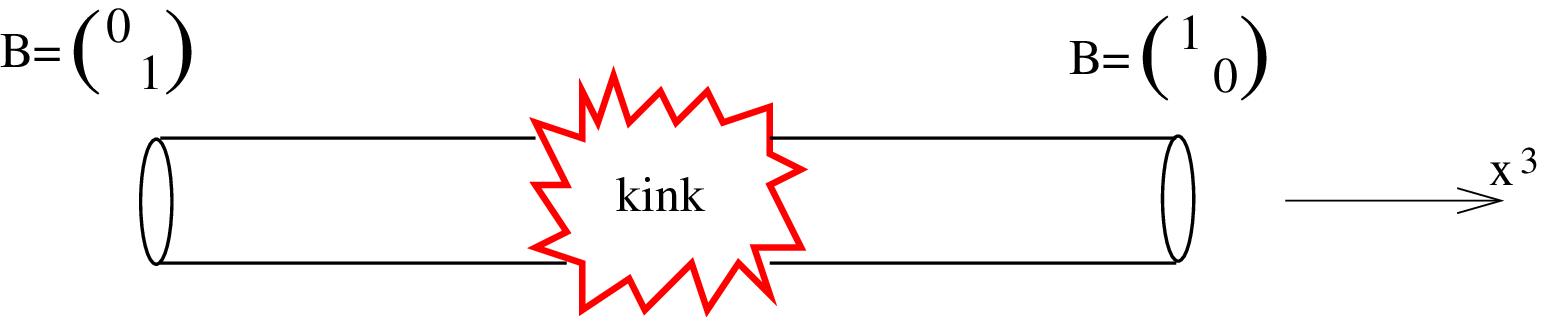}
\end{center}
\caption{The kink on the vortex string.}
\end{figure}
So whatever the kink is, it must soak up magnetic field $B={\rm
diag}(0,1)$ and spit out magnetic field $B={\rm diag}(1,0)$. In
other words, it is a source for the magnetic field $B={\rm
diag}(1,-1)$. This is precisely the magnetic field sourced by an
$SU(2)$ 't Hooft-Polyakov monopole.

\para
What's happening here? We are dealing with a theory with a mass
gap, so any magnetic monopole that lives in the bulk can't emit a
long-range radial magnetic field since the photon can't propagate.
We're witnessing the Meissner effect in a non-abelian
superconductor. The monopole is confined, its magnetic field
departing in two semi-classical flux tubes. This effect is, of
course, well known and it is conjectured that a dual effect leads
to the confinement of quarks in QCD. Here we have a simple,
semi-classical realization in which to explore this scenario.

\para
Can we find the monopole in the $d=3+1$ dimensional bulk? Although
no solution is known, it turns out that we can write down the
Bogomoln'yi equations describing the configuration \cite{monflux}.
Let's go back to our action \eqn{dwlag} and complete the square in
a different way. We now insist only that $\partial_0=A_0=0$, and
write the Hamiltonian as,
\be {\cal H}&=&\int d^3x\ \frac{1}{e^2}\Tr\
\left[(\D_1\phi+B_1)^2+(\D_2+B_2)^2+(\D_3\phi+B_3-\frac{e^2}{2}
(\sum_{i=1}^{N}q_iq_i^\dagger-v^2))^2\right]\nn\\
&&\ \ \ \
+\sum_{i=1}^N|(\D_1-i\D_2)q_i|^2+\sum_{i=1}^{N}|\D_3q_i-(\phi-m_i)q_i|^2+
\Tr\ [-v^2B_3-\frac{2}{e^2}\partial_i(\phi B_i)]\nn\\ &\geq &
\left(\int dx^3\ T_{\rm vortex}\right)+M_{\rm mono}\ee
where the inequality is saturated when the terms in the brackets
vanish,
\be \D_1\phi+B_1=0\ \ \ \ \ &,&\ \ \ \ \ \D_1q_i=i\D_2q_i\nn\\
\D_2\phi+B_2=0\ \ \ \ \ &,&\ \ \ \ \ \D_3q_i=(\phi-m_i)q_i\label{master2}\\
\D_3\phi+B_3&=&\frac{e^2}{2}(\sum_{i=1}^Nq_iq_i^\dagger-v^2)
\nn\ee
As you can see, these are an interesting mix of the monopole
equations and the vortex equations. In fact, they also include the
domain wall equations --- we'll see the meaning of this when we
come to discuss the applications. These equations should be
thought of as the master equations for BPS solitons, reducing to
the other equations in various limits. Notice moreover that these
equations are over-determined, but it's simple to check that they
satisfy the necessary integrability conditions to admit solutions.
However, no non-trivial solutions are known analytically. (Recall
that even the solution for a single vortex is not known in closed
form). We expect that there exist solutions that look like figure
28.
\begin{figure}[htb]
\begin{center}
\epsfxsize=3.2in\leavevmode\epsfbox{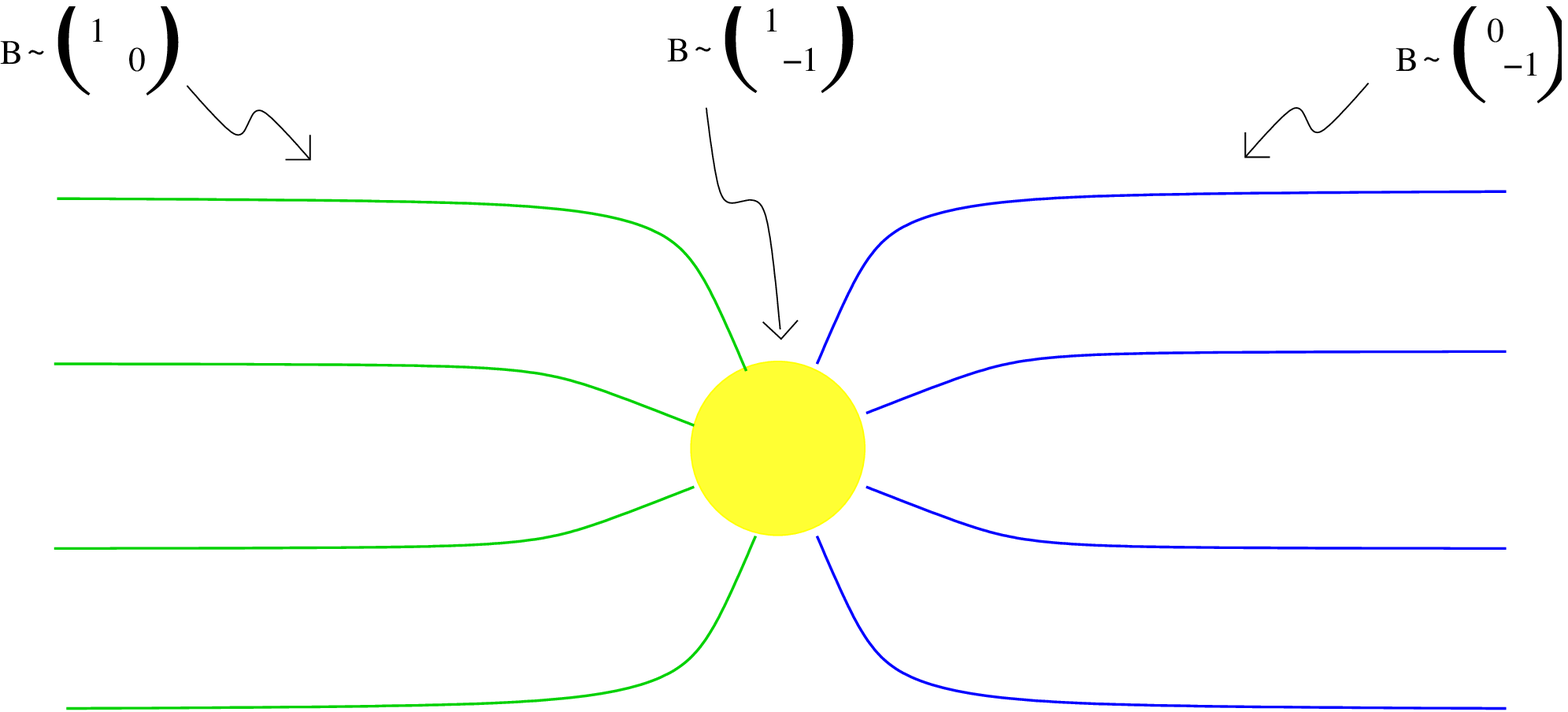}
\end{center}
\caption{The confined magnetic monopole.}
\end{figure}

\para
The above discussion was for $k=1$ and $N_f=N_c$. Extensions to
$k\geq 2$ and also to $N_f\geq N_c$ also exist, although the
presence of the adjoint scalar $Z$ on the vortex worldvolume means
that the kinks on the string aren't quite the same as the domain
wall equations \eqn{dw}. But if we set $Z=0$, so that the strings
lie on top of each other, then the discussion of domain walls in
four-dimensions carries over to kinks on the string. In fact, it's
not hard to check that we've chosen our notation wisely: magnetic
monopoles of charge $\vg$ descend to kinks on the vortex strings
with topological charge $\vg$.

\para
In summary,
\be\mbox{Kink on the Vortex String} = \mbox{Confined Magnetic
Monopole}\nn\ee
The BPS confined monopole was first described in \cite{monflux},
but the idea that kinks on string should be interpreted as
confined monopoles arose previously in \cite{zstring} in the
context of $Z_N$ flux tubes. More recently, confined monopoles
have been explored in several different theories
\cite{kneipp,austring,markov,evslin}. We'll devote Section 4.7 to
more discussion on this topic.

\subsubsection{Instantons}

We now ask what became of instantons. At first glance, it doesn't
look promising for the instanton! In the bulk, the FI term $v^2$
breaks the gauge group, causing the instanton to shrink. And the
presence of the masses means that even in the center of various
solitons, there's only a $U(1)$ restored, not enough to support an
instanton. For example, an instanton wishing to nestle within the
core of the vortex string shrinks to vanishing size and it looks
as if the theory \eqn{dwlag} admits only singular, small
instantons.

\para
While the above paragraph is true, it also tells us how we should
change our theory to allow the instantons to return: we should
consider non-generic mass parameters, so that the $SU(N_f)$ flavor
symmetry isn't broken to the maximal torus, but to some
non-abelian subgroup. Let's return to the example discussed in
Section 4.5: $U(2)$ gauge theory with $N_f=4$ flavors. Rather than
setting all masses to be different, we chose $m_1=m_2=m$ and
$m_3=m_4=-m$. In this limit, the breaking of the flavor symmetry
is $SU(4)\rightarrow S[U(2)\times U(2)]$, and this has interesting
consequences.

\para
To find our instantons, we look at the domain wall which
interpolates between the two vacua $\phi=m{\bf 1}_2$ and
$\phi=-m{\bf 1}_2$. When all masses were distinct, this domain
wall had 8 collective coordinates which had the interpretation of
the position and phase of 4 elementary domain walls \eqn{expos}.
Now that we have non-generic masses, the domain wall retains all 8
collective coordinates, but some develop a rather different
interpretation: they correspond to new orientation modes in the
unbroken flavor group. In this way, part of the domain wall theory
becomes the $SU(2)$ chiral Lagrangian \cite{sydbrane2}.

\para
Inside the domain wall, the non-abelian gauge symmetry is
restored, and the instantons may safely nestle there, finding
refuge from the symmetry breaking of the bulk. One can show that,
from the perspective of the domain wall worldvolume theory, they
appear as Skyrmions \cite{skyrme}. Indeed, closer inspection
reveals that the low-energy dynamics of the domain wall also
includes a four derivative term necessary to stabilize the
Skyrmion, and one can successfully compare the action of the
instanton and Skyrmion. The relationship between instantons and
Skyrmions was first noted long ago by Atiyah and Manton
\cite{atiyahman}, and has been studied recently in the context of
deconstruction \cite{hill,hill2,son}.

\subsection{The Quantum Vortex String}

So far our discussion has been entirely classical. Let's now turn
to the quantum theory. We have already covered all the necessary
material to explain the main result. The basic idea is that
$d=1+1$ worldsheet theory on the vortex string captures quantum
information about the $d=3+1$ dimensional theory in which it's
embedded.  If we want certain information about the 4d theory, we
can extract it using much simpler calculations in the 2d
worldsheet theory.

\para
I won't present all the calculations here, but instead simply give
a flavor of the results \cite{sy,vstring}. The precise
relationship here holds for $\N=2$ theories in $d=3+1$,
corresponding to $\N=(2,2)$ theories on the vortex worldsheet. The
first hint that the 2d theory contains some information about the
4d theory in which its embedded comes from looking at the
relationship between the 2d FI parameter and the 4d gauge
coupling,
\be r=\frac{4\pi}{e^2}\label{rtwopie}\ee
This is a statement about the classical vortex solution. Both
$e^2$ in 4d and $r$ in 2d run at one-loop. However, the
relationship \eqn{rtwopie} is preserved under RG flow since the
beta functions computed in 2d and 4d coincide,
\be
r(\mu)=r_0-\frac{N_c}{2\pi}\log\left(\frac{\mu_{UV}}{\mu}\right)
\ee
This ensures that both 4d and 2d theories hit strong coupling at
the same scale $\Lambda=\mu\exp(-2\pi r/N_c)$.

\para
Exact results about the 4d theory can be extracted using the
Seiberg-Witten solution \cite{sw}. In particular, this allows us
to determine the spectrum of BPS states in the theory. Similarly,
the exact spectrum of the 2d theory can also be determined by
computing the twisted superpotential \cite{nick,nick2}. The
punchline is that the spectrum of the two theories coincide. Let's
see what this means. We saw in \eqn{kmequal} that the classical
kink mass coincides with the classical monopole mass
\be M_{\rm kink}=M_{\rm mono}\label{wonder}\ee
This equality is preserved at the quantum level. Let me stress the
meaning of this. The left-hand side is computed in the $d=1+1$
dimensional theory. When $(m_i-m_j) \gg \Lambda$, this theory is
weakly coupled and $M_{\rm kink}$ receives a one-loop correction
(with, obviously, two-dimensional momenta flowing in the loop).
Although supersymmetry forbids higher loop corrections, there are
an infinite series of worldsheet instanton contributions. The
final expression for the mass of the kink schematically of the
form,
\be M=M_{\rm clas}+M_{\rm one-loop}+\sum_{n=1}^\infty M_{\rm
n-inst}\ee
The right-hand-side of \eqn{wonder} is computed in the $d=3+1$
dimensional theory, which is also weakly coupled for
$(m_i-m_J)\gg\Lambda$. The monopole mass $M_{\rm mono}$ receives
corrections at one-loop (now integrating over four-dimensional
momenta), followed by an infinite series of Yang-Mills instanton
corrections. {\it And term by term these two series agree}!

\para
The agreement of the worldsheet and Yang-Mills instanton
expansions apparently has its microscopic origin in the results if
the previous lecture. Recall that performing an instanton
computation requires integration over the moduli space (${\cal V}$
for the worldsheet instantons; ${\cal I}$ for Yang-Mills).
Localization theorems hold when performing the integrals over
${\cal I}_{k,N}$ in ${\cal N}=2$ super Yang-Mills, and the final
answer contains contributions from only a finite number of points
in ${\cal I}_{k,N}$ \cite{nekr}. It is simple to check that all of
these points lie on ${\cal V}_{k,N}$ which, as we have seen, is a
submanifold of ${\cal I}_{k,N}$.

\para
The equation \eqn{wonder} also holds in strong coupling regimes of
the 2d and 4d theories where no perturbative expansion is
available. Nevertheless, exact results allow the masses of BPS
states to be computed and successfully compared. Moreover, the
quantum correspondence between the masses of kinks and monopoles
is not the only agreement between the two theories. Other results
include:

\begin{itemize}
\item The elementary internal excitations of the string can be
identified with W-bosons of the 4d theory. When in the bulk, away
from the string, these W-bosons are non-BPS. But they can reduce
their mass by taking refuge in the core of the vortex whereupon
they regain their BPS status. \end{itemize}

This highlights an important point: the spectrum of the 4d theory,
both for monopoles and W-bosons, is calculated in the Coulomb
phase, when the FI parameter $v^2=0$. However, the vortex string
exists only in the Higgs phase $v^2\neq 0$. What's going on? A
heuristic explanation is as follows: inside the vortex, the Higgs
field $q$ dips to zero and the gauge symmetry is restored. The
vortex theory captures information about the 4d theory on its
Coulomb branch.

\begin{itemize}
\item As we saw in Sections 2.3 and 4.4, both the  4d theory and
the 2d theory contain dyons. We've already seen that the spectrum
of both these objects is given by the "square-root" formula (2.41)
and \eqn{dwdyonspec}. Again, these agree at the quantum level.

\item Both theories manifest the Witten effect: adding a theta
angle to the 4d theory induces an electric charge on the monopole,
shifting its mass. This also induces a theta angle on the vortex
worldsheet and, hence, turns the kinks into dyons.

\item We have here described the theory with $N_f=N_c$. For
$N_f>N_C$, the story can be repeated and again the spectrum of the
vortex string coincides with the spectrum of the 4d theory in
which it's embedded.
\end{itemize}

\para
In summary, we have known for over 20 years that gauge theories in
4d share many qualitative features with sigma models in 2d,
including asymptotic freedom, a dynamically generated mass gap,
large $N$ expansions, anomalies and the presences of instantons.
However, the vortex string provides a quantitative relationship
between the two: in this case, they share the same quantum
spectrum.

\subsection{The Brane Construction}

In Lecture 3, we derived the brane construction for $U(N_c)$ gauge
theory with $N_f$ hypermultiplets. To add masses, one must
separate the hypermultiplets in the $x^4$ direction. One can now
see the number of vacua \eqn{nvac} since each of the $N_c$
D4-branes must end on one of the $N_f$ D6-branes.
\begin{figure}[htb]
\begin{center}
\epsfxsize=4.7in\leavevmode\epsfbox{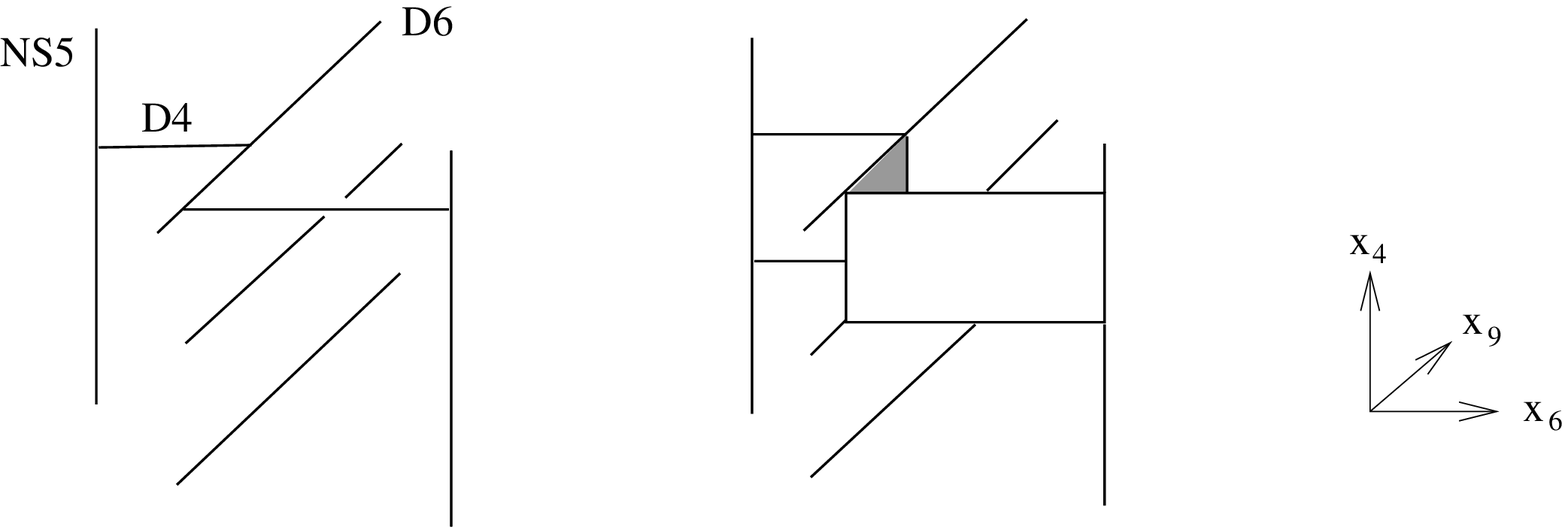}
\end{center}
\caption{The D-brane configuration for an elementary
$\vg=\valpha_1$ domain wall when $N_c=1$ and $N_f=3$.}
\end{figure}

\para
To describe a domain wall, the D4-branes must start in one vacua,
$\Xi_-$ at $x_3\rightarrow-\infty$, and interpolate to the final
vacua $\Xi_+$ as $x^3\rightarrow +\infty$. Viewing this integrated
over all $x^3$, we have the picture shown in figure 29.
To extract the dynamics of domain walls, we need to understand the
worldvolume theory of the curved D4-brane. This isn't at all
clear. Related issues have troubled previous attempts to extract
domain wall dynamics from D-brane set-up \cite{kinky,sakaidbrane},
although some qualitative features can be seen. However, we can
make progress by studying this system in the limit
$e^2\rightarrow\infty$, so that the two NS5-branes and the $N_f$
D6-branes lie coincident in the $x^6$ direction \cite{mdw}. The
portions of the D4-branes stretched in $x^6$ vanish, and we're
left with D4-branes with worldvolume $01249$, trapped in squares
in the $49$ directions where they are sandwiched between the NS5
and D6-branes. Returning to the system of domain walls in an
arbitrary topological sector $\vg=\sum_i n_i\valpha_i$, we have
the system drawn in figure 30.
\begin{figure}[htb]
\begin{center}
\epsfxsize=4.7in\leavevmode\epsfbox{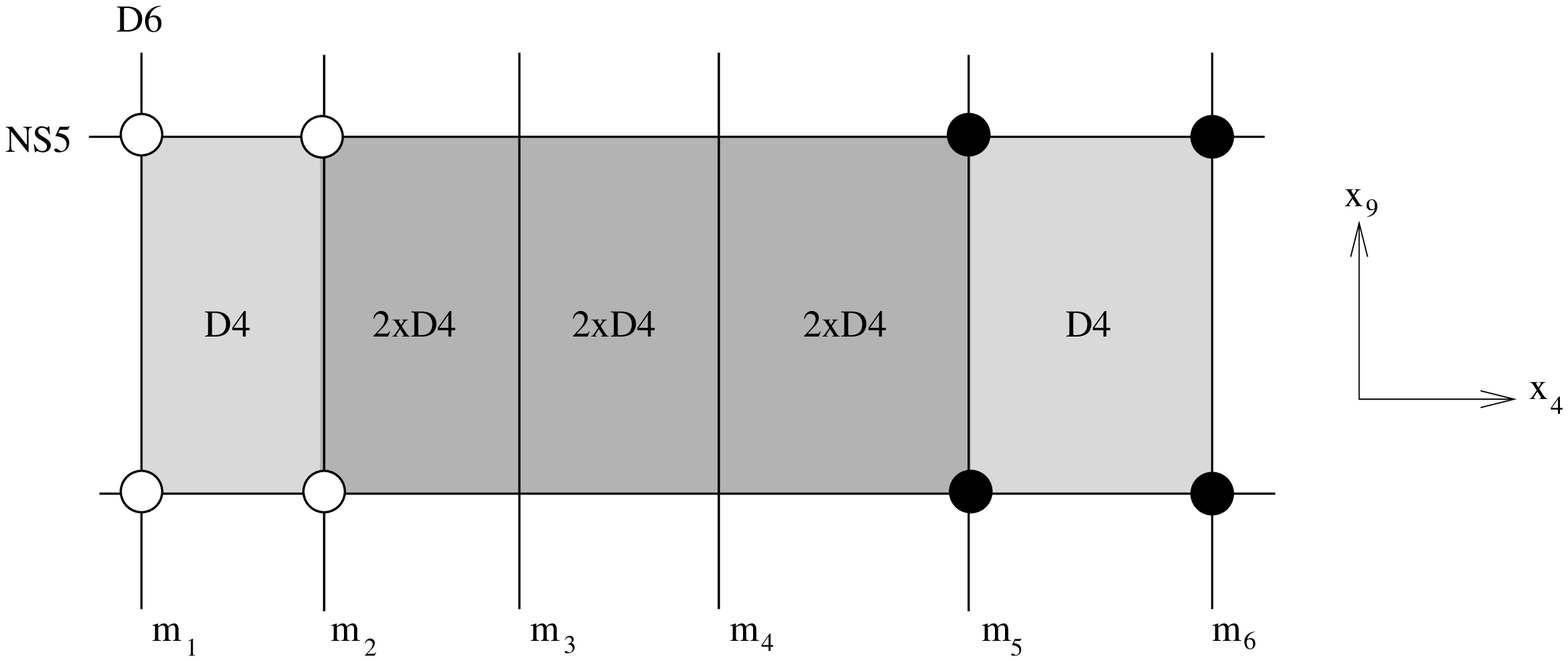}
\end{center}
\caption{The D-brane configuration in the $e^2\rightarrow\infty$
limit.}
\end{figure}

\para
We can now read off the gauge theory living on the D4-branes. One
might expect that it is of the form $\prod_i U(n_i)$. This is
essentially correct. The NS5-branes project out the $A_9$
component of the gauge field, however the $A_4$ component survives
and each $U(n_a)$ gauge theory lives in the interval $m_i\leq
x_4\leq m_{i+1}$. In each segment, we have $A_4$ and $X_3$, each
an $n_i\times n_i$ matrix. These fields satisfy
\be \frac{dX_3}{dx^4}-i[A_4,X_3]=0 \label{dwnahm}\ee
modulo $U(n_i)$ gauge transformations acting on the interval
$m_i\leq x_4\leq m_{i+1}$, and vanishing  at the boundaries. These
equations are kind of trivial: the interesting details lie in the
boundary conditions. As in the case of monopoles, the interactions
between neighbouring segments depends on the relative size of the
matrices:

\para
\underline{\ $n_i=n_{i+1}$:\ } The $U(n_i)$ gauge symmetry is
extended to the interval $m_i\leq x_4\leq m_{i+2}$ and an impurity
is added to the right-hand-side of Nahm's equations, which now
read
\be \frac{dX_3}{dx_4}-i[A_4,X_3]=
\psi\psi^\dagger\delta(x_4-m_{i+1}) \label{dwimp}\ee
where the impurity degree of freedom $\psi$ transforms in the
fundamental representation of the $U(n_i)$ gauge group, ensuring
the combination $\psi\psi^\dagger$ is a $n_i\times n_i$ matrix
transforming, like $X_1$, in the adjoint representation. These
$\psi$ degrees of freedom are chiral multiplets which survive the
NS5-brane projection.

\para
\underline{\ $n_i=n_{i+1}-1$:\ } In this case $X_3\rightarrow
(X_3)_-$, a $n_i\times n_i$ matrix, as $x_4\rightarrow (m_i)_-$
from the left. To the right of $m_i$, $X_3$ is a $(n_i+1)\times
(n_i+1)$ matrix obeying
\be X_3\rightarrow \left(\begin{array}{cc} y & a^\dagger \\ a &
(X)_-
\end{array}\right)\ \ \ \ \ {\rm as\ } x_4\rightarrow (m_i)_+
\label{bc}\ee
where $y_\mu\in\R$ and each $a_\mu$ is a complex $n_i$-vector. The
obvious analog of this boundary condition holds when
$n_i=n_{i+1}+1$.

\para
These boundary conditions are obviously related to the Nahm
boundary conditions for monopoles that we met in Lecture 2.

\subsubsection{The Ordering of Domain Walls Revisited}

We now come to the important point: the ordering of domain walls.
Let's see how the brane construction captures this. We can use the
gauge transformations to make $A_4$ constant over the interval
$m_i\leq x^4\leq m_{i+1}$. Then \eqn{dwnahm} can be trivially
integrated in each segment to give
\be X_3(x^4)=e^{iA_4x^4}\hat{X}_3e^{-iA_4x^4}\ee
Then the positions of the $\valpha_i$ domain walls are given by
the eigenvalues of $X_3$ restricted to the interval $m_i\leq
x_4\leq m_{i+1}$. Let us denote this matrix as $X_3^{(i)}$ and the
eigenvalues as $\lambda^{(i)}_m$, where $m=1,\ldots n_i$. We have
similar notation for the $\valpha_{i+1}$ domain walls. Suppose
first that $n_i=n_{i+1}$. Then the impurity \eqn{dwimp} relates
the two sets of eigenvalues by the jumping condition
\be X_1^{(i+1)}=X_1^{(i)}+\psi\psi^\dagger \label{jump}\ee
We will now show that this jumping condition \eqn{jump} correctly
captures the interlacing nature of neighboring domain walls.

\para
To see this, consider firstly the situation in which
$\psi^\dagger\psi\ll \Delta\lambda^{(i)}_{m}$ so that the matrix
$\psi\psi^\dagger$ may be treated as a small perturbation of
$X_1^{(i)}$. The positivity of $\psi\psi^\dagger$ ensures that
each $\lambda_m^{(i+1)}\geq \lambda_m^{(i)}$. Moreover, it is
simple to show that the $\lambda_m^{(i+1)}$ increase monotonically
with $\psi^\dagger\psi$. This leaves us to consider the other
extreme, in which $\psi^\dagger\psi\rightarrow \infty$. It this
limit $\psi$ becomes one of the eigenvectors of $X_1^{(i+1)}$ with
corresponding eigenvalue $\lambda^{(i+1)}_{n_i}=\psi^\dagger\psi$,
corresponding to the limit in which the last domain wall is taken
to infinity. What we want to show is that the remaining $n_i-1$
$\valpha_{i+1}$ domain walls are trapped between the $n_i$
$\valpha_i$ domain walls as depicted in figure 26. Define the
$n_i\times n_i$ projection operator
\be P=1-\hat{\psi}\hat{\psi}^\dagger \ee
where $\hat{\psi}=\psi/\sqrt{\psi^\dagger\psi}$. The positions of
the remaining $(n_i-1)$ $\valpha_{i+1}$ domain walls are given by
the (non-zero) eigenvalues of $PX_1^{(i)}P$. We must show that,
given a rank $n$ hermitian matrix $X$, the eigenvalues of $PXP$
are trapped between the eigenvalues of $X$. This well known
property of hermitian matrices is simple to show:
\be
\det(PXP-\mu)&=&\det(XP-\mu) \nn\\
&=&\det(X-\mu-X\hat{\psi}\hat{\psi}^\dagger)\nn\\ &=&\det(X-\mu)
\det(1-(X-\mu)^{-1}X\hat{\psi}\hat{\psi}^\dagger)\nn\ee
Since $\hat{\psi}\hat{\psi}^\dagger$ is rank one, we can write
this as
\be \det(PXP-\mu)&=&
\det(X-\mu)\,[1-\Tr((X-\mu)^{-1}X\hat{\psi}\hat{\psi}^\dagger)]\nn\\
&=&
-\mu\,\det(X-\mu)\,\Tr((X-\mu)^{-1}\hat{\psi}\hat{\psi}^\dagger)
\nn\\ &=&
-\mu\left[\prod_{m=1}^n(\lambda_m-\mu)\right]\,\left[\sum_{m=1}^n
\frac{|\hat{\psi}_m|^2}{\lambda_m-\mu}\right] \ee where
$\hat{\psi}_m$ is the $m^{\rm th}$ component of the vector $\psi$.
We learn that $PXP$ has one zero eigenvalue while, if the
eigenvalues $\lambda_m$ of $X$ are distinct, then the eigenvalues
of $PXP$ lie at the roots the function
\be R(\mu)=\sum_{m=1}^n\frac{|\hat{\psi}_m|^2}{\lambda_m-\mu}\ee
The roots of $R(\mu)$ indeed lie between the eigenvalues
$\lambda_m$. This completes the proof that the impurities
\eqn{dwimp} capture the correct ordering of the domain walls.

\para
The same argument shows that the boundary condition \eqn{bc} gives
rise to the correct ordering of domain walls when $n_{i+1}=n_i+1$,
with the $\valpha_{i}$ domain walls interlaced between the
$\valpha_{i+1}$ domains walls. Indeed, it is not hard to show that
\eqn{bc} arises from \eqn{dwimp} in the limit that one of the
domain walls is taken to infinity.

\subsubsection{The Relationship to Monopoles}

You will have noticed that the brane construction above is closely
related to the Nahm construction we discussed in Lecture 2. In
fact, just as the vortex moduli space $\vkn$ is related to the
instanton moduli space $\ikn$, so the domain wall moduli space
$\wg$ is related to the monopole moduli space $\mg$. The domain
wall theory is roughly a subset of the monopole theory.
Correspondingly, the domain wall moduli space is a complex
submanifold of the monopole moduli space. To make this more
precise, consider the isometry rotating the monopoles in the
$x^1-x^2$ plane (mixed with a suitable $U(1)$ gauge action). It we
denote the corresponding Killing vector as $h$, then
\be \W_{\vg}\cong\left.\M_{\vg}\right|_{h=0} \ee
This is the analog of equation (3.36), relating the vortex and
instanton moduli spaces.

\para
Nahm's equations have appeared previously in describing domain
walls in the ${\cal N}=1^\star$ theory \cite{boris}. I don't know
how those domain walls are related to the ones discussed here.

\subsection{Applications}

We've already seen one application of kinks in section 4.7,
deriving a relationship between 2d sigma models and 4d gauge
theories. I'll end with a couple of further interesting
applications.

\subsubsection{Domain Walls and the 2d Black Hole}

Recall that we saw in Section 4.3.2 that the relative moduli space
of a two domain walls with charge $\vg=\valpha_1+\valpha_2$ is the
cigar shown in  figure 22. Suppose we consider domain walls as
strings in a $d=2+1$ dimensional theory, so that the worldvolume
of the domain walls is $d=1+1$ dimensional. Then the low-energy
dynamics of two domain walls is described by a sigma-model on the
cigar.

\para
There is a very famous conformal field theory with a cigar target
space. It is known as the two-dimensional black hole \cite{cigar}.
It has metric,
\be ds^2_{BH}=k^2[dR^2+\tanh^2R\ d\theta^2]\label{bhmetric}\ee
The non-trivial curvature at the tip of the cigar is cancelled by
a dilaton which has the profile
\be \Phi=\Phi_0-2\cosh R\ee
So is the dynamics of the domain wall system determined by this
conformal field theory? Well, not so obviously: the metric on the
domain wall moduli space $\W_{\valpha_1+\valpha_2}$ does not
coincide with \eqn{bhmetric}. However, $d=1+1$ dimensional theory
is not conformal and the metric flows as we move towards the
infra-red. There is a subtlety with the dilaton which one can
evade by endowing the coordinate $R$ with a suitable anomalous
transformation under RG flow. With this caveat, it can be shown
that the theory on two domain walls in $d=2+1$ dimensions does
indeed flow towards the conformal theory of the black hole with
 the identification $k=2v^2/m$ \cite{mmotw}.

\para
The conformal field theory of the 2d black hole is dual to
Liouville theory \cite{fzz,kkk}. If we deal with supersymmetric
theories, this $\N=(2,2)$ conformal field theory has Lagrangian
\be L_{Liouville}=\int d^4\theta\ \frac{1}{2k}|Y|^2
+\frac{\mu}{2}\int d^2\theta\ e^{-Y} + {\rm
h.c.}\label{liouville}\ee
and the equivalence between the two theories was proven using the
techniques of mirror symmetry in \cite{horikap}.  In fact, one can
also prove this duality by studying the dynamics of domain walls.
Which is rather cute. We work with the $\N=4$ (eight supercharges)
$U(1)$ gauge theory in $d=2+1$ with $N_f$ charged hypermultiplets.
As we sketched above, if we quantize the low-energy dynamics of
the domain walls, we find the $\N=(2,2)$ conformal theory on the
cigar. However, there is an alternative way to proceed: we could
choose first to integrate out some of the matter in three
dimensions. Let's get rid of the charged hypermultiplets to leave
a low-energy effective action for the vector multiplet. As well as
the gauge field, the vector multiplet contains a triplet of real
scalars $\bphi$, the first of which is identified with the $\phi$
we met in \eqn{dwlag}. The low-energy dynamics of this effective
theory in $d=2+1$ dimensions can be shown to be
\be
L_{eff}=H(\bphi)\,\partial_\mu\bphi\cdot\partial^\mu\bphi+H^{-1}(\bphi)
(\partial_\mu\sigma + \bomega\cdot
\partial_\mu\bphi)^2-v^4H^{-1}\ee
Here $\sigma$ is the dual photon  (see (2.62)) and
$\nabla\times\bomega=\nabla H$, while the harmonic function $H$
includes the corrections from integrating out the $N_f$
hypermultiplets,
\be H(\bphi) = \frac{1}{e^2}+\sum_{i=1}^{N_f}\frac{1}{|\bphi-{\bf
m}_i|}\ee
where each triplet ${\bf m}_i$ is given by ${\bf m}_i=(m_i,0,0)$.
We can now look for domain walls in this $d=2+1$ effective theory.
Since we want to study two domain walls, let's set $N_f=3$. We see
that the theory then has three, isolated vacua, at
$\bphi=(\phi,0,0)=(m_i,0,0)$.

\para
We now want to study the domain wall that interpolates between the
two outer vacua $\phi=m_1$ and $\phi=m_3$. It's not hard to show
that, in contrast to the microscopic theory \eqn{dwlag}, there is
no domain wall solutions interpolating between these vacua. One
can find a $\valpha_1$ domain wall interpolating between
$\phi=m_1$ and $\phi=m_2$. There is also a $\valpha_2$ domain wall
interpolating between $\phi=m_2$ and $\phi=m_3$. But no
$\valpha_1+\valpha_2$ domain wall between the two extremal vacua
$\phi=m_1$ and $\phi=m_3$. The reason is essentially that only a
single scalar, $\phi$, changes in the domain wall profile, with
equations of motion given by flow equations,
\be\partial_3\phi=v^2 H^{-1}(\phi)\ee
But since we have only  a single scalar field, it must actually
pass through the middle vacuum (as opposed to merely getting
close) at which point the flow equations tell us
$\partial_3\phi=0$ and it doesn't move anymore.

\para
Although there is no solution interpolating between $\phi=m_1$ and
$\phi=m_2$, one can always write down an approximate solution
simply by superposing the $\valpha_1$ and $\valpha_2$ domain walls
in such a way that they are well separated. One can then watch the
evolution of this configuration under the equations of motion and,
from this, extract an effective force between the domain wall
\cite{manforcedw}. For the case in hand, this calculation was
performed in \cite{mmotw}, where it was shown that the force is
precisely that arising from the Liouville Lagrangian
\eqn{liouville}. In this way, we can use the dynamics of domain
walls to derive the mirror symmetry between the cigar and
Liouville theory.

\subsubsection{Field Theory D-Branes}

As we saw in Section 4.3.2 of this lecture, the moduli space of a
single domain wall is $\W\cong\R\times \S^1$. This means that the
theory living on the $d=2+1$ dimensional worldvolume of the domain
wall contains a scalar $X$, corresponding to fluctuations of the
domain wall in the $x^3$ direction, together with a periodic
scalar $\theta$ determining the phase of the wall.
%
%
But in $d=2+1$ dimensions, a periodic scalar can be dualized in
favor of a photon living on the wall $4\pi v^2
\partial_\mu\theta=\epsilon_{\mu\nu\rho}F^{\nu\rho}$.
Thus the low-energy dynamics of the wall can alternatively be
described by a free $U(1)$ gauge theory with a neutral scalar $X$,
\be L_{\rm wall}=\ft12T_{\rm wall}\left((\partial_\mu X)^2 +
\frac{1}{16\pi^2v^4}F_{\mu\nu}F^{\mu\nu}\right)\ee
This is related to the mechanism for gauge field localization
described in \cite{dvali}.

\para
As we have seen above, the theory also contains vortex strings.
These vortex strings can end on the domain wall, where their ends
are electrically charged. In other words, the domain walls are
semi-classical D-branes for the vortex strings. These D-branes
were first studied in \cite{dbrane,sydbrane,sydbrane2}.
(Semi-classical D-brane configurations in other theories have been
studied in \cite{moreed,moreed1, moreed2} in situations without
the worldvolume gauge field). The simplest way to see that the
domain wall is D-brane is using the BIon spike described in
Section 2.7.4,  where we described monopole as D-branes in $d=5+1$
dimensions.

\para
We can also see this D-brane solution from the perspective of the
bulk theory. In fact, the solution obeys the  equations
\eqn{master2} that we wrote down before. To see this, let's
complete the square again but we should be more careful in keeping
total derivatives. In a theory with multiple vacua, we have
\be {\cal H}&=&\int d^3x\ \frac{1}{2e^2}\Tr\
\left[(\D_1\phi+B_1)^2+(\D_2+B_2)^2+(\D_3\phi+B_3-e^2(\sum_{i=1}^{N}
q_iq_i^\dagger-v^2))^2\right]\nn\\ &&
+\sum_{i=1}^N|(\D_1-i\D_2)q_i|^2+\sum_{i=1}^{N}|\D_3q_i-(\phi-m_i)q_i|^2+
\Tr\ [-v^2B_3-\frac{1}{e^2}\partial_i(\phi B_i)+v^2\partial_3\phi]\nn\\
&\geq & \left(\int dx^1dx^2\ T_{\rm wall}\right)+\left(\int dx^3\
T_{\rm vortex}\right)+M_{\rm mono}\label{lastone}\ee
and we indeed find the central charge appropriate for the domain
wall. In fact these equations were first discovered in abelian
theories to describe D-brane objects \cite{sydbrane}.

%
%

\para
These equations have been solved analytically in the limit
$e^2\rightarrow\infty$ \cite{dbrane,allquarter}. Moreover, when
multiple domain walls are placed in parallel along the line, one
can construct solutions with many vortex strings stretched between
them as figure 31, taken from \cite{allquarter}, graphically
illustrates.
\begin{figure}[htb]
\begin{center}
\epsfxsize=3.2in\leavevmode\epsfbox{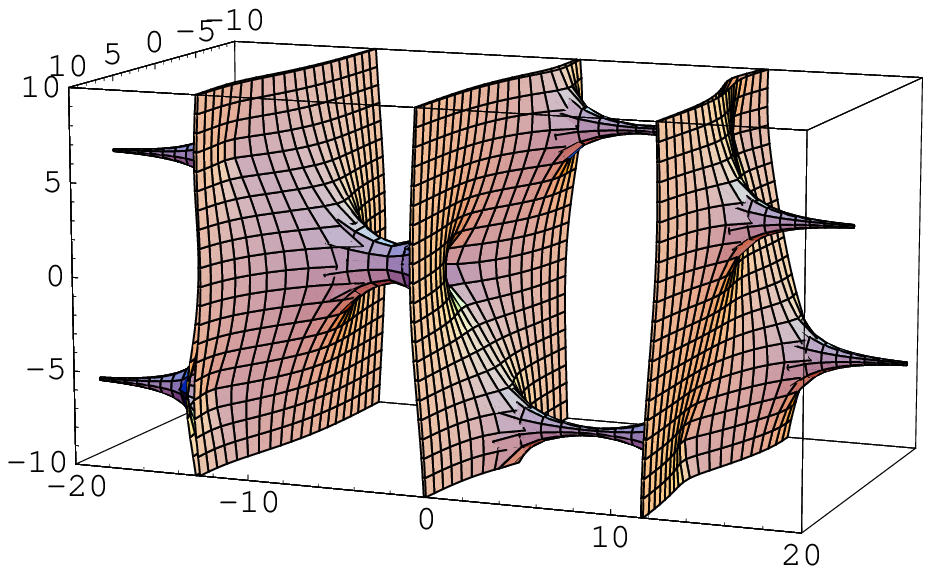}
\end{center}
\caption{Plot of a field theoretic D-brane configuration
\cite{allquarter}.} \label{}
\end{figure}

\para
Some final points on the field theoretic D-branes
\begin{itemize}
\item In each vacuum there are $N_c$ different vortex strings. Not
all of them can end on the bordering domain walls. There exist
selection rules describing which vortex string can end on a given
wall. For the $\valpha_i$ domain wall, the string associated to
$q_i$ can end from the left, while the string associated to
$q_{i+1}$ can end from the right \cite{boojum}. \item For finite
$e^2$, there is a negative binding energy when the string attaches
itself to the domain wall, arising from the monopole central
charge in \eqn{lastone}. Known as a boojum, it was studied in this
context in \cite{boojum,sybooj}. (The name boojum was given by
Mermin to a related configuration in superfluid ${}^3$He
\cite{mermin}).

\item One can develop an open string description of the domain
wall dynamics, in which the motion of the walls is governed by the
quantum effects of new light states that appear as the walls
approach. Chern-Simons interactions on the domain wall worldvolume
are responsible for stopping the walls from passing. Details can
be found in \cite{elsewhere}.
\end{itemize}

\newpage

\section*{Acknowledgements}

These lectures were given in June 2005 at TASI, the Theoretical
Advanced Study Institute held at the University of Colorado in
Boulder. It was a fun week, made especially enjoyable by the sharp
questions and interest of the students. I'd also like to
thank the organizers, Shamit Kachru, K.T. Mahanthappa and Eva
Silverstein, for inviting me to speak. I'm grateful to Jacob
Bourjaily, Mohammad Edalati and Bonnie Parker for offering many comments.
Finally, my thanks to Allan Adams, Nick
Dorey, Greg Moore, Arvind Rajaraman, Moshe Rozali, Paul Townsend
and Erick Weinberg for recent discussions on some of the issues mentioned in
these notes.

\end{document}